\documentclass[%
 reprint,
 amsmath,amssymb,
 aps,
]{revtex4-1}
\usepackage{dcolumn}
\usepackage{bm}
\usepackage[pdftex]{graphicx}
\usepackage{amsmath}
\usepackage{amsfonts}
\usepackage{amssymb}
\usepackage{subfigure}
\usepackage{color}
\usepackage{float}
\usepackage[dvipsnames]{xcolor}
\usepackage{environ}
\NewEnviron{resize}{%
	\scalebox{0.9}{$\BODY$}
}
\usepackage{url}
\usepackage[linktocpage=true]{hyperref}
\usepackage{placeins}
\usepackage{soul}
\usepackage{float}
\usepackage{appendix}
\usepackage{indentfirst}
\begin{document}

\preprint{APS/123-QED}

\title{Non-Uniform Superlattice Magnetic Tunnel Junctions}

\author{Sabarna Chakraborti}
\author{Abhishek Sharma} 
\email{abhishek@iitrpr.ac.in }
\affiliation{Department of Electrical Engineering, Indian Institute of Technology Ropar\\ Nangal Rd, Hussainpur, Rupnagar, Punjab 140001
}


\begin{abstract}
 We propose a new class of non-uniform superlattice magnetic tunnel junctions (Nu-SLTJs) with the Linear, Gaussian, Lorentzian, and P\"oschl-teller width and height based profiles manifesting a sizable enhancement in the TMR($\approx 10^4-10^6\%$) with a significant suppression in the switching bias($\approx$9 folds) owing to the physics of broad-band spin filtering. By exploring the negative differential resistance region in the current-voltage characteristics of the various Nu-SLTJs, we predict the Nu-SLTJs offer fastest spin transfer torque switching in the order of a few hundred picoseconds. We self-consistently employ the atomistic non-equilibrium Green's function formalism coupled with the Landau–Lifshitz–Gilbert–Slonczewski equation to evaluate the device performance of the various Nu-SLTJs. We also present the design of minimal three-barrier Nu-SLTJs having significant TMR($\approx 10^4\%$) and large spin current for ease of device fabrication. We hope that the class of Nu-SLTJs proposed in this work may lay the bedrock to embark on the exhilarating voyage of exploring various non-uniform superlattices for the next generation of spintronic devices.


\end{abstract}

\maketitle


\section{\label{sec:level1}INTRODUCTION}
\begin{figure}[h]
	\centering
		\includegraphics[width=0.85\linewidth]{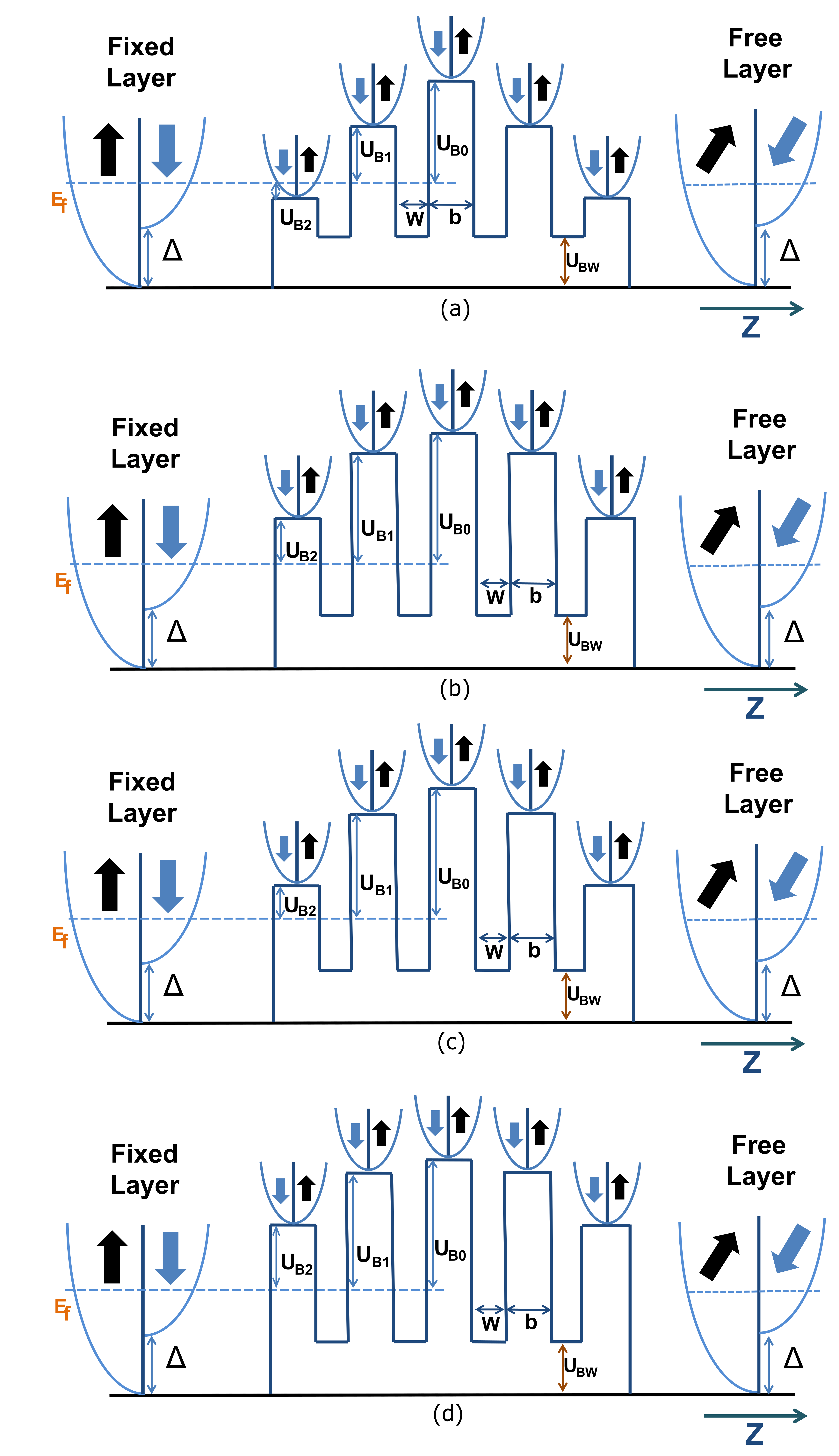}
	\caption{Band diagram of the  Nu-SLTJs with (a) Linear height (LH-SLTJ), (b) Gaussian height (GH-SLTJ), (c) Lorentzian height (LRH-SLTJ) and  (d)P\"oschl-Teller height (PH-SLTJ) profile.}
	\label{fig:BDHS}
\end{figure}

The research in the ambit of magnetic devices has witnessed an augmented proliferation with the development of magnetic tunnel junctions (MTJs) owing to their versatile applications in nanoscale spintronics.
Ever since the pioneering discovery of tunnel magnetoresistance(TMR)\cite{parkin2004giant} and spin transfer torque(STT)\cite{berger1996emission,slonczewski1996current}, MTJs are being explored as the building blocks for energy-efficient integrated circuits of the beyond Moore era. Along with the STT- magnetoresistive random access memory (MRAM) and the spin torque nano oscillators(STNO), the utility of MTJs have been extended across the diverse disciplines of current imaging\cite{le2006tunnel}, magnetoencephalography\cite{kanno2022scalp}, weapon detection\cite{rudin1997investigative}, wireless communication\cite{chen2016spin}, magnetic recording\cite{braganca2010nanoscale} and biomedical imaging\cite{robbes2006highly}. 
\begin{figure}[h]
	\centering
	\subfigure[]{\includegraphics[width=1.6 in]{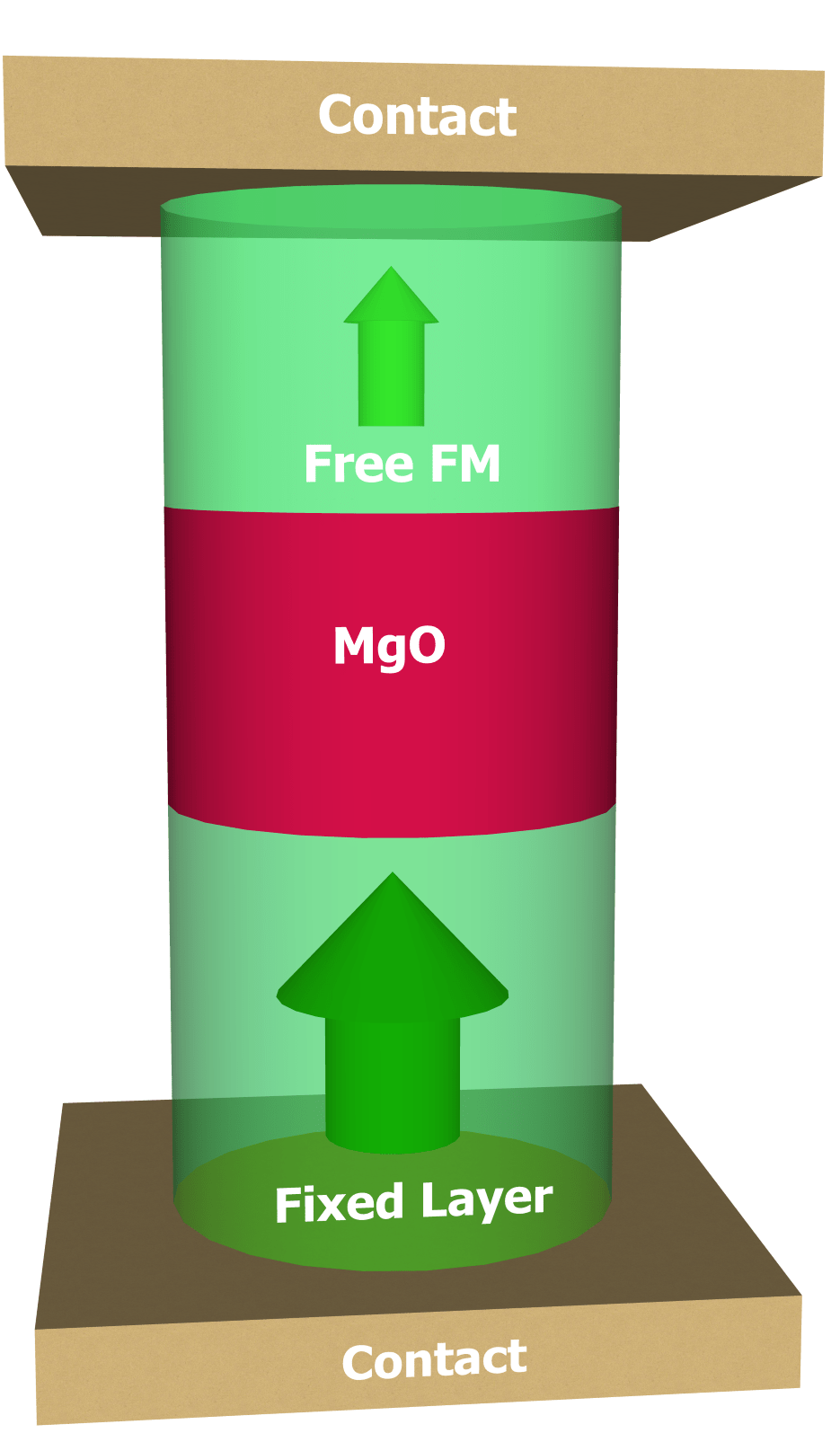}}
	\subfigure[]{\includegraphics[width=1.55 in]{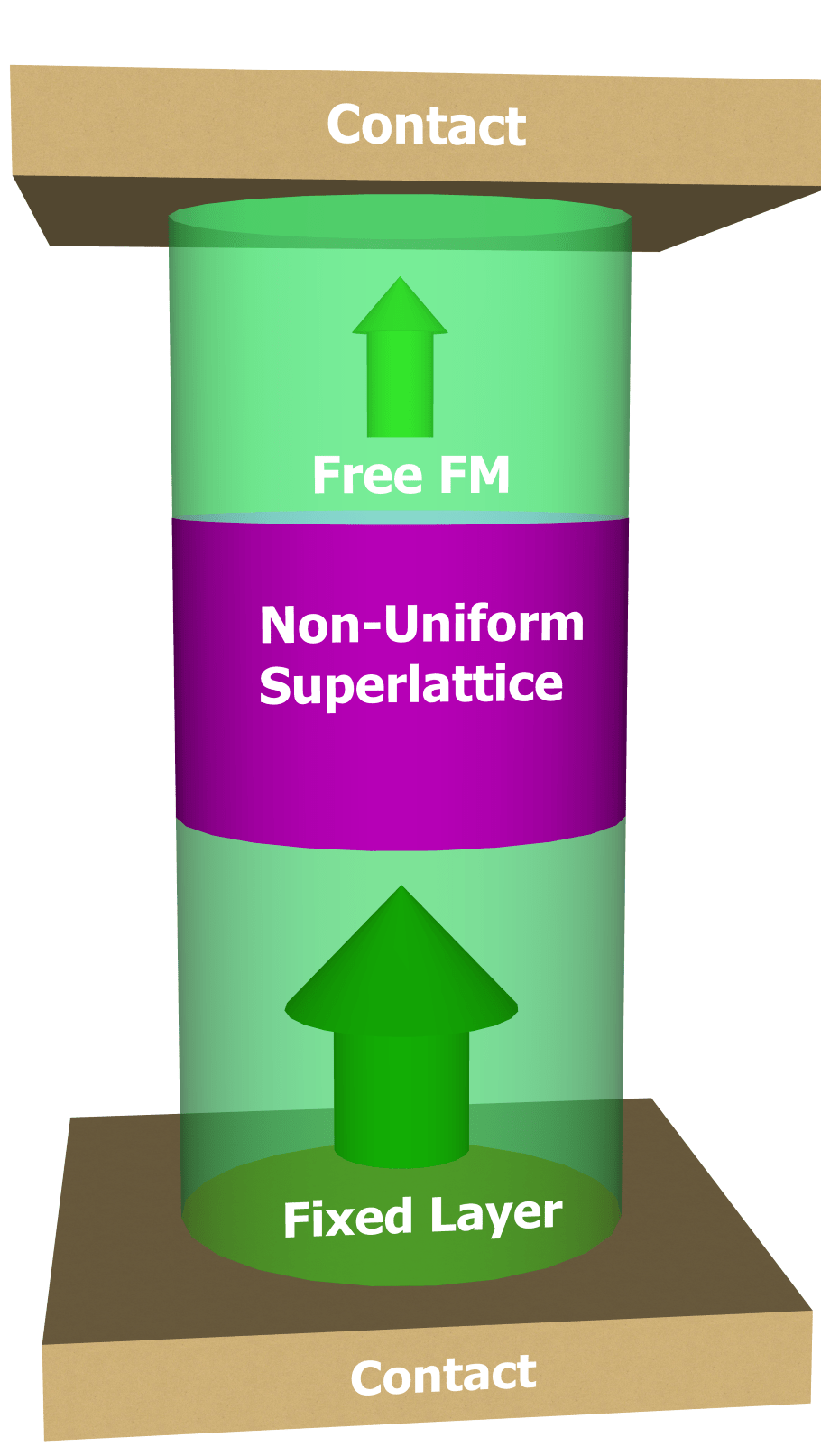}}		
	
	\caption{Device schematics of a (a) trilayer MTJ with MgO as an insulator and (b) a non-uniform superlattice magnetic tunnel junction (Nu-SLTJ) featuring a non-uniform superlattice instead of the oxide layer.}
	\label{fig:PMA-MTJ}
\end{figure}
Besides, the research of materializing the STT-MRAMs and the spin torque nano oscillators(STNOs) have predominantly anchored the novel literature of spintronic devices for their potential applications in the immediate future\cite{krivorotov2005time,choi2014spin,cheng2016terahertz,bhattacharjee2018neel,jiang2004substantial,mangin2006current,diao2007spin,santos2020ultrathin}. A confluence of high-speed operation, non-volatility, durability, and high density along with CMOS compatibility have eventuated the STT-MRAM as a viable candidate for state-of-the-art storage devices\cite{ikeda2007magnetic,apalkov2013spin,khvalkovskiy2013basic,bhatti2017spintronics,prenat2007cmos}. Apart from that, the broad tuning range\cite{bonetti2010experimental}, nanoscopic footprints\cite{villard2009ghz}, and effortless integration with cutting-edge CMOS technology have conferred the STNO a pivotal role in various frequency modulation schemes, wireless radio-frequency communication, and futuristic read heads\cite{zeng2013spin,mizushima2010signal,braganca2010nanoscale}.\\
\begin{figure}[h]
	\centering
	\includegraphics[width=3 in]{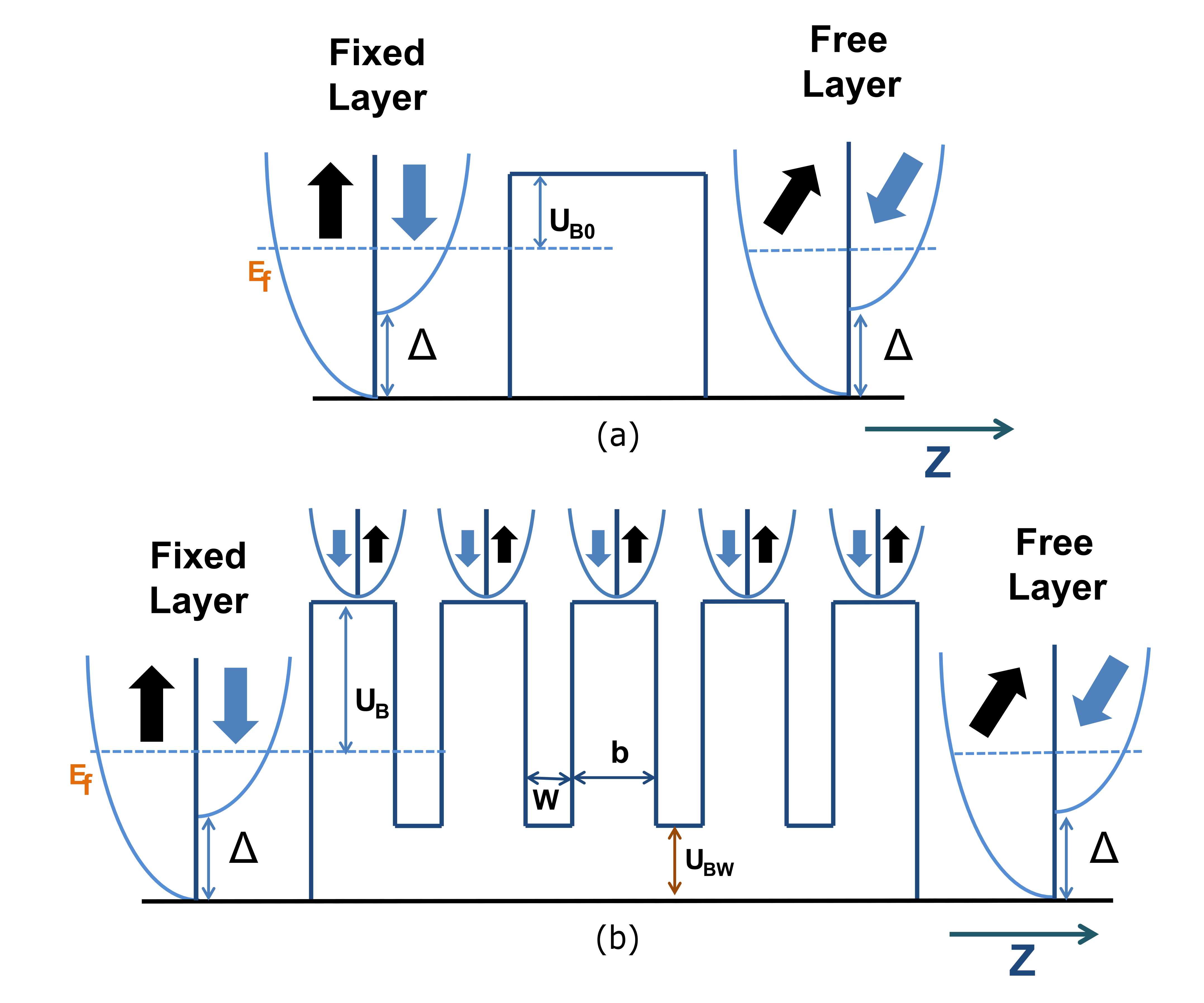}
	\caption{(a) Band diagram of a conventional tri-layer MTJ and (b)a regular superlattice-based MTJ. }
		\label{fig:MTJ_RSLTJ_BD}
\end{figure}
A typical MTJ comprises a free and fixed ferromagnet(FM) separated by an insulating material(MgO)\cite{butler2001spin} as depicted in Fig. \ref{fig:PMA-MTJ}(a). An MTJ offers different resistances in the parallel($R_{PC}$) and anti-parallel($R_{APC}$) configurations of the magnetization of the fixed and free FMs due to spin-dependent tunneling of the electrons, which is quantified by the TMR(\%)$=(R_{PC}-R_{APC})/(R_{APC}) \times 100$\%.
\begin{figure}[h]
	\centering
		\includegraphics[width=0.85\linewidth]{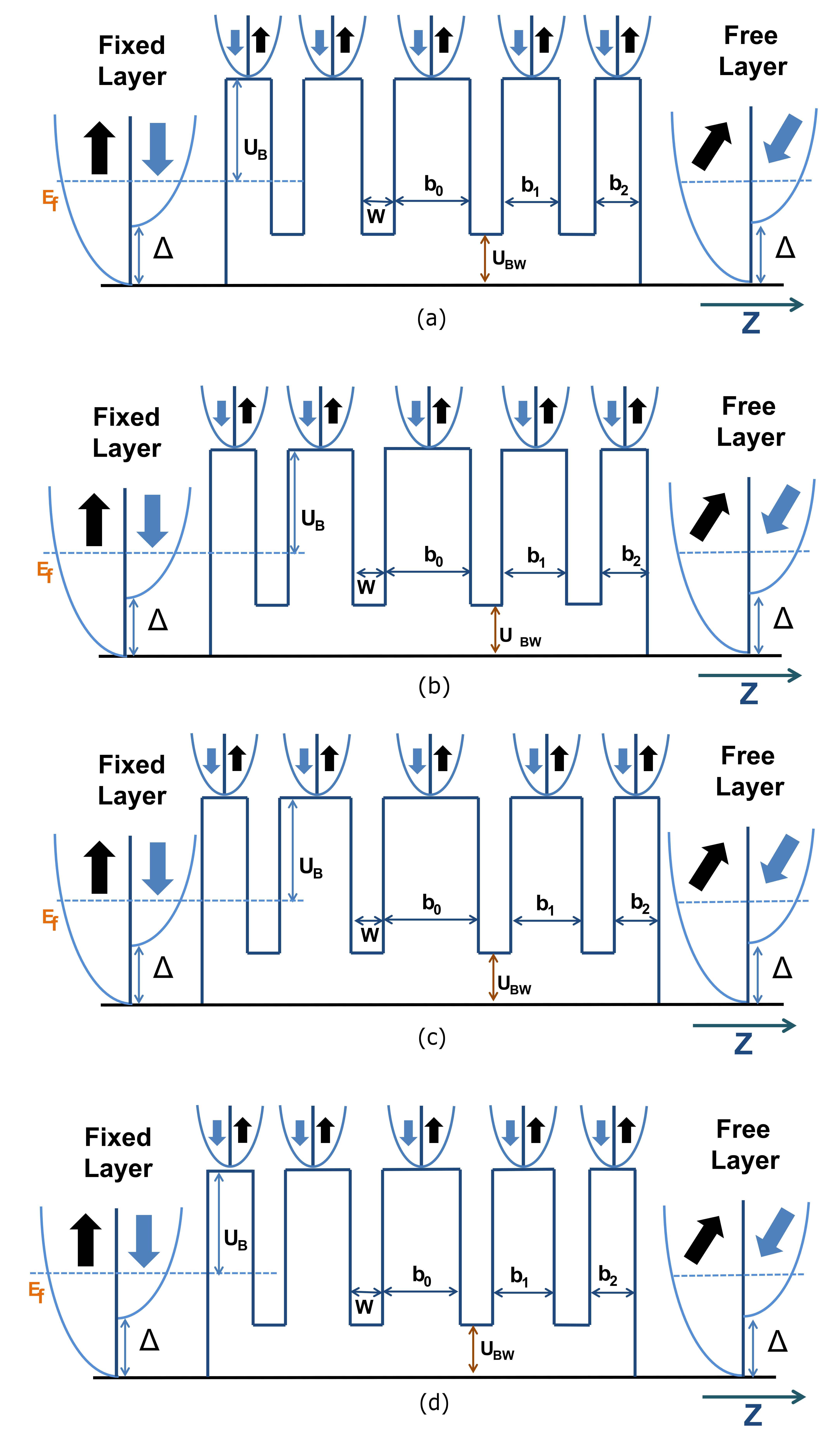}
	\caption{Band diagram of the  Nu-SLTJs with (a) Linear width (LW-SLTJ),(b) Gaussian width (GW-SLTJ), (c) Lorentzian width (LRW-SLTJ) and  (d)P\"oschl-Teller width (PW-SLTJ) profiles.}
	\label{fig:BDWS}
\end{figure}
The magnetization of the free FM can be switched using the STT exerted by the spin-polarized electrons coming from the fixed FM\cite{berger1996emission,slonczewski1996current,meo2022magnetisation}. Perpendicular magnetic anisotropy (PMA) facilitates the magnetization of the fixed and free layers of the MTJ to align perpendicularly to the film plane. A typical PMA-MTJ (p-MTJ) yields fast switching in the order of a few nanoseconds along with lower power consumption and superior thermal stability compared to in-plane MTJs\cite{slaughter2016technology}. However, a faster switching demands a high current density that undermines the aspirations of energy-efficient writing\cite{apalkov2013spin,endoh2020recent}. One may address this problem while applying the spin-orbit torque(SOT) to the free FM, where the switching current does not tunnel through the oxide layer of the MTJ, resulting in a significant reduction in power consumption\cite{meo2022magnetisation}. But, SOT-switching is not deterministic in the p-MTJs, and one may need to apply some external magnetic field for deterministic switching\cite{byun2021switching}. At the same time, the separate reading and writing path in the SOT-driven MTJs increases the complexity of the fabrication process. Further, neither type of the devices (STT and SOT driven p-MTJs) yield an ultra-high sensitivity due to their inadequate TMR(\%) in comparison to the proposed devices.\\
\indent Technologically relevant applications based on the MTJ devices require a high TMR(\%) and a low switching bias\cite{PhysRevApplied.8.064014,7571106,sharma2018role}. 
There have been consistent efforts to propose regular superlattice-based MTJs (SLTJs) to provide an ultra-high TMR($\approx10^5\%$) and reduction in switching bias\cite{chen2014enhancement,chen2015ultrahigh} owing to the physics of spin selective resonant filtering. Various 2D in-plane and van der waal(vdw) heterostructures have been explored in the quest of achieving a high TMR and a decimated power consumption. For example, the first principle calculation of vdw heterostructures such as Co/MoS2/graphene/MoS2/Co\cite{devaraj2021large} and Au/graphene /2 monolayer of CrI3 / graphene /Au\cite{zhang2022electronic} have predicted to engender a TMR of 1270\% and 2$\times10^4\%$ respectively. On the other hand, ab-initio calculations of in-plane heterostructures like VS$_2$/MoS$_2$/VS$_2$ have revealed a TMR of $4\times10^3\%$\cite{zhao2018designing}. Along with these proposals, there have been consistent experimental efforts to realize superlattice-based MTJs(SLTJs) after the successful fabrications of various multiple-quantum well-based magnetic tunnel junctions\cite{tao2019coherent,bhattacharjee2016effect,tseng2020superlattice}. 
Relying on this patronage, we propose a compendium of various nonuniform-SLTJs(Nu-SLTJs) for spintronic applications with the Gaussian, Linear, Lorentzian, and P\"oscl-Teller width(Fig. \ref{fig:BDWS}) and height(Fig. \ref{fig:BDHS}) profiles\cite{diez2000gaussian,gomez1999electron,sanchez2019non}. The physics of spin selective broad-band filtering (see Appendix A) in these proposed Nu-SLTJs culminates furtherance in the TMR(\%) and a significant reduction in the switching bias of the proposed devices along with minimal layered Nu-SLTJs design.\\
In order to benchmark the switching biases of the various proposed devices (PDs) against the regular MTJ, we define a performance index as the suppression in the switching bias(SSB) given by
\begin{equation}
SSB=\frac{|SB_{MTJ}^{PC\rightarrow APC}|+|SB_{MTJ}^{APC\rightarrow PC}|}{|SB_{PD}^{PC\rightarrow APC}|+|SB_{PD}^{APC\rightarrow PC}|} 
\end{equation} 
where PC(APC) denotes the parallel(anti-parallel) configuration, $SB_{MTJ(PD)}^{APC\rightarrow PC}$ and $SB_{MTJ(PD)}^{PC\rightarrow APC}$ represents the switching biases applied to the trilayer MTJ(PD) for the APC to PC and the PC to APC switching, and SB represents the switching bias as discussed in Appendix \ref{Appendix:A}.\\
\indent We chronicle this article by first describing the tight-binding device Hamiltonian and the self-consistent coupling of the Landau–Lifshitz–Gilbert–Slonczewski equation with the Non-equilibrium Green's Function formalism in section \ref{sec:2}, thereby providing a theoretical foundation behind our computational framework. Subsequently, we lay down the detailed design of various width(W) and height(H)-based Nu-SLTJs(W-SLTJs and H-SLTJs) along with the simulation results of each device in the section \ref{sec:3}. It incorporates the spin and charge current characteristics along with the switching dynamics of the various Nu-SLTJs comprised of Linear width(\ref{LWSLTJ}), Gaussian width(\ref{GWSLTJ}), Lorentzian width(\ref{LRWSLTJ}), P\"oschl-Teller width(\ref{PWSLTJ}), Linear height(\ref{LHSLTJ}), Gaussian height(\ref{GHSLTJ}), Lorentzian height(\ref{LRHSLTJ}) and P\"oschl-Teller height(\ref{PHSLTJ}) profiles including the trilayer MTJ(\ref{MTJ}). We also demonstrate the relevance of the Nu-SLTJs in order to address the shortcomings of a regular SLTJ. Thereafter, we analyse the performances of the aforementioned devices at higher voltages in section \ref{sec:4} while envisioning their possible applications in the STNOs. Thenceforth in section.\ref{sec:5} we compare various performance indices of the Nu-SLTJs in Tab.(\ref{Table:1}), followed by the conclusion in section \ref{sec:6}, thereby abridging a synopsis of the study.
\section{\label{sec:2}Mathematical Modelling and Simulation Details}
We begin with the description of the device schematics of the tri-layer MTJ and the Nu-SLTJ depicted in Figs. \ref{fig:PMA-MTJ}(a) and \ref{fig:PMA-MTJ}(b). The Nu-SLTJ is realized by sandwiching a non-uniform heterostructure comprised of insulator(I)-normal metal(NM)- - -I-NM-I- - -NM-I layers between the fixed and free FMs as shown in Figs.\ref{fig:BDWS}$\&$\ref{fig:BDHS}. We have used MgO as an insulator to implement the W-SLTJs, whereas H-SLTJs have been realized by utilizing a stoichiometrically substituted Mg$_x$Zn$_{1-x}$O\cite{tian2014miscibility} as barrier region. One can estimate the band gap of such stoichiometrically substituted semiconductors via the first-order interpolation method\cite{li2014controlling}. The quantum(q)-well regions of the non-uniform superlattice may consist of an NM (Ru,Cu,Ti etc.) such that the dimension of the q-wells remain significantly smaller than spin coherent length\cite{sharma2021proposal,chen2014enhancement,sharma2018band,chen2015ultrahigh}.\\
\indent To analyse the quantum transport through mesoscopic devices such as an MTJ or an Nu-SLTJ, we first describe the relationship of the Hamiltonian matrix [H] with the energy-resolved spin-dependent retarded Green’s function matrix [G(E)], given by\cite{datta2018lessons}
\begin{eqnarray}
	[G(E)] = [EI-\Sigma-H]^{-1}\\
	\label{green_fun}
	[\Sigma]=[\Sigma_F]+[\Sigma_f]
	\label{Sigma}
\end{eqnarray}
where the Hamiltonian matrix [H]=[H$_{0}$]+[U$_{0}$] comprises the nearest neighbour tight-binding Hamiltonian [H$_{0}$] and the applied potential matrix [U$_{0}$], while $I$ represents an identity matrix with dimensions of the device Hamiltonian [H]. In equation.\ref{Sigma}, we formulate the coupling of the Hamiltonian matrix corresponding to the device region [H] with the self-energy matrices of the fixed and the free FMs described via  $\Sigma_F$ and $\Sigma_f$, respectively. We describe the ferromagnetic behavior of the contacts by the Stoner model \cite{ralph2008spin}, with conduction band exchange splitting ($\Delta$), Fermi energy (E$_f$), and effective mass (m$_{fm}$). Based on the above-mentioned formulation, we compute the Hamiltonian matrix[H], the self-energy matrix ($\Sigma$), and the energy-resolved spin-dependent retarded Green's function matrix[G(E)] for the tight binding framework\cite{sharma2021proposal}. In order to take care of the current conduction at various modes, we include the respective transverse mode energies($E_t$) to the diagonal elements of the device Hamiltonian given by Et=$\hbar ^2k^2_{x}/2m^*+\hbar ^2k^2_{y}/2m^*$, where $k_x$ and $k_y$ represents wave vectors along the $x$ and $y$ axis, while the electron transport takes place along the $z$ direction. \cite{sharma2021proposal}.\\
\begin{figure}[h]
		\centering
		\includegraphics[width=0.8\linewidth]{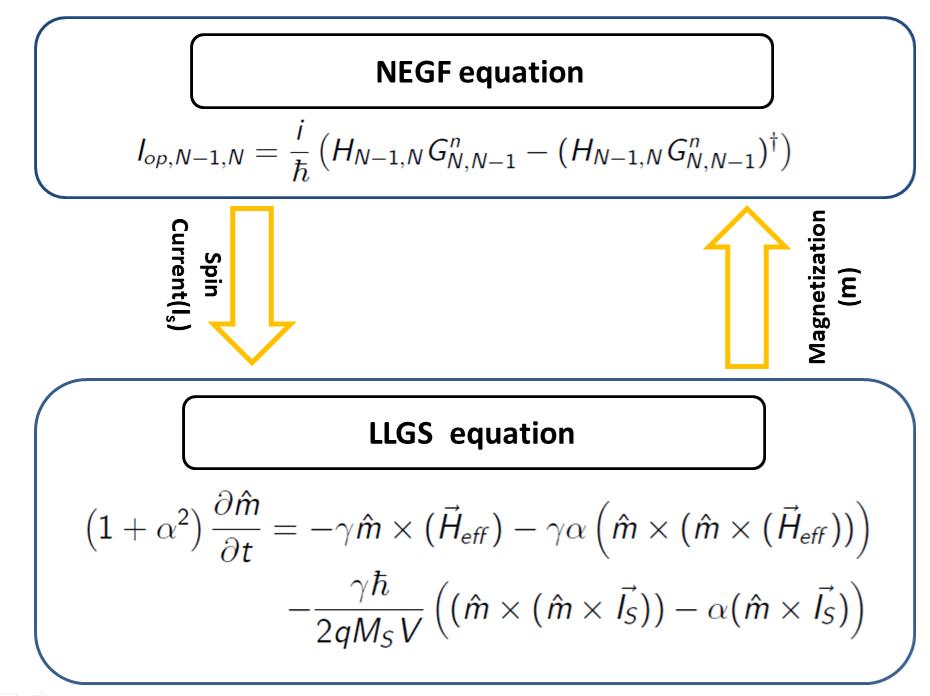}
		\caption{The self-consistent coupling of the non-equilibrium Green's function with the Landau–Lifshitz–Gilbert–Slonczewski equation.}
		\label{fig:SelfConsistent}
\end{figure}
\indent In the next step, we describe the density of electrons in the channel region by the diagonal elements of the energy resolved electron correlation matrix [$G^n(E)$], given by
\begin{equation}
[G^n]= \int [G(E)] [\Sigma^{in}(E)] [G(E)]^{\dagger}dE 	\label{G_n}
\end{equation}
\begin{equation}
[\Sigma^{in}(E)]=[\Gamma_F(E)]f_F(E)+[\Gamma_f(E)]f_f(E) \label{Sigma_in}
\end{equation}
where the quantities $[\Gamma_F(E)]=i\left ([\Sigma_F(E)]- [\Sigma_F(E)]^{\dagger} \right )$ and $[\Gamma_f(E)]=i\left( [\Sigma_f(E)]-[\Sigma_f(E)]^{\dagger} \right )$ denotes the spin-dependent broadening matrices \cite{datta2005quantum} of the fixed and free FM contacts, respectively. To construct the applied potential matrix[U$_0$], we assume a linear drop across the oxide region and no drop across the metal layers. The Fermi-Dirac distribution functions related to the free(f) and fixed(F) FMs are denoted as $f_f(E)$ and $f_F(E)$, respectively. We apply the boundary conditions by assuming $U_{FixedFM}=-qV/2$ and $U_{FreeFM}=qV/2$, where $V$ is the applied voltage.
\indent The current between $N-1$th and $N$th lattice points of the device is described by the current operator I$_{op}$ \cite{datta1997electronic}, given by
\begin{equation}
{I}_{op,N-1,N}=\frac{i}{\hbar}\left(H_{N-1,N}G^{n}_{N,N-1}-(H_{N-1,N}G^{n}_{N,N-1})^{\dagger}\right)
\end{equation}
 The charge current through the device $I$ is defined as  
\begin{equation}
I =q \int  \text{ Real [Trace(}\hat{I}_{op}\text{)]}dE
\end{equation}
and spin current $I_S$ is given by
\begin{equation}
I_{S_i} =q \int  \text{ Real [Trace(}\sigma_i\cdot\hat{I}_{op}\text{)]}dE
\end{equation}
where $\sigma_i$ denotes the Pauli spin-matrices along $\hat{x}$, $\hat{y}$, $\hat{z}$ directions.\\
\indent The magnetization dynamics of the free layer in the presence of an applied magnetic field and spin current is calculated using the self consistent coupling of the NEGF formalism with the Landau–Lifshitz–Gilbert–Slonczewski (LLGS) equation as described in Fig. \ref{fig:SelfConsistent}. The LLGS equation\cite{slonczewski1996current} is given by
\begin{equation}
\begin{split}
\left( 1+\alpha^{2}\right) \frac{\partial \hat{m}}{\partial t} = -\gamma \hat{m} \times \vec{H}_{eff} \\- \gamma \alpha \left( \hat{m} \times ( \hat{m} \times \vec{H}_{eff})\right)  \\
- \frac{\gamma\hbar}{2qM_SV} \left((\hat{m}\times(\hat{m}\times\vec{I_S}))-\alpha(\hat{m}\times\vec{I_S})\right) \label{eq:8}
\end{split}	
\end{equation}
where $\hat{m}$ represents the direction of magnetization of the free layer, $\alpha$ represents the Gilbert damping parameter and $\gamma $ is called the gyromagnetic ratio of  the electron, $M_s$ represents the saturation magnetization, $V$ represents the volume of the FM contacts. The effective magnetic field is given by $\vec{H_{eff}}=\vec{H_{applied}}+\vec{H_{k\perp}}m_z \hat{z}$, with $\vec{H_{applied}}$ being the applied field which is taken as zero for our simulations, and $\vec{H_{k\perp}} \vec{m_z}$ represents the uniaxial magnetic anisotropy of the FMs with PMA.\\
\indent The spin current is decomposed into three components such that  
 $$\vec{I_{S}}=I_{s,m}\hat{m}+I_{s,||}\hat{M}+I_{s,\perp}\hat{M}\times\hat{m}$$
 where $\hat{M}$ and $\hat{m}$ represents a unit vector along the direction of the fixed and free FM, respectively and apart from that the I$_{s||}$ and I$_{s\perp}$ denotes the Slonczewski and field like term. 

 Further, the equation.\ref{eq:8} can be divided in two factors such that 
 \begin{equation}
 (1+\alpha^2)\frac{\partial \hat m}{\partial t}=\ Spinning factor + \ Damping factor
 \end{equation}
where, the spinning factor(SF) describes the angular frequency of the magnetization around the easy axis and the damping factor(DF) represents the rate of change of the angle between the easy axis and the magnetization.

After plugging in the value of the $\vec{I_s}$ in equation.\ref{eq:8} the spinning factor(SF) can be re-casted as 
\begin{equation}
    SF=-\gamma\left(\hat m \times \vec{H}_{eff}+\frac{\hbar}{2qM_sV}(I_{s,\perp}-\alpha I_{s,||})\hat{m}\times \hat{M}\right) 
\end{equation}

Similarly, the expression of the damping factor(DF) can be simplified as
 \begin{equation}
 \begin{split}
     DF=-\gamma\alpha(\hat{m}\times \hat{m}\times\vec{{H}_{eff}})+ \\ 
     \gamma\frac{\hbar}{2qM_sV}(I_{s,||}+\alpha I_{s,\perp})\hat{m}\times \hat{m}\times \hat{M}
\end{split} 
\label{eq:DF}
 \end{equation}
We can safely ignore the term $\alpha I_{s,\perp}$ in Eq.~\ref{eq:DF} as it is much smaller than $I_{s,||}$. Hence, the critical current for switching is given as $I_{s||,c}=\frac{2qM_sV\alpha}{\hbar} H_{eff}$. The value of the critical switching current for an in plane MTJ under the macro-spin assumption is given by I$_{s||,c}$=$\frac{2\alpha e}{\hbar}VM_s(H_K+H_d/2)$, where $H_K$ and $H_d$ denotes the in-plane anisotropy and the demagnetizing field, respectively\cite{ralph2011spin}. It is worthwhile mentioning that the value of the $H_d$ is mostly one order higher compared to the $H_K$ which causes a significant hike in power consumption. One may reduce the impact of the $H_d$ while utilizing FMs with PMA\cite{tudu2017recent}, where the direction of the magnetization aligns perpendicularly to the thin film plane\cite{prakash2017effect}. The effective magnitude of the perpendicular anisotropy is given as $H_{K\perp}=2\frac{\sigma}{tM_s}-H_d$, where $t$ and $\sigma$ denotes the thickness and the interface PMA factor of an FM thin film, respectively. The inter-facial PMA dominates the $H_d$ when the width of the FM is made smaller beyond a critical thickness\cite{ikeda2010perpendicular}. The value of the critical current is given by $I_{s||,c}=\frac{2qM_sV\alpha}{\hbar} H_{K\perp}$.\\ 
\indent In this work, we have used CoFeB\cite{miao2006inelastic} for the FM contacts with the Fermi energy $E_f=2.25$ eV and conduction band exchange splitting $\Delta=2.15$ eV\cite{sharma2018band} while assuming the effective masses of an electron in the MgO barriers, in the NM quantum wells and in the FM contacts as $m_{ox}=0.18m_e$, $m_{nm}=0.9m_e$ and $m_{fm}=0.8m_e$, respectively \cite{datta2011voltage}, where $m_e$ denotes the free-electron-mass. We have taken the barrier height between CoFeB and MgO as U$_B=0.76$ eV above Fermi level \cite{datta2012behin} and considered the conduction band offset between the FM and the NM layer as U$_{BW}$=0.4 eV. The NM q-wells have the same thickness of $W=0.35$ nm to keep them within the fabrication limits \cite{tao2019coherent,yang2015domain}. We have considered the minimum oxide width as $0.6$ nm which is the least thickness of the MgO that can be deposited reliably \cite{deac2008bias}. We have taken the thickness of the free FM as 1.3 nm \cite{ikeda2010perpendicular} and have considered the cross-sectional area of the FM contacts as 0.25$\pi\times 30^2$ nm$^2$ so that the magnetization dynamics of the free FM can be described by a macro-spin model \cite{sato2014properties}. The value of the saturation magnetization is considered as $M_S=1150$ emu/cc and the perpendicular uni-axial magnetic anisotropy is taken as $H_{k\perp}$=3.3 k Oe\cite{gajek2012spin}. The thermal stability factor is given by $\Delta_{EB}=H_KM_sV/2k_BT\approx42$ . The value of the damping parameter and gyromagnetic ratio are taken as $\alpha=0.01$ and $\gamma=17.6$ MHz/Oe, respectively.\\
\indent After evaluating all the parameters in the equation of the critical current given by I$_{s||,c}=\frac{2qM_sV\alpha}{\hbar} H_{k\perp}$, we obtain an I$_{s||,c}$ of 0.0106 mA. When the I$_{s||}$ becomes equal to the I$_{s||,c}$, the damping force(DF) acting on the free layer becomes zero and the relative angle($\theta$) between the easy axis and magnetization of the free layer becomes constant. In order to perform the switching of the free FM, we need to supply a switching current, higher than the critical current, so that the spin current($I_{s||}$) induced anti-damping force ($\frac{\hbar}{2qM_sV}I_{s,||}\hat{m}\times \hat{m}\times \hat{M}))$ dominates over the damping due to the uniaxial magnetic anisotropy($-\gamma (\alpha(\hat{m}\times \hat{m}\times\vec{{H}_{eff}})$). Whilst the I$_{s||}$ is near the critical current, the effective damping force acting on the free layer becomes negligible. Hence the switching of the magnetization takes a longer time. To avoid this, we apply a spin current(I$_{s||}$=0.0127 mA), which is 20\%  higher than the critical current (see Appendix:\ref{Appendix:A}) thereby ensuring a comfortable switching.
\section{\label{sec:3}Device Details and Simulation Results}
\begin{figure}[h]
	\centering
	\includegraphics[scale=0.38]{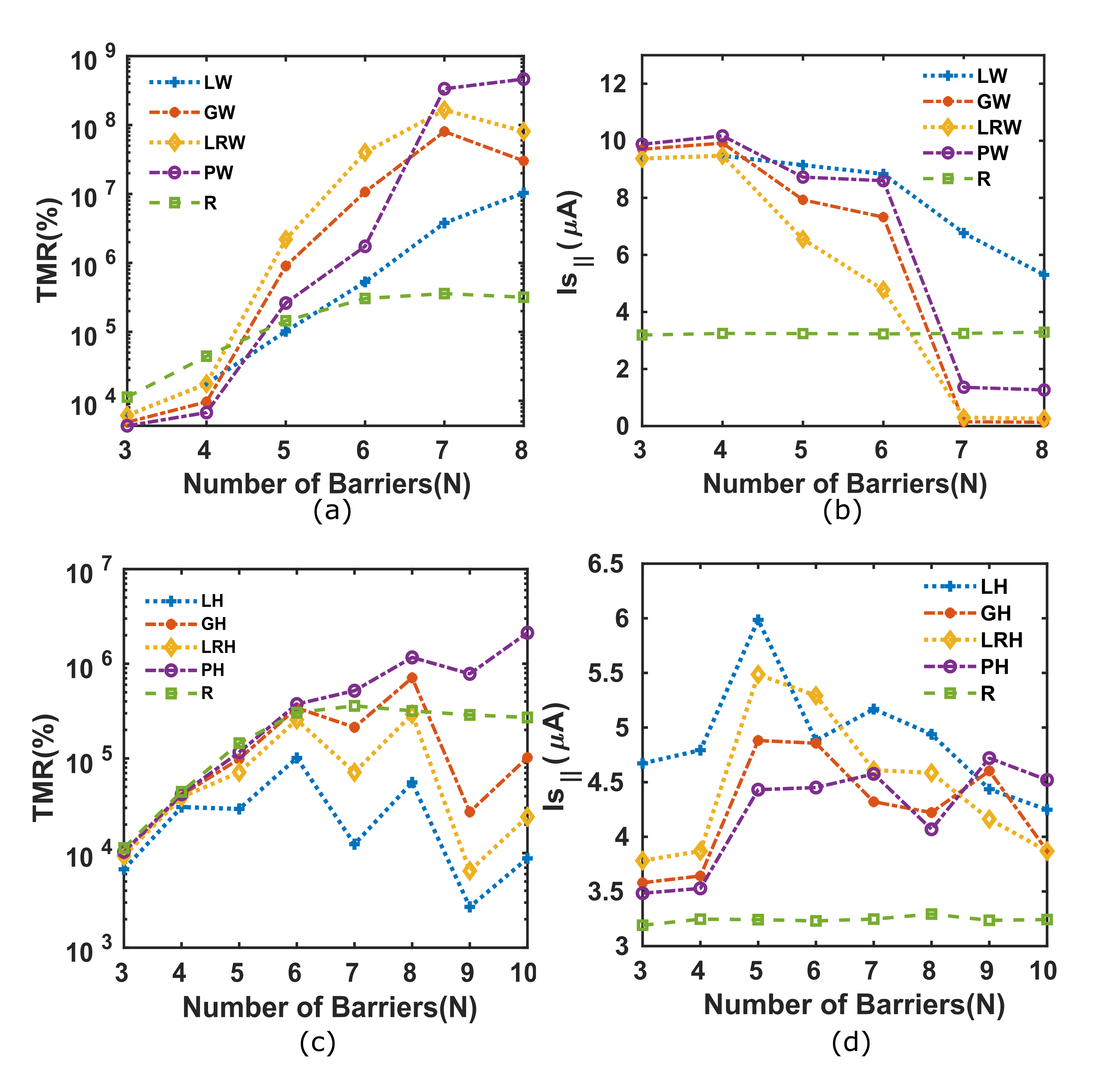}
	\caption{(a) TMR(\%) and (b) I$_{s||}$ of the W-SLTJs, (c) TMR(\%) and (d) I$_{s||}$ of the H-SLTJs with respect to the number of barriers at V=5 mV. Here, LW represents Linear width, GW represents Gaussian width, LRW represents Lorentzian width, PW represents P\"oschl-Teller width, LH represents Linear height, GH represents Gaussian height, LRH represents Lorentzian height, PH represents P\"oschl-Teller height and R represents Regular superlattice.}
	\label{fig:NOB_SLTJ}
\end{figure}

In the wide design landscape of non-uniform superlattices, we first explore the TMR(\%) and I$_{s||}$ of various Nu-SLTJs with respect to the number of barriers (oxide layers) as shown in the Fig.~\ref{fig:NOB_SLTJ} (at V=5mV). 
It can be inferred from Fig.~\ref{fig:NOB_SLTJ}(a) \& (b) that in W-SLTJs, the TMR(\%) first increases with the number of barriers and then decreases whereas I$_{s||}$ decreases monotonically. As the number of barriers(N) goes from 4 to 5, we see a significant increase in the TMR(\%) along with a respectable $I_{s||}$. At N=6, although a higher TMR(\%) is achieved but I$_{s||}$ drops even further. In our design of width-based SLTJs, the thickness of the central barriers increases with $N$ to keep the width of terminal barriers within fabrication limits \cite{deac2008bias}. The thickness of the central oxide in the GW-SLTJ reaches 2.1 nm at $N=7$ because of the exponent factor $e^{-x^2/7}$ (see Eq. \ref{eq:Gaussian_width_eq}). It vanishes the spin selective transmission through the GW-SLTJ for $N>6$, resulting in a decimated $I_{s||}$.  A similar behaviour is noticed for both the PW and LRW-SLTJs making them dysfunctional for $N>6$ except for the LW-SLTJ.\\
\indent We present the dependence of the TMR(\%) and I$_{s||}$ of the H-SLTJs with respect to the number of barriers in Fig. \ref{fig:NOB_SLTJ}. As we increase the number of barriers from three to four, we see a slight increase in the I$_{s||}$(Fig. \ref{fig:NOB_SLTJ}(d)). This can be attributed to the fact that an increase in the number of barriers increases the number of resonant peaks and the neighborhood Bloch states, which gives rise to a better broad-band spin filtering. In the H-SLTJs, when we have odd number of barriers, there is only one central barrier with a scattering potential of 3.01 eV. But for H-SLTJs with even number of barriers, there are two central barriers with a scattering potential of 3.01 eV. Therefore, for the even number of barriers, the H-SLTJs provide a larger hindrance to the electron tunneling and hence offer a poorer I$_{s||}$ and a better TMR. as we increase the number of barriers in the H-SLTJs, the terminal barriers possess lower barrier heights. Consequently, they do not hinder electron tunneling significantly and  may help to engender a better broad-band transmission while ameliorating the I$_{s||}$. Along similar lines, an improvement in the transmission might deteriorate the ratio of the transmissions in the PC and the APC within the Fermi window while reducing the TMR. Therefore, the H-SLTJs do not show monotonic behavior akin to the W-SLTJs. Figure. \ref{fig:NOB_SLTJ}(c)\&(d) illustrates that the $I_{s||}$ of the H-SLTJs peaks at N=5 along with a sizeable TMR(\%).\\
\indent Hence, in the subsequent subsections, we present detailed designs of the W and H-SLTJs with n=5 and draw a performance comparison with a regular SLTJ and a trilayer MTJ in Tab.~\ref{Table:1}. Figure. \ref{fig:NOB_SLTJ} exemplifies that the regular SLTJ(R-SLTJ) offers an  adequate TMR(\%) in comparison to the Nu-SLTJs but fails to provide a respectable $I_{s||}$. Hence, Nu-SLTJs emerge as viable alternatives for meager power consumption and faster switching.
 \subsection{\label{MTJ}Characteristics of a Tri-layer MTJ}
We first explore the electrical characteristics of a trilayer MTJ shown in Fig.~\ref{fig:PMA-MTJ}(a), along with the STT-switching of the free FM.
\begin{figure}[h]
	\centering
	\includegraphics[scale=0.33]{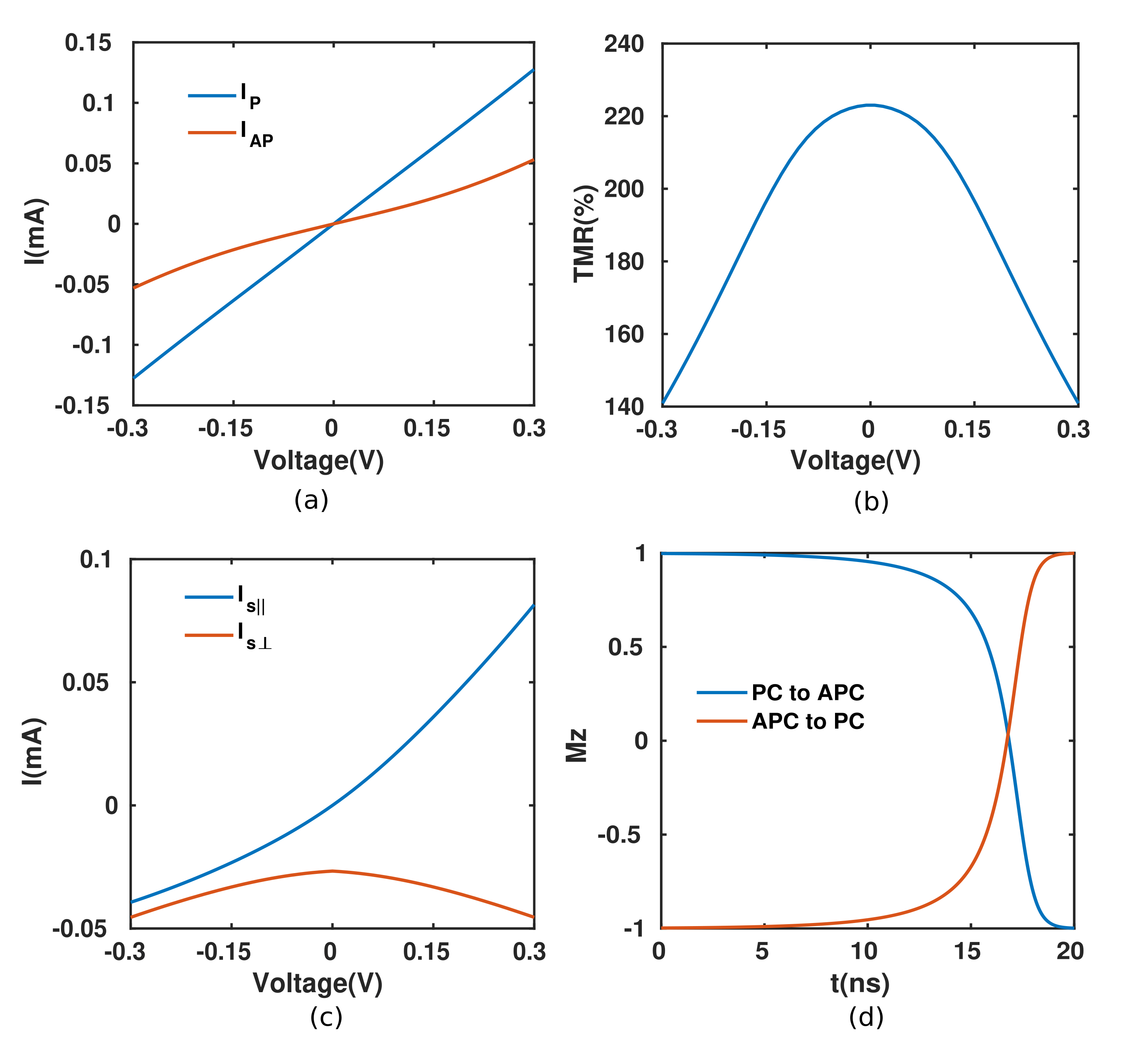}
	\caption{(a) I-V characteristics of the MTJ in the PC(I$_P$) and APC(I$_{AP}$). (b) Variation of TMR(\%), (c) I$_{s\parallel}$ and I$_{s\perp}$ with the applied voltage. (d) STT based switching  of the free layer from the PC to APC(V=-74 mV) and the APC to PC(V=60 mV).}
	\label{fig:MTJ_CTIS}
\end{figure}
The device consists of an MgO barrier of $1$ nm thickness sandwiched between fixed and free FMs. The band diagram of the device is shown in Fig. \ref{fig:MTJ_RSLTJ_BD}(a). We show the I-V characteristics of the device in the PC and the APC in Fig. \ref{fig:MTJ_CTIS}(a). The maximum TMR(\%) provided by the tri-layer MTJ is 240\%(Fig. \ref{fig:MTJ_CTIS}(b)). Figure. \ref{fig:MTJ_CTIS}(d) illustrates the STT-switching of the free layer from the APC to PC and the PC to APC at switching-bias of 60 mV and -74 mV, respectively(see Appendix: \ref{Appendix:A}).  The difference in magnitude of the $SB^{PC\rightarrow APC}$ and  $SB^{APC\rightarrow PC}$ originates from the asymmetrical $I_{s||}-V$ relationship of the device as shown in Fig.~\ref{fig:MTJ_CTIS}(c). The TMR and spin current of the trilayer MTJ can be reasoned out from spin selective transmission spectra of the PC (see Fig. \ref{fig:MTJ_RSLTJ_TM}(a)) and the APC (see Fig. \ref{fig:MTJ_RSLTJ_TM}(b)) enabled by physics of single barrier tunneling.
 
\subsection{\label{LWSLTJ}Device characteristics of Linear Width based Nu-SLTJ (LW-SLTJ)}
To construct the linear width(LW) based Nu-SLTJ shown in Fig. \ref{fig:BDWS}(a), we consider the width of n-th($b_n$) barrier as $$b_n=b_0-n\times0.15$$
where $b_2=0.6$ nm and n=0,1,2. Here $n=0$ represents the central barrier, $n=1$ represents the two neighbour barriers of the central barrier, and $n=2$ represents the two terminal barriers.
 \begin{figure}[h]
	\centering
	\includegraphics[scale=0.35]{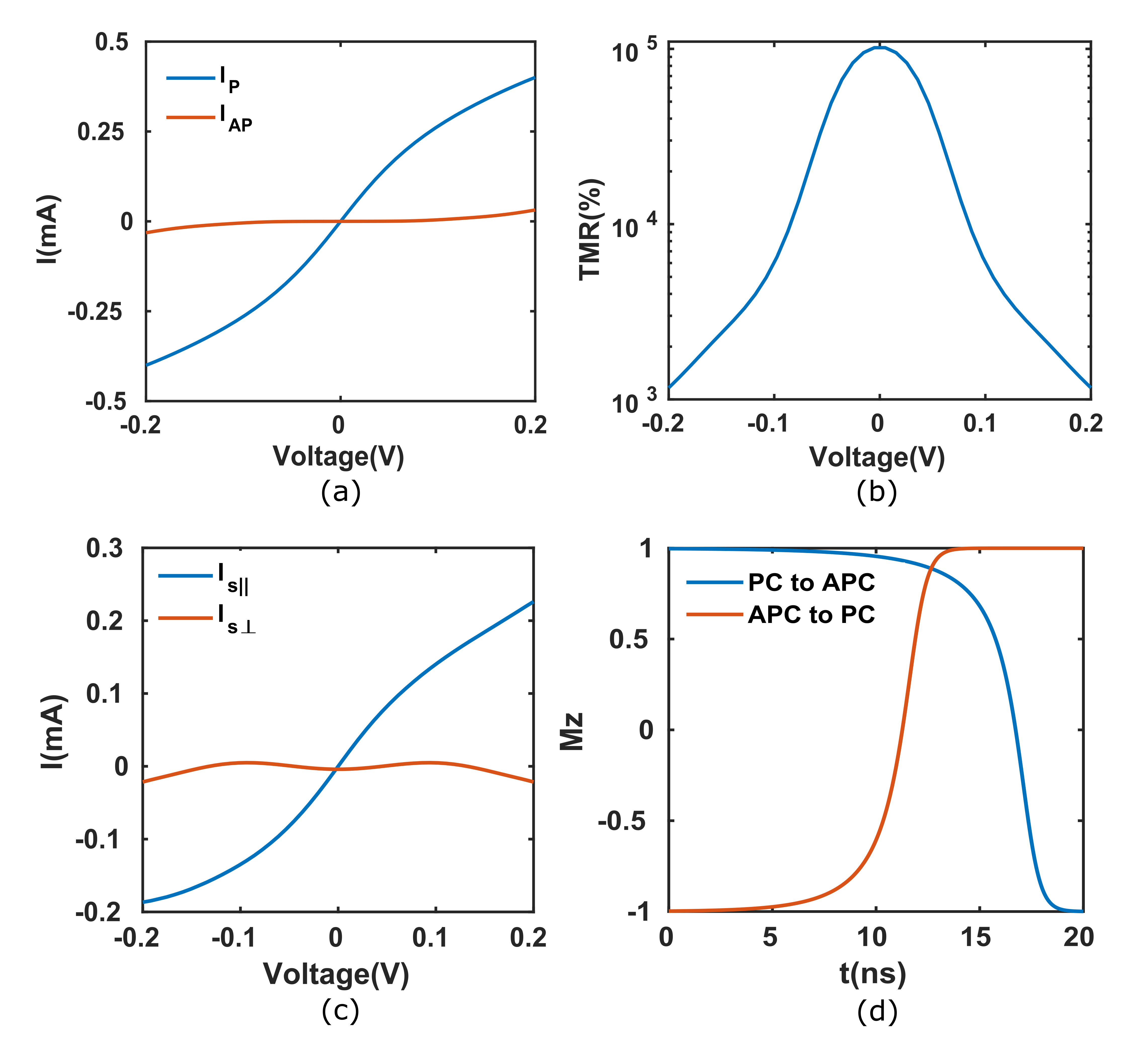}
	\caption{(a) I-V characteristics of the LW-SLTJ in the PC(I$_P$) and APC(I$_{AP}$). (b) Variation of TMR(\%), (c) I$_{s\parallel}$ and I$_{s\perp}$ with the applied voltage. (d) STT based switching  of the free layer from the PC to APC(V=-7.4 mV) and the APC to PC(V=7.4 mV).}
	\label{fig:LWSLTJ_CTIS}
\end{figure}
In Fig. \ref{fig:LWSLTJ_CTIS}(a) we present the I-V characteristics of the LW-SLTJ in the PC and the APC. The maximum TMR(\%) provided by the device is $\approx 10^5$\%.
Figure \ref{fig:LWSLTJ_CTIS}(d) shows the STT-switching of the free layer from the PC to APC and the APC to PC with SB$^{PC\rightarrow APC}$ and SB$^{APC\rightarrow PC}$ of -7.4 mV and 7.4 mV, respectively. Unlike the trilayer MTJ, the $I_{s||}$-V characteristics of the LW-SLTJ is nearly symmetrical around 0V(Fig. \ref{fig:LWSLTJ_CTIS}(c)). Hence the magnitude of switching biases remains the same in both the cases. The sizable ratio of the LW-SLTJ enabled transmissions in the PC(Fig.\ref{fig:WSLTJ_TM}(a)) and APC(Fig.\ref{fig:WSLTJ_TM}(b)) provides to a monumental enhancement in the  TMR(\%) as shown in Fig. \ref{fig:LWSLTJ_CTIS}(b). Furthermore, the broad-band spin filtering exhibited by the LW-SLTJ device provides a sizable I$_{s||}$ which reduces the switching bias substantially, thereby yielding the largest SSB of 9.05(see Tab.~\ref{Table:1}).


\subsection{\label{GWSLTJ}Device characteristics of Gaussian Width based Nu-SLTJ (GW-SLTJ)}
\begin{figure}[h]
	\centering
	\includegraphics[scale=0.34]{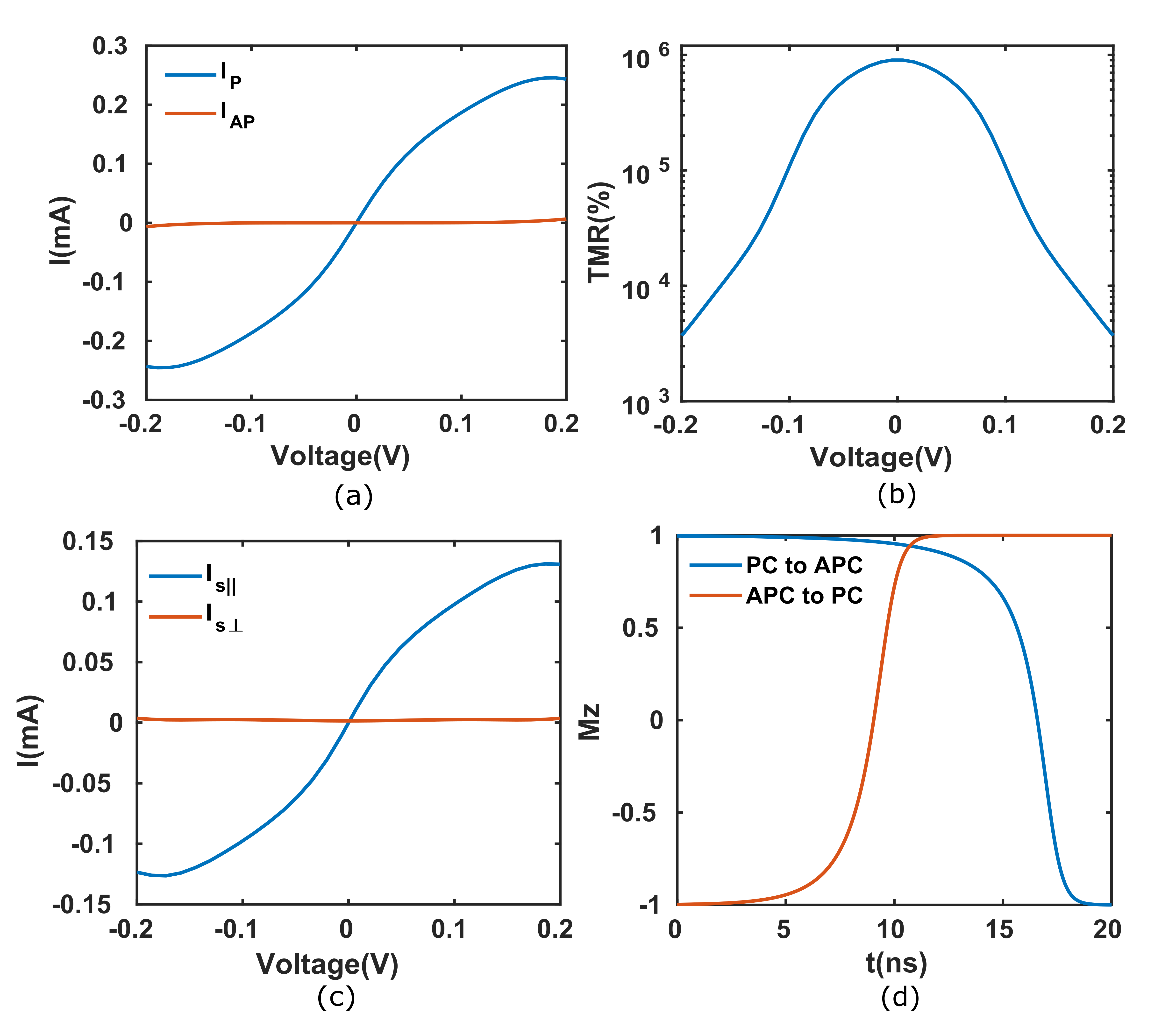}
	\caption{(a) I-V characteristics of the GW-SLTJ in the PC(I$_P$) and APC(I$_{AP}$). (b) Variation of TMR(\%), (c) I$_{s\parallel}$ and I$_{s\perp}$ with the applied voltage. (d) STT based switching  of the free layer from the PC to APC(V=-8.88 mV) and the APC to PC(V=8.88 mV).}
	\label{fig:GWSLTJ_CTIS}
\end{figure}

We characterize the thickness of the n-th barrier($b_n$) in the Gaussian width(GW) based Nu-SLTJ in such a way that 
 \begin{equation}
      b_n=b_0 e^{-n^2/7}
      \label{eq:Gaussian_width_eq}
 \end{equation}
where $b_2=0.6$ nm and $n$=0,1,2. The band diagram of the device is presented in Fig. \ref{fig:BDWS}(b). In Fig \ref{fig:GWSLTJ_CTIS}(a), we present the I-V characteristics of the GW-SLTJ in the PC and the APC. 
The maximum TMR(\%) yielded by the device is $\approx9.01\times10^5$\%(Fig. \ref{fig:GWSLTJ_CTIS}(b)). Similar to the LW-SLTJ, the $I_{s||}$-V characteristics of the GW-SLTJ is nearly symmetrical around 0V(Fig. \ref{fig:GWSLTJ_CTIS}(c)). Therefore the  magnitudes of the SB$^{PC\rightarrow APC}$ and SB$^{APC\rightarrow PC}$ remain the same. In Fig. \ref{fig:GWSLTJ_CTIS}(d) we present the STT-switching of the free layer in the GW-SLTJ for I$_{s||}$=0.0127 mA. The applied voltages for the PC to APC and the APC to PC switching are -8.88 mV and 8.88 mV, respectively. Therefore, the SSB achieved in this case is 7.61, the 3rd largest among all the Nu-SLTJs(see Tab.~\ref{Table:1}). Similar to the LW-SLTJ, the significant enhancement in the TMR(\%) and I$_{s||}$ that we accomplish via GW-SLTJ can be rationalized in light of the spin selective broad-band filtering offered by the GW-SLTJ as shown in Figs. \ref{fig:WSLTJ_TM}(c)\&(d). 
\subsection{\label{LRWSLTJ}Device characteristics of Lorentzian Width based  Nu-SLTJ(LRW-SLTJ)}

The Lorentzian-width(LRW) based Nu-SLTJ is engineered in such a way that the width of n-th barrier($b_n$) is given by $$b_n=b_0(\frac{\Gamma^2}{\Gamma^2+n^2})$$
where $b_{2}=0.6$ nm and $\Gamma$=2. 
We present the band diagram of the device in Fig. \ref{fig:BDWS}(c). 
\begin{figure}[h]
	\centering
	\includegraphics[scale=0.34]{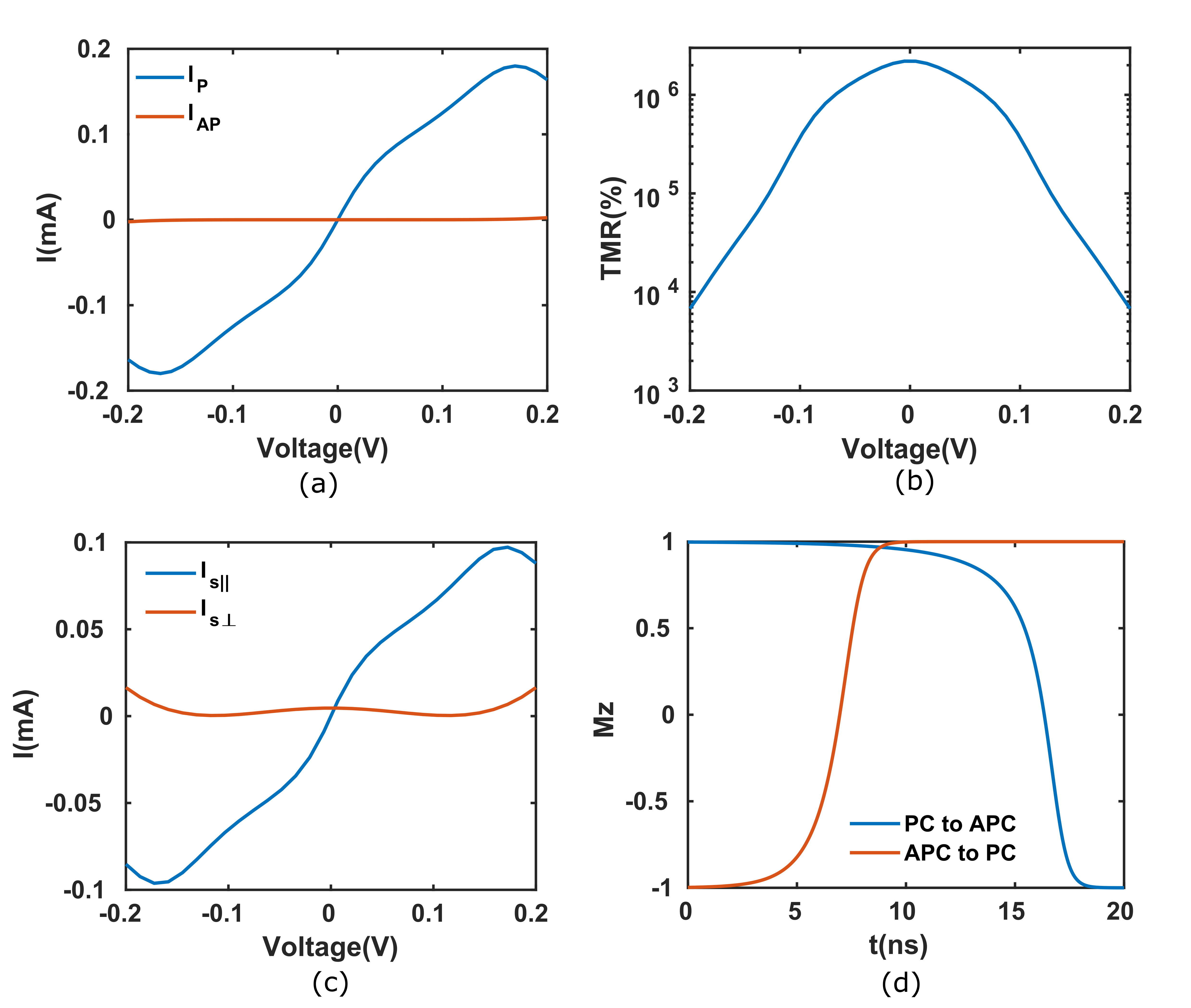}
	\caption{(a) I-V characteristics of the LRW-SLTJ in the PC(I$_P$) and APC(I$_{AP}$). (b) Variation of TMR(\%), (c) I$_{s\parallel}$ and I$_{s\perp}$ with the applied voltage. (d) STT based switching  of the free layer from the PC to APC(V=-11.7 mV) and the APC to PC(V=11.7 mV).}
	\label{fig:LRWSLTJ_CTIS}
\end{figure}
Figure. \ref{fig:LRWSLTJ_CTIS}(a) shows the I-V characteristics of the LRW-SLTJ in the PC and the APC. The maximum TMR(\%) obtained for this device is 2.19$\times10^6$(\%)(Fig. \ref{fig:LRWSLTJ_CTIS}(b)). We present the variation of the I$_{s||}$ and I$_{s\perp}$ with the applied voltage in Fig. \ref{fig:LRWSLTJ_CTIS}(c). The switching dynamics of free FM in the SLTJ is presented in Fig.\ref{fig:LRWSLTJ_CTIS}(d) with the SB$^{APC\rightarrow PC} $ and SB$^{PC\rightarrow APC}$of -11.7 mV and 11.7 mV, respectively. The LRW-SLTJ provides the 5th largest SSB with 5.73 (see Tab.~\ref{Table:1}). The superiority of the performance indices in terms of the TMR(\%) and the I$_{s||}$ that we realize through the LRW-SLTJ can be reasoned out on the basis of the heterostructure enabled broad-band spin filtering in the PC and  APC as shown in Figs. \ref{fig:WSLTJ_TM}(e)\&(f).  

\subsection{\label{PWSLTJ}Device characteristics of the P\"oschl-Teller Width based Nu-SLTJ(PW-SLTJ)}
The P\"oschl-Teller width based Nu-SLTJ is constructed in such a way that the thickness
\begin{figure}[h]
	\centering
	\includegraphics[scale=0.35]{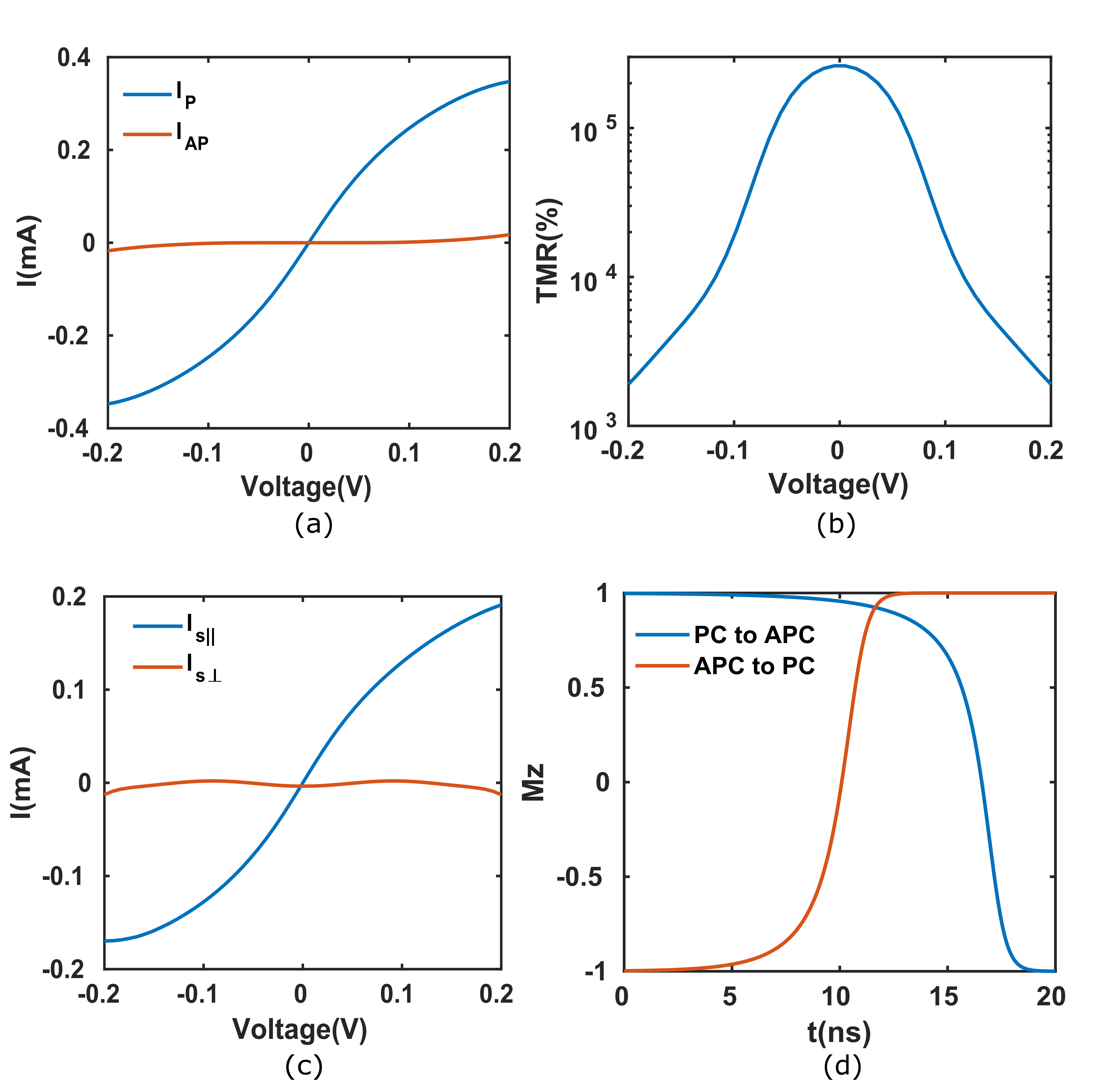}
	\caption{(a) I-V characteristics of the PW-SLTJ in the PC(I$_P$) and APC(I$_{AP}$). (b) Variation of TMR(\%), (c) I$_{s\parallel}$ and I$_{s\perp}$ with the applied voltage. (d) STT based switching  of the free layer from the PC to APC(V=-7.75 mV) and the APC to PC(V=7.75 mV).}
	\label{fig:PWSLTJ_CTIS}
\end{figure}
of n-th barrier is given by $$b_n=b_0\frac{2}{e^{n/2}+e^{-n/2}}$$
where $b_{2}$ is 0.6 nm and $n=0,1,2$. The band diagram  of the device is shown in Fig. \ref{fig:BDWS}(d). In Fig. \ref{fig:PWSLTJ_CTIS}(a) we present the I-V characteristics of the PW-SLTJ in the PC and the APC. The maximum TMR(\%) offered by the device is nearly $2.61\times10^5$(\%). Fig. \ref{fig:PWSLTJ_CTIS}(d) depicts the STT-switching of the free layer from the PC to APC and the APC to PC with the $SB^{PC\rightarrow APC}$ and $SB^{APC\rightarrow PC}$ of -7.75 mV and 7.75 mV, respectively. Hence, the SSB obtained in this case is 8.7(see Tab.~\ref{Table:1}), the second largest after the LW-SLTJ. The broad-band spin filtering in the PW-SLTJ(see Fig.~\ref{fig:WSLTJ_TM}(g)$\&$(h)) enables the device to accomplish a remarkable enhancement in the TMR(\%) and the I$_{s||}$ as shown in Fig. \ref{fig:PWSLTJ_CTIS}(b) and \ref{fig:PWSLTJ_CTIS}(c).
\subsection{\label{LHSLTJ}Device characteristics of the Linear Height based Nu-SLTJ(LH-SLTJ)}
Along with different width profiles, the Nu-SLTJs with different height profiles also substantiate their worth as potential candidates for high-performance MTJs. Hence, in the subsequent subsections, we present the height-based analogues of the above-mentioned heterostructures geared towards broad-band spin filtering.\\
\begin{figure}[h]
	\centering
	\includegraphics[scale=0.34]{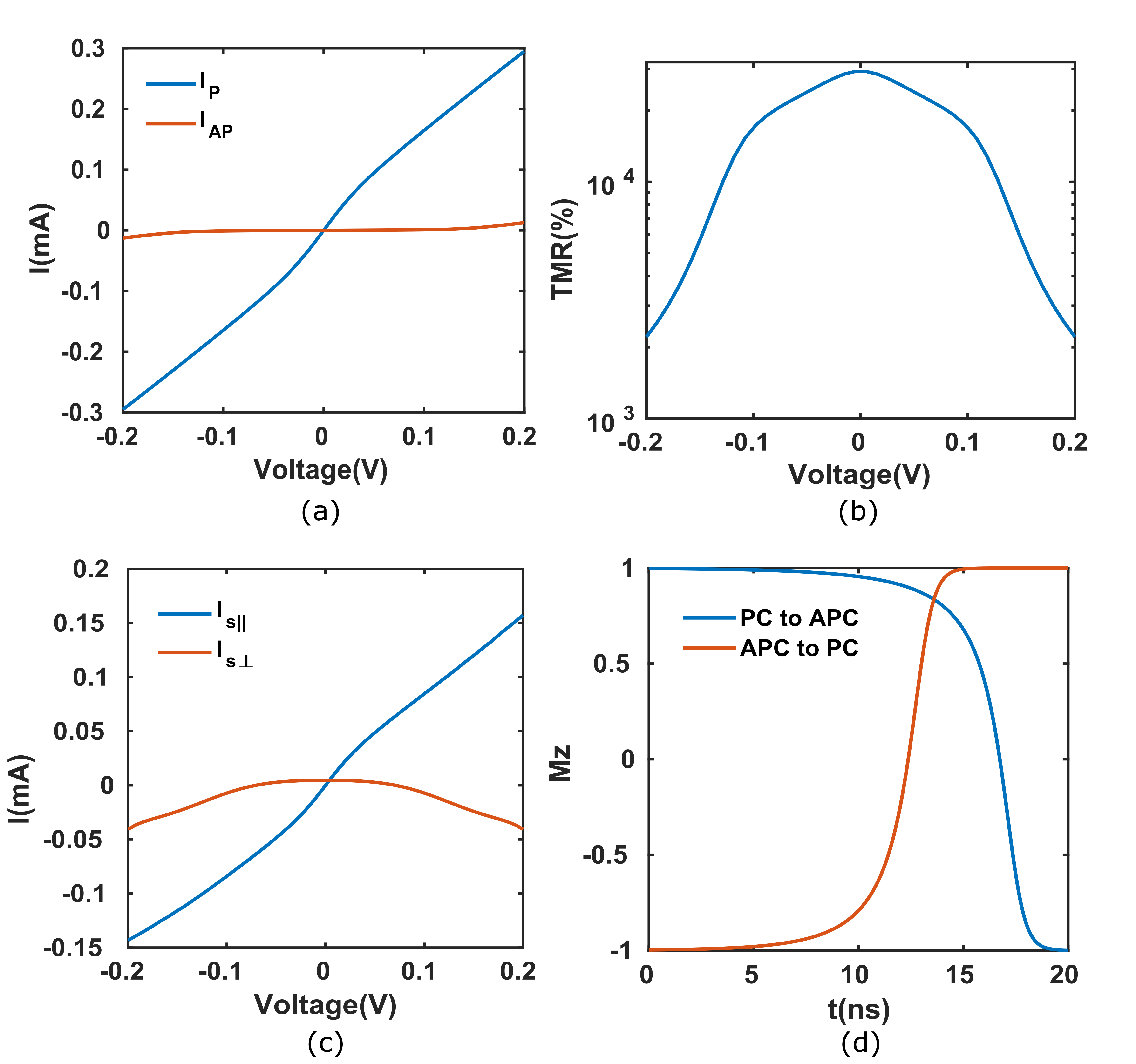}
	\caption{(a) I-V characteristics of the LH-SLTJ in the PC(I$_P$) and APC(I$_{AP}$). (b) Variation of TMR(\%), (c) I$_{s\parallel}$ and I$_{s\perp}$ with the applied voltage. (d) STT based switching  of the free layer from the PC to APC(V=-11.6 mV) and the APC to PC(V=11.6 mV).}
	\label{fig:LHSLTJ_CTIS}
\end{figure}
\indent At first we devise the Linear height(LH) based Nu-SLTJ as shown in Fig. \ref{fig:BDHS}(a), where the scattering potential of the n-th barrier($SP_n=E_f+U_{Bn}$) is given by $$E_f+U_{Bn}=(E_f+U_{B0})-0.4n$$
with $SP_0=E_f+U_{B0}=3.01$ eV representing the central barrier, $SP_1=E_f+U_{B1}=2.61$ eV representing the two neighbour barriers of the central barrier, and  $SP_2=E_f+U_{B2}=2.21$ eV representing the two terminal barriers. We have used MgO to implement the central barrier, and the rest of the barriers are realized by stoichiometrically substituted Mg$_x$Zn$_{1-x}$O where $x$ is substituted in accordance to design the respective scattering potentials.\\
\indent In Fig. \ref{fig:LHSLTJ_CTIS}(a), we present the I-V characteristics of the LHSLTJ in the PC and the APC. The maximum TMR(\%) provided by the device is approximately $\approx 2.9\times10^4$(\%)(Fig. \ref{fig:LHSLTJ_CTIS}(b)). Figure. \ref{fig:LHSLTJ_CTIS}(d) depicts the switching dynamics of the free layer with the  SB$^{PC\rightarrow APC}$ and SB$^{APC\rightarrow PC}$ of -11.6 mV and 11.6 mV, respectively. Therefore, the LH-SLTJ provides an SSB of 5.8, the 4th largest among all the NU-SLTJs(see Tab.~\ref{Table:1}). Similar to the W-SLTJs, the ratio of transmissions in the PC (Fig. \ref{fig:HSLJ_TM}(a)) and APC (Fig. \ref{fig:HSLJ_TM}(b)) is the cornerstone of the towering TMR(\%) we accomplish via LH-SLTJ. In addition to this, the large I$_{s||}$ is attributed to the LH-SLTJ enabled spin-selective broad-band transmission as shown in Fig. \ref{fig:LHSLTJ_CTIS}(c).

\subsection{\label{GHSLTJ}Device characteristics of the Gaussian Height based  Nu-SLTJ(GH-SLTJ)}
To realize the Gaussian height(GH) based Nu-SLTJ shown in Fig. \ref{fig:BDHS}(b), we engineer 
\begin{figure}[h]
	\centering
	\includegraphics[scale=0.34]{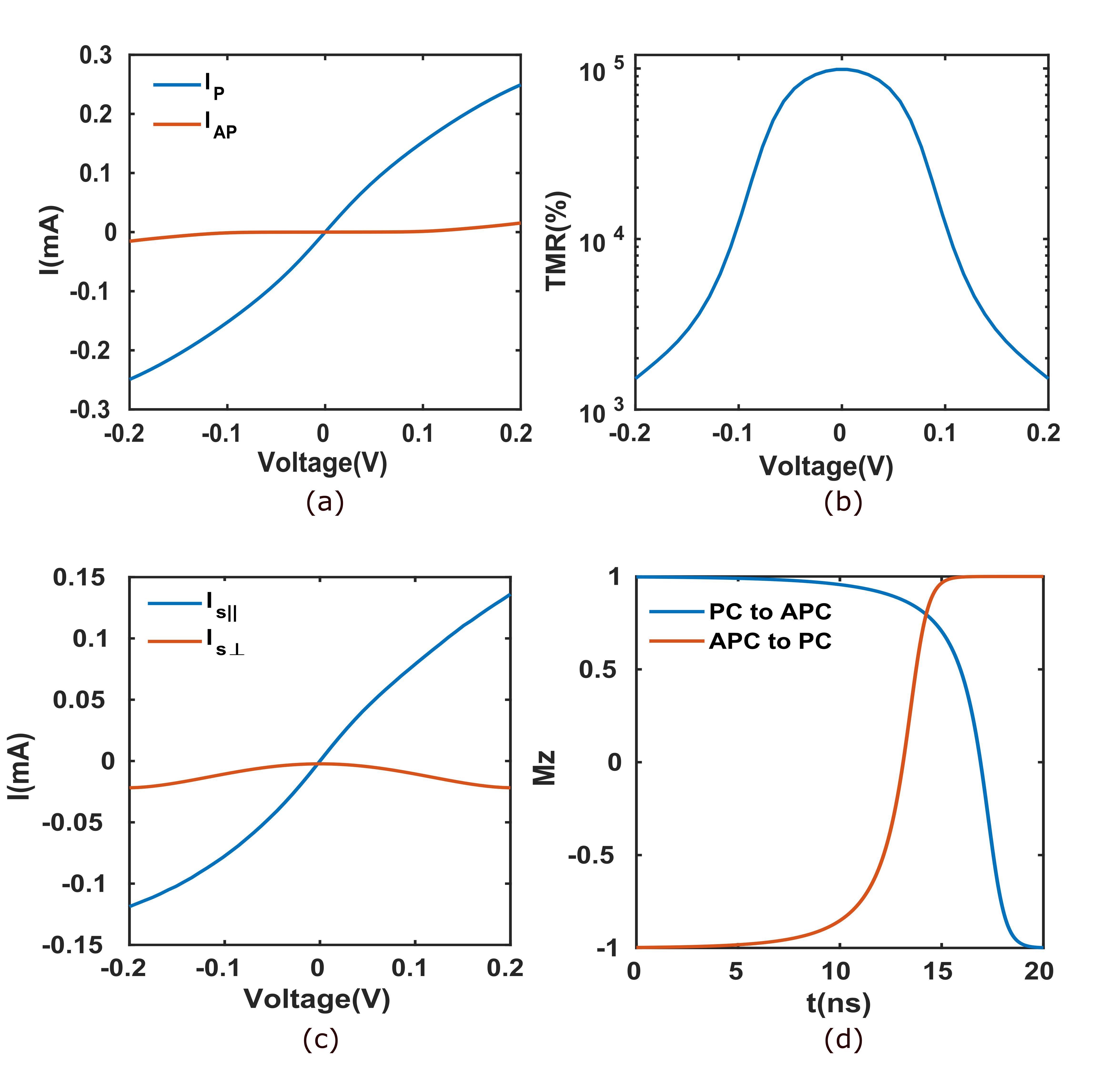}
	\caption{(a) I-V characteristics of the GH-SLTJ in the PC(I$_P$) and APC(I$_{AP}$). (b) Variation of TMR(\%), (c) I$_{s\parallel}$ and I$_{s\perp}$ with the applied voltage. (d) STT based switching  of the free layer from the PC to APC(V=-13.64 mV) and the APC to PC(V=13.64 mV).}
	\label{fig:GHSLTJ_CTIS}
\end{figure}
the scattering potential ($E_f+U_{Bn}$)  of the n-th barrier in such a way that $$E_f+U_{Bn}=(E_f+U_{B0})e^{-n^2/16}$$
where $SP_0=E_f+U_{B0}=3.01$ eV and n=0,1,2.\\
\indent Figure.\ref{fig:GHSLTJ_CTIS}(a) depicts the I-V characteristics of GH-SLTJ in the PC and the APC. The maximum TMR(\%) we notice in this case is nearly $\approx9.87\times10^4$(\%)(Fig. \ref{fig:GHSLTJ_CTIS}(b)). In Fig.\ref{fig:GHSLTJ_CTIS}(c) we present the variation of the I$_{s||}$ and I$_{s\perp}$ with applied voltage. Fig.\ref{fig:GHSLTJ_CTIS}(d) describes The STT-switching of the free layer in the GH-SLTJ where the applied voltages for the PC to APC and the APC to PC switching are -13.64 mV and 13.64 mV, respectively. The SSB observed in this case is the 7th largest with 4.9(see Tab.~\ref{Table:1}). The significant improvement in the TMR(\%) and I$_{s||}$ we achieve via GH-SLTJ can be elucidated in light of the heterostructute enabled broad-band spin filtering demonstrated by Figs. \ref{fig:HSLJ_TM}(c)$\&$(d). 
\subsection{\label{LRHSLTJ}Device characteristics of the Nu-SLTJ with Lorentzian Height Profile(LRH-SLTJ)}
We now advance to engineer the Lorentzian height(LRH) based Nu-SLTJ shown in Fig. \ref{fig:BDHS}(c), where the scattering potential($E_f+U_{Bn}$) of the n-th barrier
\begin{figure}[h]
	\centering
	\includegraphics[scale=0.36]{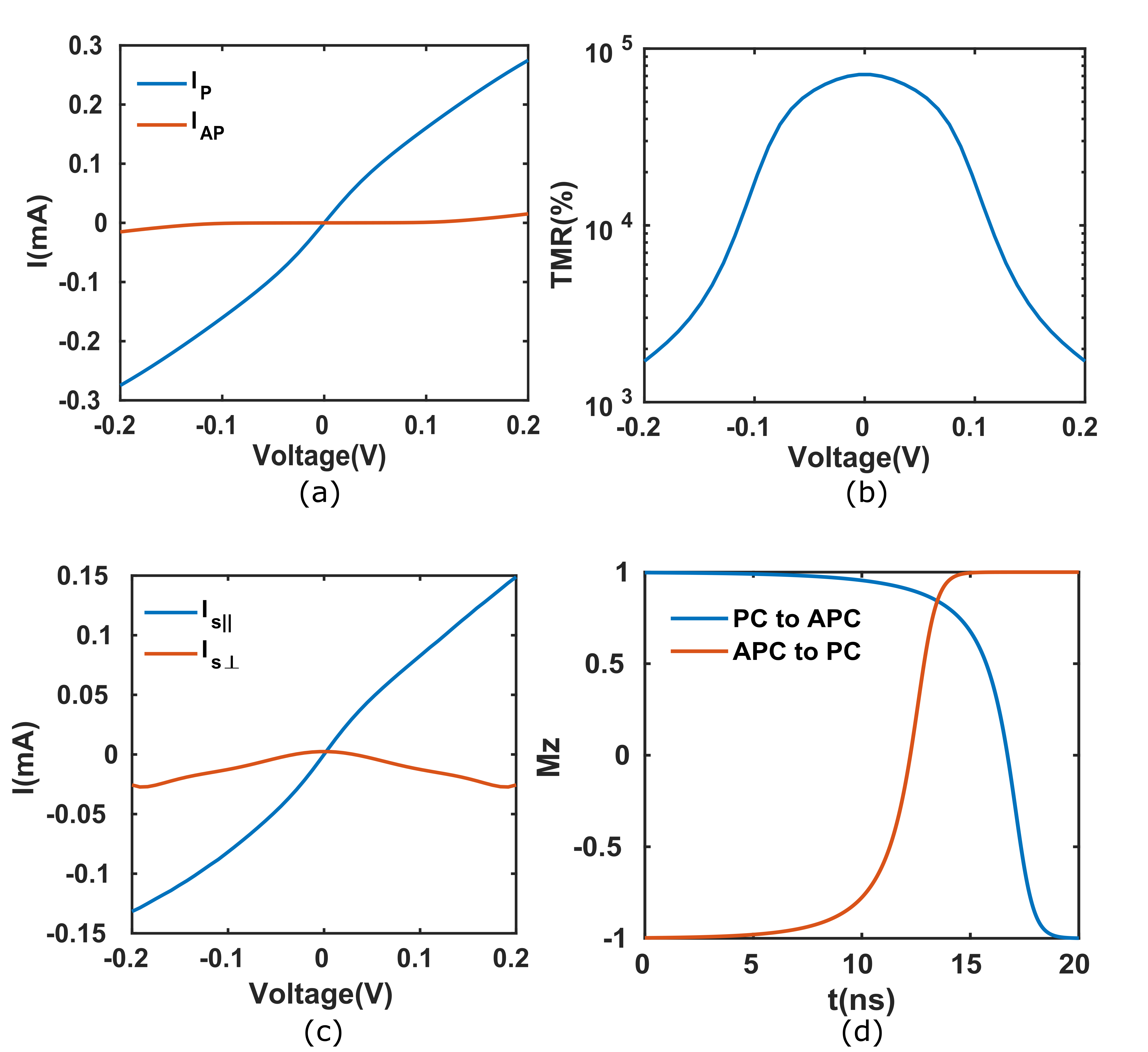}
	\caption{(a) I-V characteristics of the LRH-SLTJ in the PC(I$_P$) and APC(I$_{AP}$). (b) Variation of TMR(\%), (c) I$_{s\parallel}$ and I$_{s\perp}$ with the applied voltage. (d) STT based switching  of the free layer from the PC to APC(V=-12.3 mV) and the APC to PC(V=12.3 mV).}
	\label{fig:LRHSLTJ_CTIS}
\end{figure}
in terms of the central barrier is given by  $$E_f+U_{Bn}=(E_f+U_{B0})\frac{\Gamma^2}{n^2+\Gamma^2}$$ 
with $\Gamma=4$,$SP_0=E_f+U_{B0}=3.01$ eV and n=0,1,2.\\ 
\indent In Fig. \ref{fig:LRHSLTJ_CTIS}(a), we present the I-V characteristics of the LRHSLTJ in the PC and the APC. The maximum TMR(\%) we observe in this case is close to $\approx 7.13\times10^4$(\%)(Fig. \ref{fig:LRHSLTJ_CTIS}(b)). Figure. \ref{fig:LRHSLTJ_CTIS}(d) describes the switching dynamics of the free layer with the  SB$^{PC\rightarrow APC}$ and SB$^{APC\rightarrow PC}$ of -12.3 mV and 12.3 mV, respectively. Therefore, the LRH-SLTJ yields the 6th largest SSB with 5.4(see Tab.~\ref{Table:1}). The momentus enhancement in the TMR(\%) and I$_{s||}$(Fig. \ref{fig:LRHSLTJ_CTIS}(c)) in the LRH-SLTJ can be reasoned out on the basis of the spin selective broad-band filtering demonstrated in Figs. \ref{fig:HSLJ_TM}(e)$\&$(f).
 
\subsection{\label{PHSLTJ}Device Characteristics of the Nu-SLTJ with P\"oscl-Teller Height Profile(PH-SLTJ)}
To devise the  P\"oscl Teller height based Nu-SLTJ shown in Fig. \ref{fig:BDHS}, we consider the scattering potential($E_f+U_{Bn}$) of the n-th barrier($b_n$) as
\begin{figure}[h]
	\centering
	\includegraphics[scale=0.35]{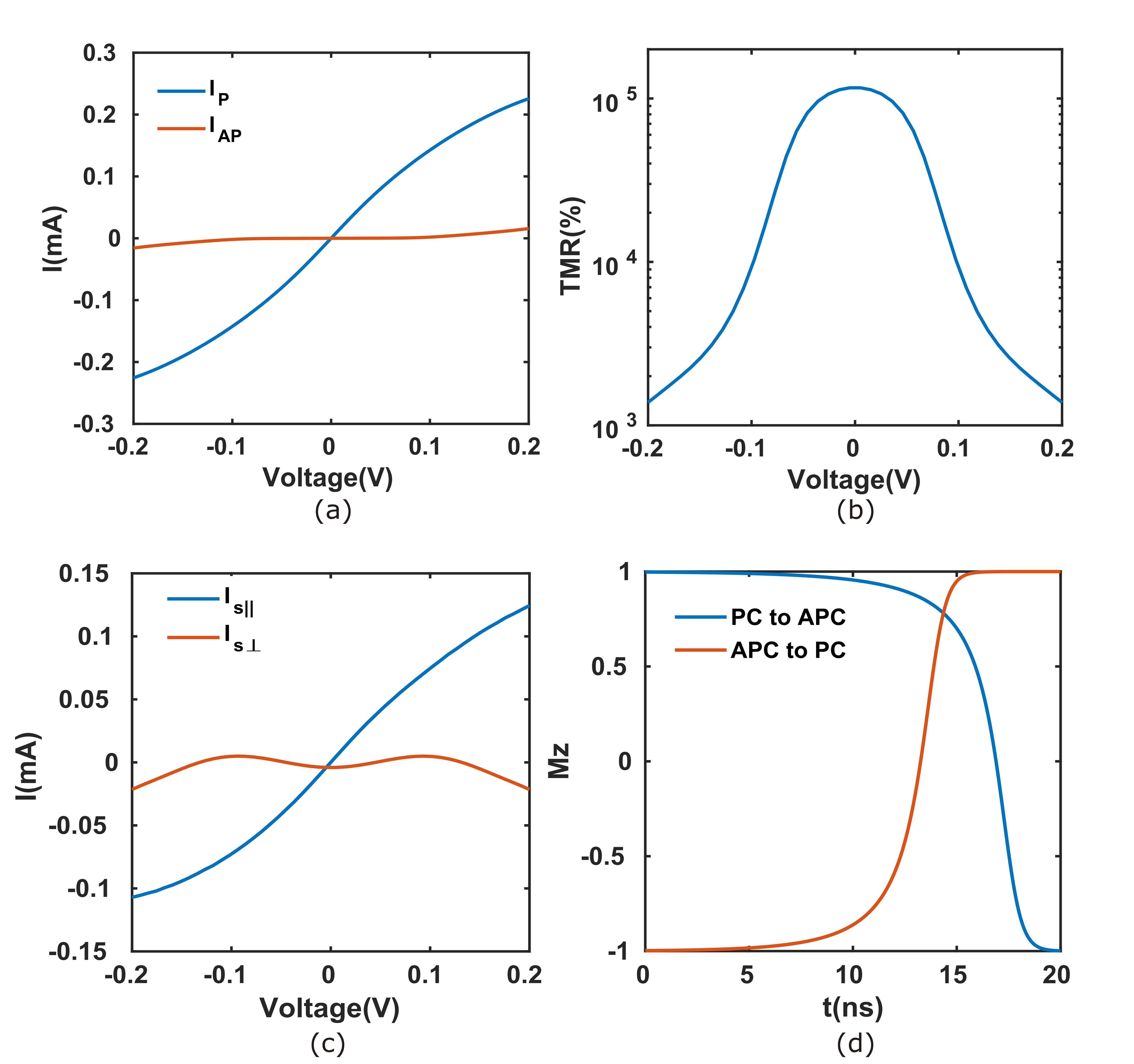}
	\caption{(a) I-V characteristics of the PH-SLTJ in the PC(I$_P$) and APC(I$_{AP}$). (b) Variation of TMR(\%), (c) I$_{s\parallel}$ and I$_{s\perp}$ with the applied voltage. (d) STT based switching  of the free layer from the PC to APC(V=-15 mV) and the APC to PC(V=15 mV).}
	\label{fig:PHSLTJ_CTIS}
\end{figure}
$$E_f+U_{Bn}=(E_f+U_{B0})\frac{2}{e^{n/4}+e^{-n/4}}$$ 
where $SP_0=E_f+U_{B0}=3.01$ eV and n=0,1,2.\\ 
\indent In Fig. \ref{fig:PHSLTJ_CTIS}(a), we present the I-V characteristics of the  PH-SLTJ in the PC and the APC. The maximum TMR(\%) provided by the device is nearly $\approx 1.16\times 10^5$\%(Fig. \ref{fig:PHSLTJ_CTIS}(b)). The variation of I$_{s||}$ and I$_{s\perp}$ with the applied voltage is plotted in Fig. \ref{fig:PHSLTJ_CTIS}(c). Figure \ref{fig:PHSLTJ_CTIS}(d) shows the STT-switching of the free layer from the PC to APC and the APC to PC with SB$^{PC\rightarrow APC}$ and SB$^{APC\rightarrow PC}$ of -15 mV and 15 mV, respectively. Thus, the PH-SLTJ provides the least SSB of 4.5(see Tab.~\ref{Table:1}). The large I$_{s||}$ and the towering TMR(\%) we get to witness in the device can be explained on the basis of the spin-selective broad-band transmissions in the PC and the APC as shown in Figs. \ref{fig:HSLJ_TM}(g)$\&$(h).

We also perform a comparative study of all these Nu-SLTJs along with a regular(R)-SLTJ (Fig. \ref{fig:MTJ_RSLTJ_BD}(b)) as shown in Fig. \ref{fig:NOB_SLTJ}. In our simulations, we have taken the thickness of all the oxide layers and the scattering potentials of the R-SLTJ as 0.8 nm and 3.01 eV respectively. 
   
\section{\label{sec:4}Performance of the Nu-SLTJs at higher voltages}

Another significant aspect of the work regarding the Nu-SLTJs encircles their performance indices at higher voltages.
 \begin{table*}
 \caption{Performance indices of all the  Nu-SLTJs, along with the R-SLTJ, and the AR-SLMTJ. Here TMR(\%)$^{Peak}$ denotes the maximum TMR(\%) provided by a device, $V_P$ represents the voltage where I$_{P}$ peaks and I$_{s||}^{Peak}$ denotes the maximum achievable spin current. The quantities SB$^{PC\rightarrow APC}$ and SB$^{APC\rightarrow PC}$ represent the switching biases for the PC to APC and the APC to PC switching, respectively. TMR(\%)$_{V_P}$ indicates the TMR at V$_P$, and t$^{FS}$ denotes the time required for the fastest switching.}
 \label{Table:1}
\begin{ruledtabular}
\begin{tabular}{cccccccccc}

 {Type of Device} & {TMR(\%)$^{Peak}$} & {SB$^{PC\rightarrow APC}$}(mV) & {SB$^{APC\rightarrow PC}$}(mV)& {SSB} &V$_P$(V) &{TMR(\%)$_{V_P}$} & {I$^{Peak}_{s||}$}(mA) & {I$_{P}^{Peak}$}(mA)& t$^{FS}$(ps)\\
		\hline
		\hline
		MTJ & 240 & -74 & 60 & 1 & -& - & - & - & -\\
		\hline
		R-SLTJ & 1.44$\times10^5$ & -20.4 & 20.4 & 3.28 & 0.24& 6.24$\times10^2$& 0.103 & 0.17 & 1061 \\
		\hline
		LW-SLTJ & 1.01$\times10^5$  & -7.4 & 7.4 & 9.05& 0.3 & 4.06$\times10^2$& 0.3 & 0.47& 380 \\
		\hline
		 GW-SLTJ& 9.01$\times10^5$&-8.88 & 8.88& 7.61 &0.19 & 4.7$\times10^3$ & 0.13 & 0.25& 882\\
		\hline
		 LRW-SLTJ & 2.19$\times 10^6$ & -11.7 & 11.7 & 5.73 & 0.18 & 1.53$\times 10^4$ & 0.1 & 0.18& 1158\\
		\hline
	     PW-SLTJ&  2.61$\times 10^5$ & -7.75 & 7.75 & 8.7& 0.22 & 1.24$\times10^3$& 0.2 & 0.35& 563\\ 
		\hline
	    LH-SLTJ& 2.9$\times 10^4$ & -11.6 & 11.6 & 5.8& 0.43 & 3.18$\times10^2$  & 0.33 & 0.49& 355\\  
		\hline
		GH-SLTJ & 9.87$\times10^4$ & -13.64 & 13.64 &4.9& 0.31 & 5$\times10^3$ & 0.19 & 0.3& 621\\
		\hline
		LRH-SLTJ & 7.13$\times 10^4$& -12.3 & 12.3 & 5.4& 0.36 & 4$\times10^2$ & 0.22 &0.35& 508\\
		\hline
		PH-SLTJ & 1.16 $\times 10^5$& -15 & 15 & 4.5& 0.29 & 5.6$\times10^2$ & 0.16 & 0.26& 672\\
		\hline
	    ARSLMTJ\cite{sharma2021proposal} & 3 $\times 10^4$ & -12 & 12 & 5.6& 0.088 &9.8$\times10^3$ & 0.06 & 0.9& 1996\\

\end{tabular}
\end{ruledtabular}
\end{table*}
Hitherto, the engineering of a robust TMR(\%) with the applied bias has appeared to be a major challenge in the novel literature\cite{sharma2018band} that reverberates while designing the Nu-SLTJs as well. Nevertheless, we demonstrate that the introduction of superlattices can accomplish a decent TMR(\%) at a higher voltage and also explore the maximum achievable I$_{s||}$ via various Nu-SLTJs to perform faster switching.\\
 \begin{figure}[h]
	\centering
	\includegraphics[scale=0.36]{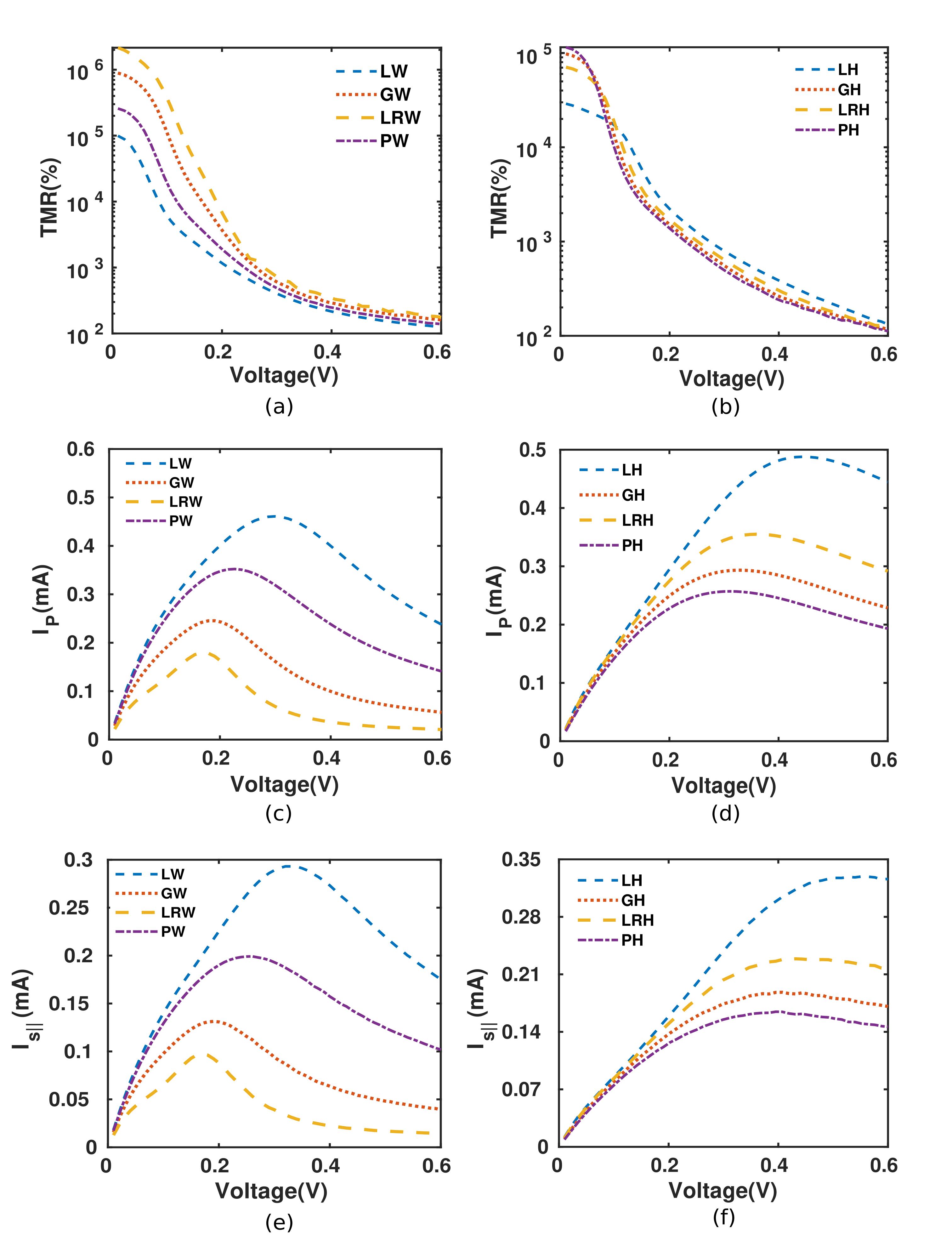}
	\caption{ High-voltage characteristics: Variation in the TMR(\%) of the (a)W-SLTJs and the (b)H-SLTJs, charge current(I$_P$) of the (c)W-SLTJs and the (d)H-SLTJs  and the I$_{s||}$ of the (e) W-SLTJs and the (f) H-SLTJs with respect to the applied bias. }
	\label{fig:HV_CTIS}
\end{figure}
\indent In the first place, we present a qualitative overview of the current flow through an Nu-SLTJ on the application of an applied bias, governed by the areas of the transmissions in the PC(A$_{PC}$) and APC(A$_{APC}$). To calculate the current for a specific voltage, we multiply the quantity $(f_F(E)-f_f(E))$ with the transmission $T(E)$, where the $f_F$ and $f_f$ represent the Fermi distribution functions of the fixed and free FMs, respectively. Along with an increase in the voltage across an Nu-SLTJ, the difference between the $f_F$ and $f_f$ goes up, resulting a greater leeway for tunneling of the electrons that leads to an upsurge in the current conduction. Besides, furtherance in the voltage after a certain point fails to engender an increment in the current flow. This can be ascribed to the bandpass nature of the transmission that inhibits any additional area of the transmission($A_T$) from falling within the Fermi window(FW) beyond an instance when the FW engulfs significant $A_T$. Moreover, as the magnitude of the applied voltage is increased even further, the transmission spectrum of a typical Nu-SLTJ becomes tapered. This results in a reduction in the current conduction that compels all the Nu-SLTJs to exhibit negative differential resistance(NDR) beyond a specific voltage($V_P$).\\ 
\indent We present the high-voltage characteristics of various Nu-SLTJs in Fig. \ref{fig:HV_CTIS} and show the variation of the TMR(\%), I$_{s||}$ and the I$_{P}$ while varying the applied voltage from 0.01 V to 0.6 V. We exclude the point at V=0, where the TMR of the devices becomes immeasurable owing to the zeroed I$_P$ and I$_{AP}$. In Fig. \ref{fig:HV_CTIS}(a),(b),(c)$\&$(d) we show that the TMR of both the H-SLTJs and the W-SLTJs exhibits a large variation with respect to the applied bias alongside an NDR of the I$_P$-V characteristics after $V_P$. Interestingly, after reaching the peak, the decline in the I$_{s||}$ of the H-SLTJs as shown in Fig. \ref{fig:HV_CTIS}(e)$\&$(f) becomes nearly insignificant, thereby providing grist to an idea for utilizing this behavior to realize a device that produces invariable I$_{s||}$ with an applied voltage.\\    
\indent The deterioration in the ratio of A$_{PC}$ and A$_{APC}$ quenches the TMR(\%) of an orthodox MTJ at a higher voltage. For instance, the trilayer MTJ we present in this work offers a TMR(\%) of $\approx$ 220\% and $\approx$ 140\% at an applied voltage of 10 mV and 300 mV as shown in Fig. \ref{fig:MTJ_CTIS}(b). We stumble upon a similar problem while engineering the Nu-SLTJs as well. Yet, the heterostructure-enabled broad-band spin filtering inhibits the TMR(\%) from falling below $\approx$10$^{3}$-10$^4$(\%) while the applied bias ranges from -0.2V to 0.2V. This provides a decent solution regarding the sensitivity of the MTJs within this range. Another question, that may lie dormant is the importance of the TMR(\%) at a higher voltage while the switching biases of all the SLTJs lie fairly below 50 mV, is answered by highlighting the applicability of an MTJ in the STNOs. A high TMR(\%) at V$_{P}$ offers a bigger difference between the I$_P$ and I$_{AP}$, which ensures a surge in the microwave output power. Therefore the Nu-SLTJs, while providing a far superior TMR(\%) and I$_P$  compared to an MTJ, emerge as more suitable alternatives in order to design large frequency STNOs with high microwave output power\cite{PhysRevApplied.8.064014}. Furthermore, we can infer from Tab. \ref{Table:1} that the LW-SLTJ provides an I$_{P}$ of 0.47 mA at the V$_P$ along with a TMR(\%) of 4.06$\times 10^2$(\%) which entitles the device as a much-coveted candidate for the aforementioned application.

 \section{\label{sec:5}Comparison of the Performance Indices of the Various Nu-SLTJs}
In this section, we  discuss the performance indices of the various SLTJs presented in Tab. \ref{Table:1} and carry out a comparative study in reference to the transmission profiles described in Appendix \ref{Appendix:B}.
We attribute the monumental TMR(\%) profiles accomplished via different heterostructures to the towering ratio of the transmissions in the PC($T_{PC}$) and APC(T$_{APC}$), and elucidate the significant enhancement in the I$_{s||}$ by the spin selective nature of the broad-band T$_{PC}$(see Appendix \ref{Appendix:B}).\\
\indent Figure. \ref{fig:WSLTJ_TM} exemplifies that the LW-SLTJ yields the largest T$_{PC}$ and hence offers a sizeable I$_{s||}$ that leads to the highest SSB  (see Tab. \ref{Table:1}) among all the Nu-SLTJs. Thereafter, the PW-SLTJ, GW-SLTJ and LRW-SLTJ comprise the list of the W-SLTJs in the descending order with reference to the A$_{PC}$ as shown in Fig. \ref{fig:WSLTJ_TM}. A similar order of the I$_{s||}$ and SSB of the respective W-SLTJs corroborates the potential impact of the A$_{PC}$ on the Slonczewski term(I$_{s||}$). We also get to witness an impressive consistency in the order of the TMR(\%) with the ratio of the A$_{PC}$ and A$_{APC}$ while analysing the relative performances of the various W-SLTJs. Fig. \ref{fig:WSLTJ_TM} conveys the LRW-SLTJ to possess the largest A$_{PC}$/ A$_{APC}$ which enables the device to exhibit the tallest TMR(\%). Interestingly, the ascending order of the W-SLTJs with respect to the I$_{s||}$ constitutes the descending order with reference to the TMR(\%).\\ 
\indent Akin to the W-SLTJs, the heterostructure-enabled spin-selective transmissions play a significant role in governing the performance indices of the H-SLTJs as well. The LH-SLTJ, LRH-SLTJ, GH-SLTJ, and PH-SLTJ comprise the descending order of the H-SLTJs with respect to the A${_{PC}}$ as shown in Fig. \ref{fig:HSLJ_TM}. The I$_{s||}$ we observe in these devices are congruent with their transmission spectra. For instance, the GH-SLTJ yields an exacerbated T$_{PC}$  compared to the LH-SLTJ resulting in a decimated I$_{s||}$. In addition to this, we also find the variation in the  TMR(\%) to be proportional with the ratio of A${_{PC}}$ and A$_{APC}$. As we move from the PH-SLTJ to LH-SLTJ, the depression in the value of A$_{PC}$/A$_{APC}$ reduces the magnitude of the TMR(\%). Similar to the W-SLTJs, the order of the H-SLTJs with respect to I$_{s||}$ and TMR(\%) are mutually opposite.\\
\indent It seems pretty outlandish that despite having a significant T$_{PC}$, the H-SLTJs tend to fall behind in accomplishing a  mammoth SSB, unlike the W-SLTJs. We can ascribe such behaviour to the widespread finger-like profile of the transmissions(see Fig \ref{fig:WSLTJ_TM} \& \ref{fig:HSLJ_TM}), which reduces effective area within the Fermi window near the switching bias and scales down the steepness of the I$_{s||}$-V characteristics(see Fig. \ref{fig:HV_CTIS}(e)\&(f)). On the contrary, although the W-SLTJs exploits their compact T$_{PC}$ to accomplish a sharp slope in the I$_{s||}$-V curve, the narrower width in the T$_{PC}$ reduces the V$_{P}$ of the devices. As a consequence, the W-SLTJs tend to exhibit the NDR at a lower voltage.\\  
\begin{figure}[h]
	\centering
	\includegraphics[scale=0.33]{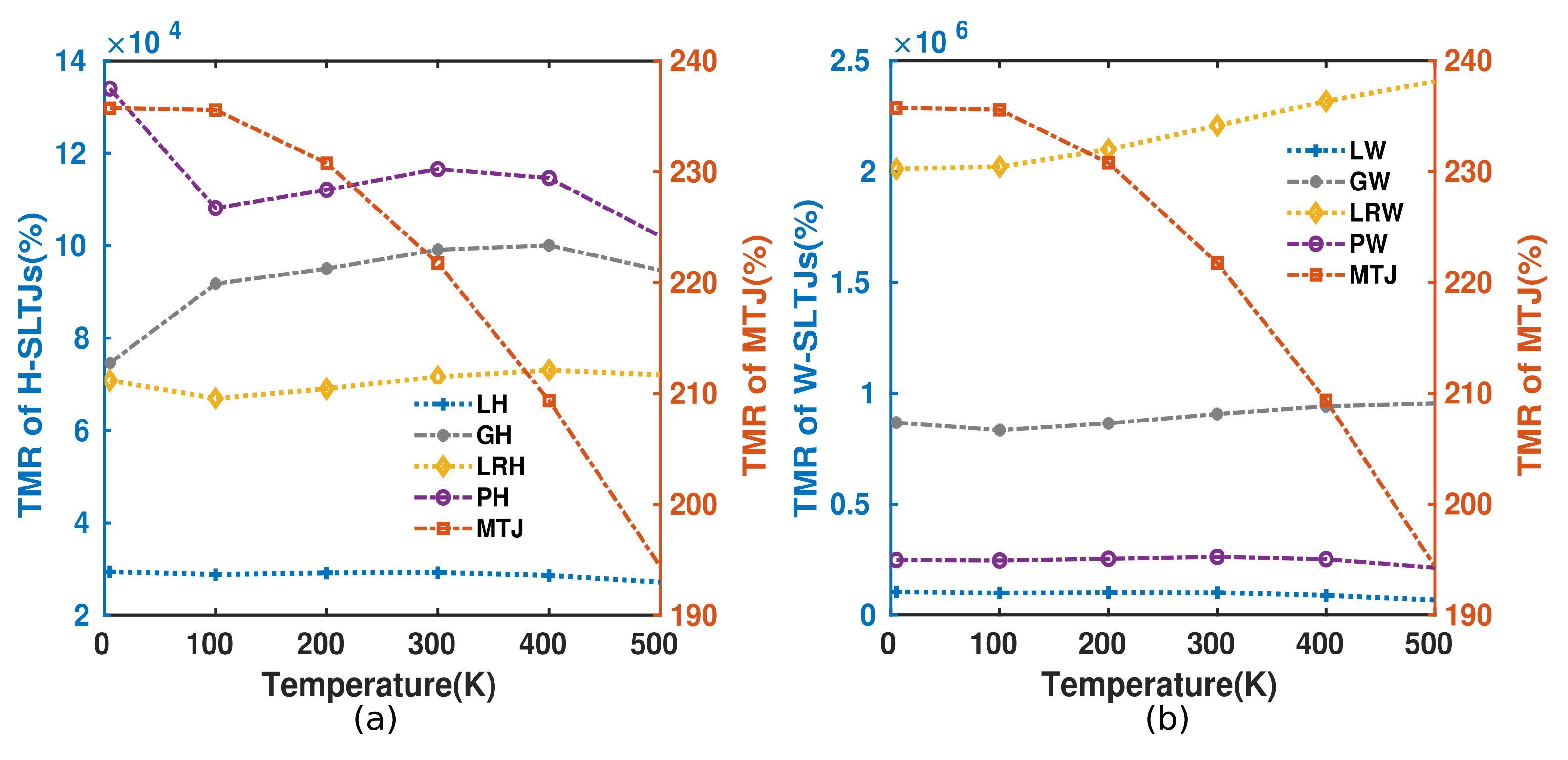}
	\caption{ Variation in the TMR(\%) of the (a)H-SLTJs and the (b)W-SLTJs in reference to a trilayer MTJ with temperature.} 
	\label{fig:Temperature_TMR}
\end{figure}
\indent The compendium of Nu-SLTJs devised in this work offer fairly superior performances compared to the anti-reflective superlattice magnetic tunnel junction (ARSLMTJ) \cite{sharma2021proposal} in many ways. In the first place, we can infer from Tab. \ref{Table:1} that the SSB of the ARSLMTJ is lower than that of all the W-SLTJs, including the LH-SLTJ. Other than that, the maximum I$_{s||}^{peak}$ and the TMR$_{V_P}$ that  we accomplish via LH-SLTJ and LRW-SLTJ are 600\% and 150\% higher than same.  Importantly, both the W-SLTJs and the H-SLTJs may comprise minimal device structures of three oxide barriers (see Fig. \ref{fig:NOB_SLTJ}) whereas the ARSLMTJ requires minimum five barriers for operation. Thus, the resounding superiority in the performance indices takes place in the Nu-SLTJS along with a reduction in fabrication complexity.\\
\indent It is worth mentioning that despite having a near box-cart transmission, the ARSLMTJ fails to outperform the W-SLTJs and the LH-SLTJ while providing a lower SSB. This can be accounted to the narrowness of it's transmission, which reduces the A$_{PC}$ engulfed by the FW near the switching bias. These findings do not underpin the necessity of the box-cart-shaped transmissions and establish the importance of a higher A$_{PC}$ within the FW as the keystone for accomplishing a lower switching bias. Another dimension that we unveil in this work is the high TMR(\%)(Fig. \ref{fig:NOB_SLTJ}(a)$\&$(b)) of the R-SLTJ devoid of a box-cart T$_{PC}$(Fig. \ref{fig:MTJ_RSLTJ_TM}(c)), which demonstrates the A$_{PC}$/A$_{APC}$ to be the predominant factor in order to achieve a high TMR(\%).
\section{\label{sec:6} Temperature dependence of the TMR of the Nu-SLTJs}

The TMR of a CoFeB/MgO/CoFeB-based MTJ has typically been observed to show a monotonic decline with the temperature\cite{zhao2022temperature,kou2006temperature}. In Fig.~\ref{fig:Temperature_TMR}
we have shown the TMR dependence of the Nu-SLTJs in reference to a typical MTJ. The decline in the TMR of a trilayer MTJ can be elucidated by the fact that an increase in the temperature widens the FW,  which allows the transmissions at higher energies to get indulged in electron transport. Unfortunately, as we move towards higher energies, the ratio of the A$_{PC}$/A$_{APC}$ gets reduced(Fig.~\ref{fig:MTJ_RSLTJ_TM}(a)), which in turn declines the TMR\%. Although earlier works have ascribed the TMR roll-off with temperature to the magnons\cite{zhao2022temperature,PhysRevB.77.014440,WANG20151649,8532138}, the same can also be explained by the analysis of the spin-dependent transmission spectra within the FW. The TMR of the Nu-SLTJs exhibits notable robustness to temperature variations. This can be attributed to the broadband nature of the transmissions, which doesn't allow the ratio of A$_{PC}$ and A$_{APC}$ at higher energies to head downhill. The non-monotonic behavior of the TMR, we observe in the Nu-SLTJs can be explained on the premise of broadened FW at higher temperatures and the transmission that gets engulfed by it at various transverse modes. The analysis of the TMR rendered by the Nu-SLTJs while subjected to magnon interactions\cite{PhysRevB.77.014440} is beyond the scope of this work and shall be addressed in future.
\section{\label{sec:7}Conclusion}
  Low TMR(\%) and high switching bias\cite{PhysRevApplied.8.064014,7571106,sharma2018role} are the two foremost drawbacks that gridlock the possibility a typical MTJ to become a superior alternative for state-of-the-art storage devices. This study explores the broad-band spin filtering in various Nu-SLTJs that manifests an appreciable enhancement in the TMR and a significant SSB, while endowing the LW-SLTJ with the best TMR(\%)-$I_{s||}$ trade-off(TMR(\%)$\approx$10$^5$, SSB$\approx$9.05 and t$^{FS}\approx 380$ ps). Since the Nu-SLTJs accomplish a fast STT-switching in the order of a few hundreds of picoseconds, they successfully address the shortcoming of non-deterministic switching in the SOT devices while keeping up with the speed\cite{meo2022magnetisation,prenat2015ultra}. We unveil that a large $A_{PC}$ and a high  A$_{PC}$/A$_{APC}$ within the FW overrides the necessity for the box-cart shape of a transmission in order to obtain a low switching bias and a high TMR(\%). Apart from that, we anticipate the Nu-SLTJs devised to find possible applications in the STNOs due to their large I$_P$ and considerable TMR(\%) at V$_P$. 
  Moreover, we foresee the recent development of double metallic quantum well-based MTJs\cite{tao2019coherent} to add further fuel in the investigation of heterostructure-based memory devices. Finally, we conclude the theoretical analysis of Nu-SLTJs with lower switching bias and high TMR(\%) proposed in this article to hopefully pave the avenues for cutting-edge heterostructure-based magnetic tunnel junctions.
\section*{Data availability statement}
The data that support the findings of this study are available upon reasonable request from the authors.
 \appendix
 \section{{ Switching of the free layer}
\label{Appendix:A}}
\begin{figure}[h]
	\centering
	\includegraphics[scale=0.34]{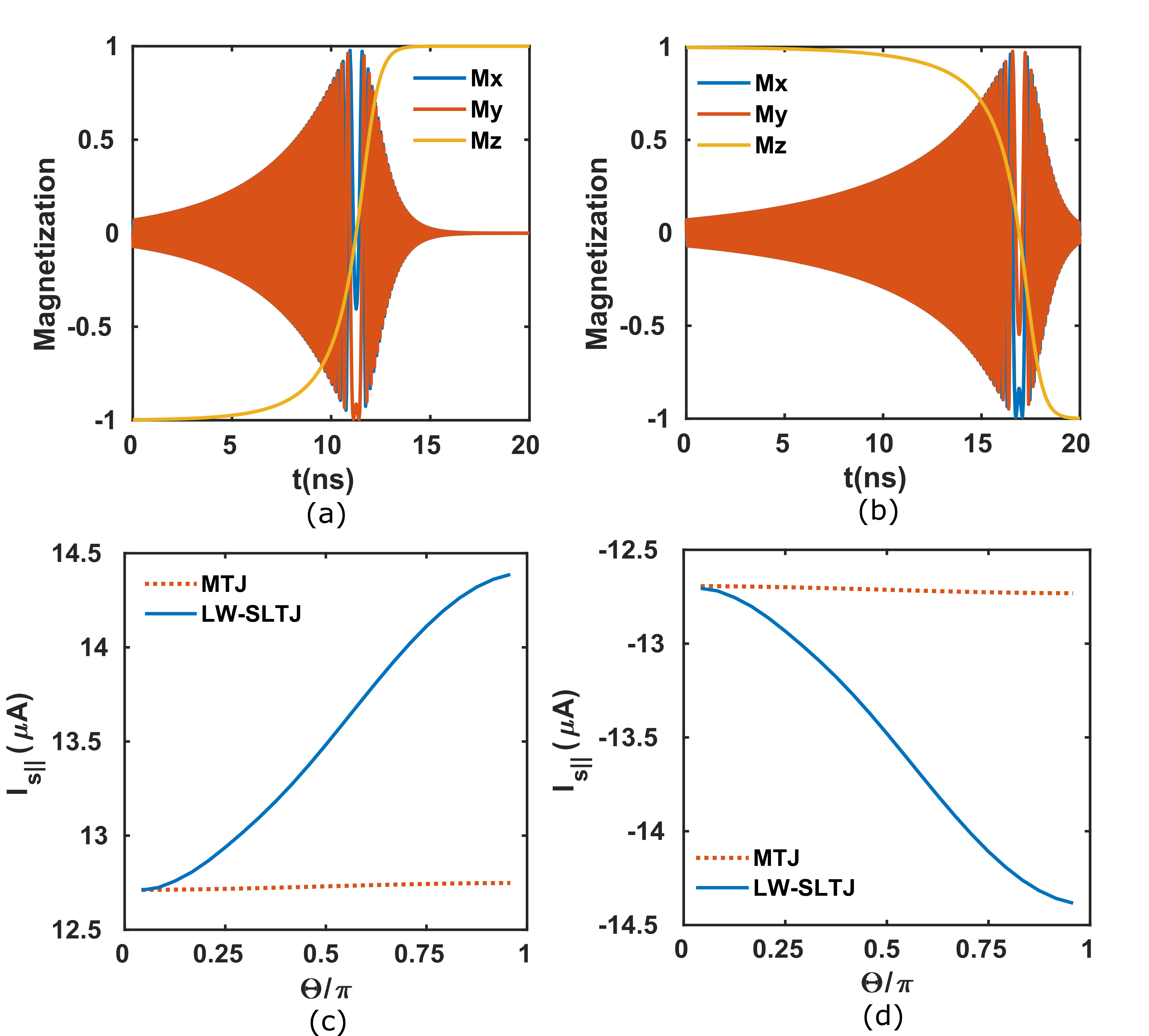}
	\caption{ Switching dynamics of the LW-SLTJ: (a) APC to PC switching (V=7.4 mV) (b) PC to APC switching (V=-7.4 mV). Variation of the $I_{s||}$ with $\theta$ for the LW-SLTJ(MTJ) at (c) SB$^{APC\rightarrow PC}_{LW-SLTJ(MTJ)}$ and (d)SB$^{PC\rightarrow APC}_{LW-SLTJ(MTJ)}$.}
	\label{fig:SWD}
\end{figure}

This section deals with the STT-switching of the free FM in the LW-SLTJ, as shown in Fig. \ref{fig:SWD}.  
The analysis of the switching dynamics involves the self-consistent coupling of the NEGF formalism with the LLGS equation described in Fig. \ref{fig:SelfConsistent}. As the applied I$_{s||}$ and the magnetization of the free FM resides in the same direction at equilibrium, it is necessary to impart a slight misalignment in the magnetization to produce a non-zero torque\cite{meo2022magnetisation}. Since it is not possible to switch the free FM from either 0$^{\circ}$  or 180$^{\circ}$ for the above-mentioned reasons, we assume the relative angle between the magnetization of the free and the fixed FM($\Theta$) as 4$^{\circ}$ and 176$^{\circ}$ in the PC and the APC, respectively.
 Apart from that, a switching current near the I$_{s||,c}$ induces an insignificant damping to the magnetization. Hence in order to perform a comfortable switching, we apply an I$_{s||}$ that exceeds the critical current I$_{s||,c}$ by 20\%. The desired I$_{s||}$ is achieved with an $SB^{PC\rightarrow APC}$ and $SB^{APC\rightarrow PC}$ of -7.4 mV and 7.4 mV, respectively. In contrast to a typical MTJ, we observe a considerable variation in the I$_{s||}$ of the LW-SLTJ with respect to the $\Theta$ as shown in Fig. \ref{fig:SWD}(c)$\&$(d). Therefore, the time required for the PC to APC(Fig. \ref{fig:SWD}(a)) and the APC to PC(Fig. \ref{fig:SWD}(b)) switching  becomes dissimilar. Analogous to the LW-SLTJ, the I$_{s||}$ of all the Nu-SLTJs devised in this work exhibit a similar behavior with the $\Theta$.
 \begin{figure*}
\includegraphics[scale=0.38]{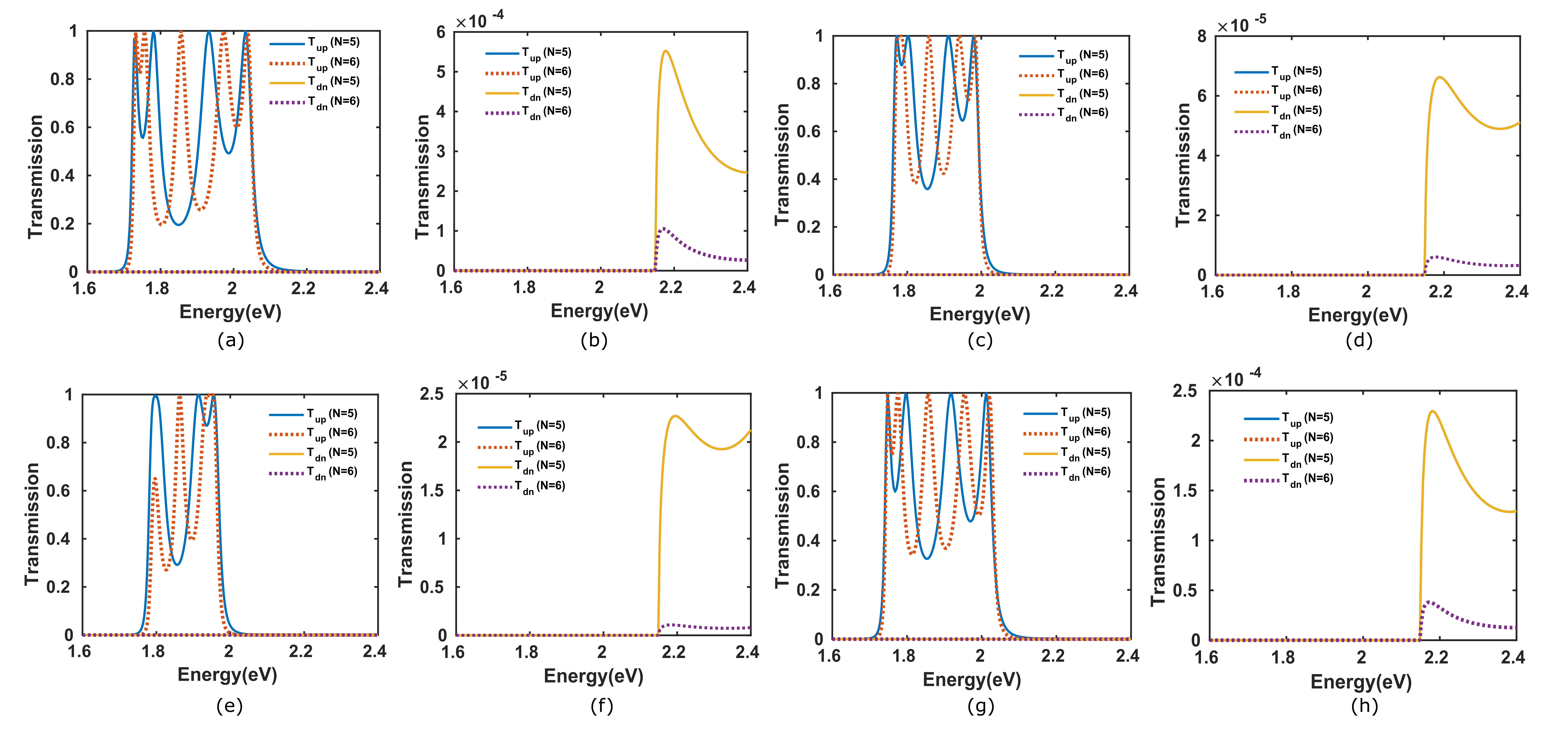}
\caption{Transmission profies of the 
(a) LW-SLTJ in the PC and (b) the APC, (c) GW-SLTJ in the PC and (d) the APC,
(e) LRW-SLTJ in the PC and (f)the APC (g) PW-SLTJ in the PC and
(h) the APC at the lowest available transverse energy with an applied bias of V=0V. Here T$_{up}$ denotes up-spin transmission and T$_{dn}$ represents down-spin transmission. The transmissions are given for number of oxide barriers(N)=5 and N=6.}
\label{fig:WSLTJ_TM}
\end{figure*}

 The magnitude of the $SB^{PC\rightarrow APC}$ and the $SB^{APC\rightarrow PC}$ remains the same for all the Nu-SLTJs due to symmetrical variation of the I$_{s||}$ with respect to the applied voltage near 0V. 
 \section{{Analysis of the Transmission Spectra}
 \label{Appendix:B}}
The analysis of the transmission spectra provide vivid insights to engineer the heterostructure-based magnetic tunnel junctions.
\begin{figure}[h]
	\centering
	\includegraphics[scale=0.34]{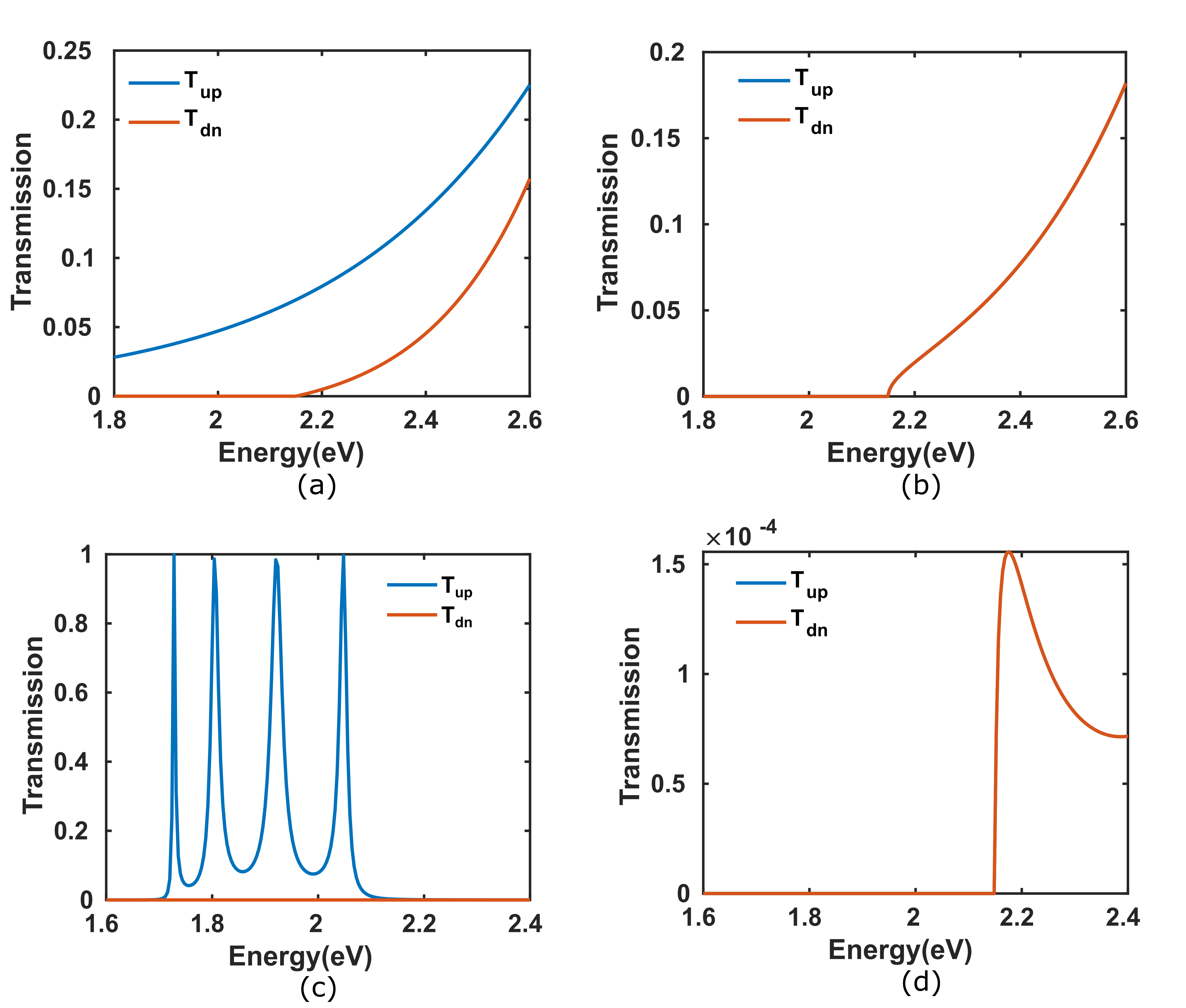}
	\caption{(a) Transmission profile of the trilayer MTJ in the PC and (b) APC, (c) transmission profile of the  R-SLTJ in the PC and (b) APC at the lowest available transverse energy(E$_t$) with the applied voltage V=0V. Here T$_{up}$ denotes upspin transmission and T$_{dn}$ represents downspin transmission.}
	\label{fig:MTJ_RSLTJ_TM}
\end{figure}
In order to determine the current through a mesoscopic device such as an MTJ and/or Nu-SLTJ, we  multiply the Fermi level difference of the fixed and free FMs
\begin{figure*}
\includegraphics[scale=0.39]{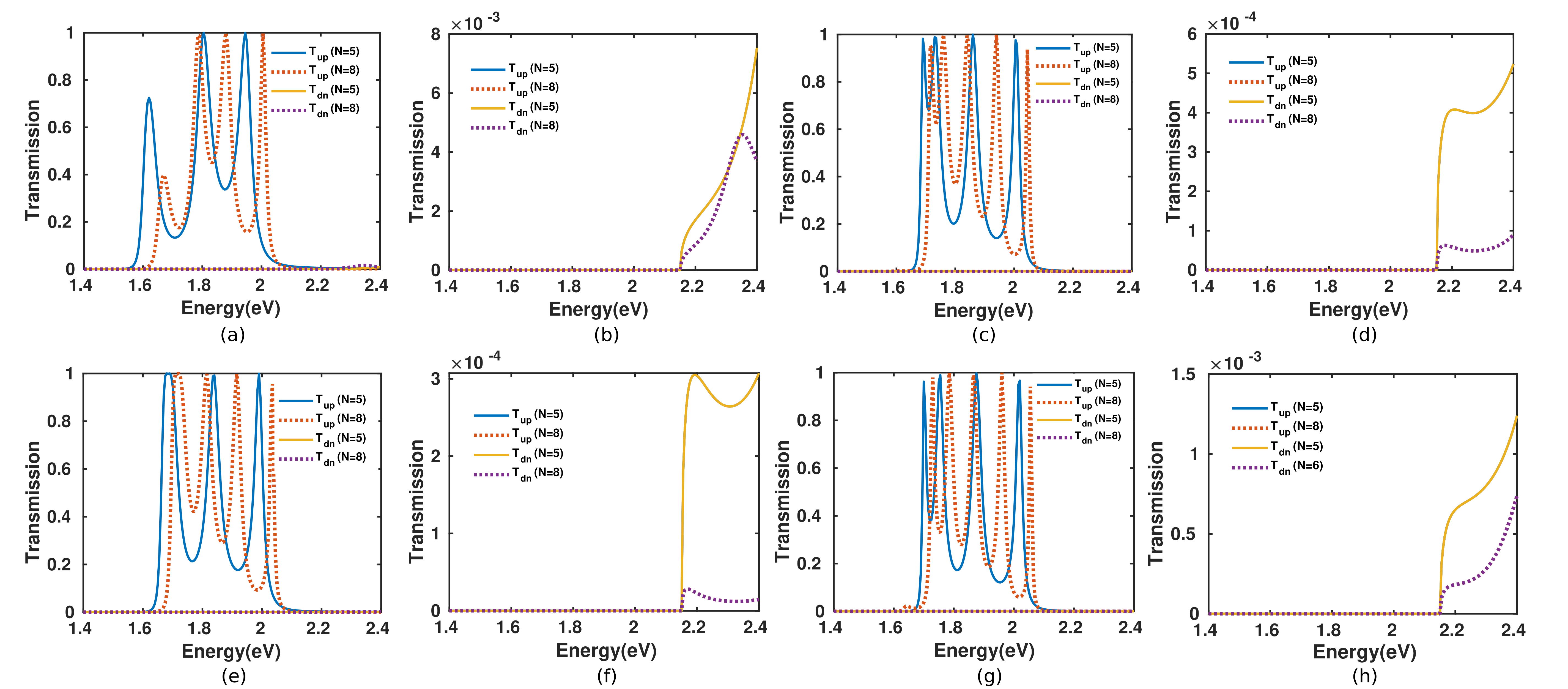}
\caption{Transmission profile: (a) LH-SLTJ in the PC and (b) APC, (c) GH-SLTJ in the PC and (d) APC, (e) LRH-SLTJ in the PC and (f) APC, (g) PH-SLTJ in the PC and (h) APC at the lowest available transverse energy with an applied bias of V=0V. Here T$_{up}$ denotes upspin transmission, and T$_{dn}$ represents downspin transmission.  The transmissions are given for the number of oxide barriers(N)=5 and N=6.}
\label{fig:HSLJ_TM}
\end{figure*}
to the transmission and integrate the product over the energy.
\begin{figure}[h]
	\centering
	\includegraphics[scale=0.4]{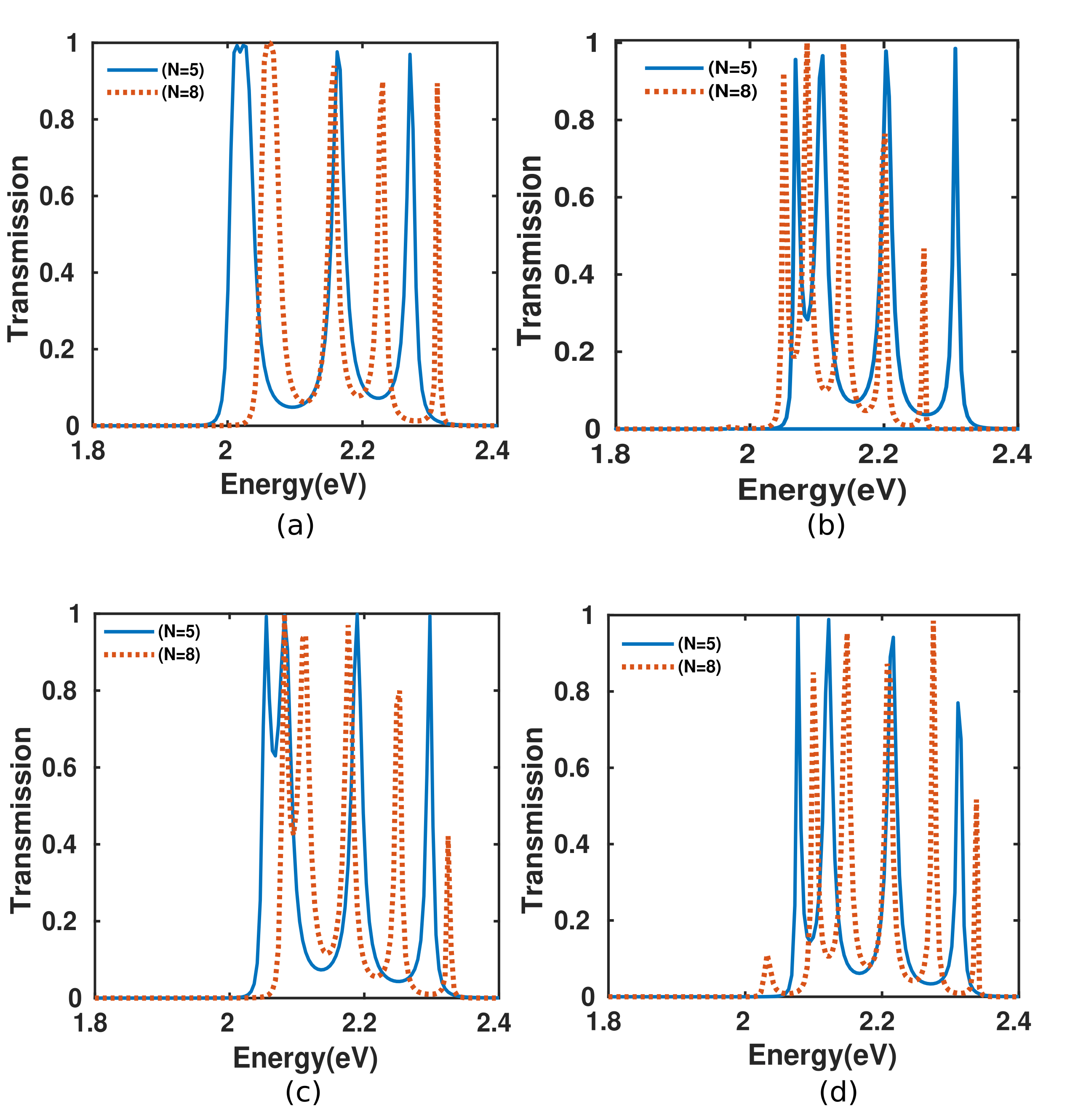}
	\caption{Transmission profiles of the (a) LH-SLTJ, (b) GH-SLTJ, (c) LRH-SLTJ and (d) PH-SLTJ in the PC for N=5 and N=8 at an E$_t$=0.2 eV with the applied voltage V=0.}
	\label{fig:HSLTJ_F_window}
\end{figure}
\indent Hence, the ratio of the A$_{PC}$ and A$_{APC}$ within the Fermi-window(FW) plays a pivotal role in governing the TMR(\%).  Apart from that, the up-spin transmission(T$_{up}$) in a spin-selective heterostructure heavily dominates its down-spin counterpart(T$_{dn}$).
Such compelling spin filtering causes the spin(I$_{z}$=I$_{up}$-I$_{dn}$) and the charge current(I$_P$) in the $z$ direction to become nearly equal in magnitude. Further, the strong correlation of the I$_z$ with the Slonczewski term (I$_{s||}$) in the spin selective SLTJs ensures a high I$_P$ to eventuate a large I$_{s||}$. Based on these heuristic arguments, we utilize the T$_{PC}$ to  predict the behavior of the I$_{s||}$ with the applied bias.\\
\indent We embark on elucidating the spintronic characteristics of various Nu-SLTJs with the T$_{PC}$ and  T$_{APC}$ of both the trilayer MTJ and the regular SLTJ depicted in Fig. \ref{fig:MTJ_RSLTJ_TM}.
The modest opposite spin transmission shown in Fig. \ref{fig:MTJ_RSLTJ_TM}(a) declines the spin filtering capability of the trilayer MTJ, which restricts the TMR to a trifling 240\% compared to the Nu-SLTJs.
In addition to this, the value of the T$_{PC}$ lies well below 1 within the FW, which exacerbates the A$_{PC}$ resulting in a low I$_{s||}$. On the other hand, despite offering a commendable spin filtering, the R-SLTJ exhibits a finger-like T$_{PC}$ which significantly reduces the A$_{PC}$ and thereby yields a feeble I$_{s||}$ along with a meager SSB. Therefore, we propose the idea of Nu-SLTJs, where the presence of the electronic Bloch states in the neighbourhood of the resonant peaks gives rise to a  spin selective broad-band T$_{PC}$, resulting an admirable I$_{s||}$.
The broad-band spin filtering is referred to the widened T$_{PC}$ that allows a broad transmission window for the electrons aligned to the FMs in the PC, while the opposite spin transmission is nearly suppressed to zero as shown in Figs. \ref{fig:HSLJ_TM},\ref{fig:WSLTJ_TM}. Further, the towering ratio of the   A$_{PC}$ and the A$_{APC}$ witnessed in the Nu-SLTJs (Figs. \ref{fig:HSLJ_TM},\ref{fig:WSLTJ_TM}) leads to an ultra-high boost in the TMR(\%).\\
\indent In Fig. \ref{fig:WSLTJ_TM} we present the transmissions of various W-SLTJs in the PC and the APC for both $N=5$ and $N$=6, where $N$ denotes the number barriers comprising the Nu-SLTJs. We find the quantity A$_{PC}$/A$_{APC}$ and the T$_{PC}$ to elucidate the TMR(\%) and the I$_{s||}$ with a noble consistency. For example, the ratio of the A$_{PC}$ and A$_{APC}$ is able to coherently predict the comparative performance of the TMR(\%) in the various W-SLTJs. In addition to this, the relative magnitude of the I$_{s||}$ corroborates the decisive impact of the T$_{PC}$ on it. Regardless of the width based profiles, the passband in the T$_{PC}$ exhibits remnant oscillations as described in  Fig. \ref{fig:WSLTJ_TM}. The T$_{PC}$ of the W-SLTJs display a slight reduction in the area  with a surge in the oscillations as the number of barriers are increased from $N=5$ to $N=6$. Fig. \ref{fig:NOB_SLTJ}(a) shows a steep upsurge in the TMR(\%) as we increase the number of barriers from five to six, which can be categorically explained by the sharp dip in the A$_{APC}$ at $N=6$ as depicted in Fig. \ref{fig:WSLTJ_TM}. At the same time, the comparable I$_{s||}$ of the LW-SLTJ, GW-SLTJ and the PW-SLTJ at $N=5$ and $N=6$ maintain a precise consistency with their T$_{PC}$. The considerable reduction in the A$_{PC}$ of the LRW-SLTJ from $N=5$ to $N=6$ reverberates in it's I$_{s||}$ as well.\\ 
\indent In  Fig. \ref{fig:HSLJ_TM}, we show both the T$_{PC}$ and T$_{APC}$ of all the H-SLTJs for $N=5$ and $N=8$. We find H-SLTJs display more intricate results in comparison to their width-based counterparts. As we increase the number of the barriers from $N=5$ to $N=8$, even though the A$_{PC}$ of the GH-SLTJ, LRH-SLTJ, and the PH-SLTJ shows an upsurge, the I$_{s||}$ of the respective devices suffers a decline in contrast to our earlier discussion. This might appear to be a clear dichotomy in the first place but we find the explanation of the bewilderment to be concealed behind the transmissions at higher transverse modes(E$_t$)\cite{sharma2021proposal}.
As we move towards higher E$_t$, the gross bulk of transmission spectra approaches the Fermi window and then eventually moves beyond it. Hence the transmissions at a higher E$_t$ around the fermi window plays a more conclusive role in order to determine the I$_{s||}$ and the TMR(\%). In Fig. \ref{fig:HSLTJ_F_window}(b),(c)$\&$(d) we demonstrate that for E$_t$=0.2 eV the A$_{PC}$ of the GH-SLTJ, LRH-SLTJ, and the PH-SLTJ at $N=5$ offers a higher area than that of at $N=8$, thereby drawing a coherent conclusion in this regard. Apart from that, the increase in the TMR(\%) of the H-SLTJs from $N=5$ to $N=8$ is clearly interpretable from Fig. \ref{fig:HSLJ_TM}. 
\section{{Guidelines for Designing the Nu-SLTJs}
 \label{Appendix:C}}
 In this leg of the article, we describe the impact of modulating
 \begin{figure}[h]
	\centering
	\includegraphics[scale=0.35]{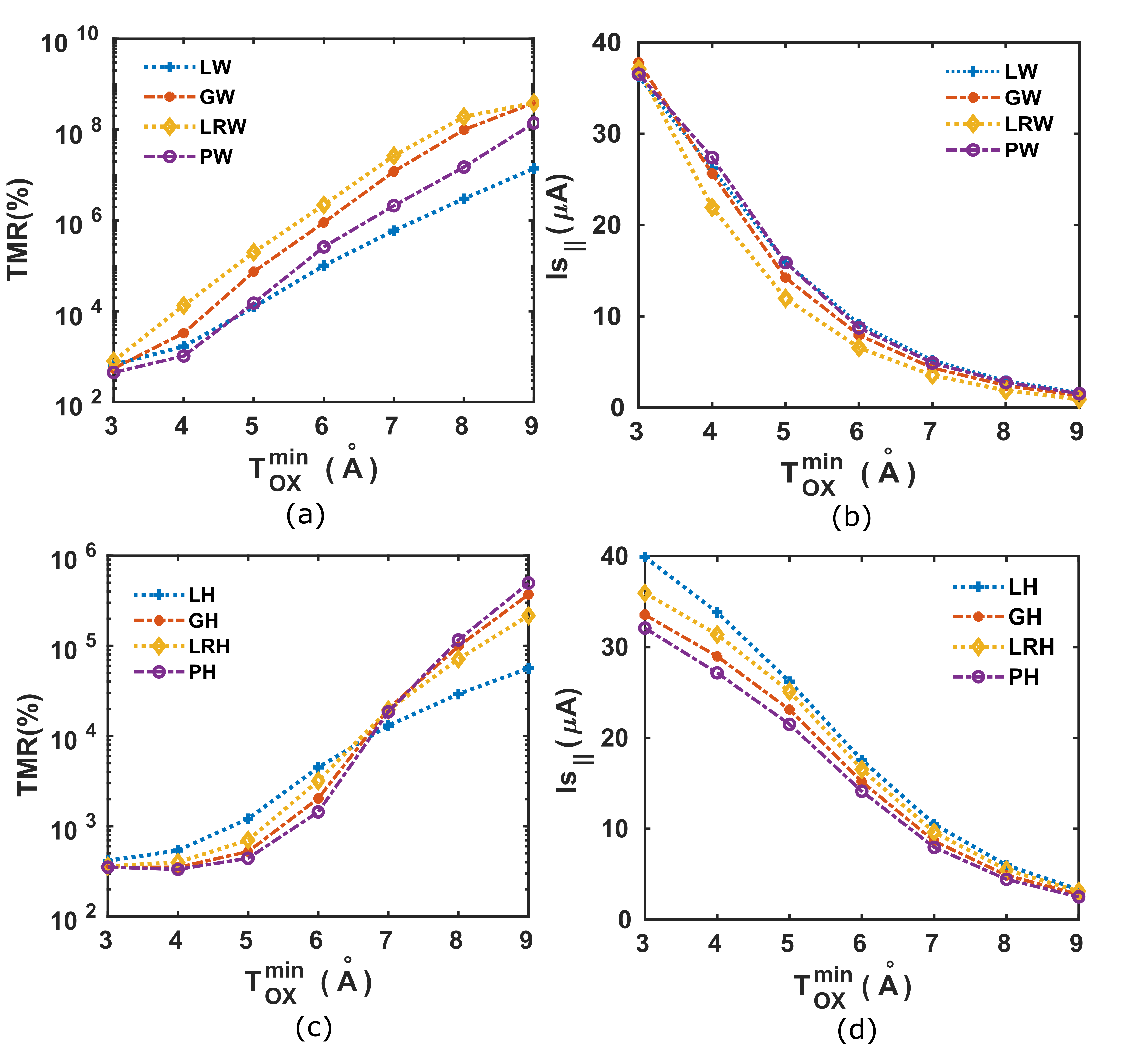}
	\caption{Device characteristics: The variation in the TMR(\%) of the (a)W-SLTJs and the (b)H-SLTJs, I$_{s||}$ of the (c)W-SLTJs and the (d)H-SLTJs with the T$_{OX}^{mix}$ at an applied bias of V=5 mV.}
	\label{fig:T_OX}
\end{figure}
 the scattering potential($SP$) and width($b$) of the oxide barriers to design the Nu-SLTJs and argue that altering either of them for enhancing the TMR(\%) abates the I$_{s||}$ and vice versa, thereby leading to a novel hypothesis of TMR-I$_{s||}$ trade-off. In Fig. \ref{fig:T_OX} we present the variation of the TMR and I$_{s||}$ while varying the minimum thickness of the oxide layer of Nu-SLTJs from 3\AA  to 9\AA. Besides T$_{OX}^{min}$ determines the thickness of all the other barriers in the W-SLTJs, the thickness of each barriers comprising an H-SLTJ is T$_{OX}^{min}$ itself. 
 \begin{figure}[!b]
	\centering
	\includegraphics[scale=0.35]{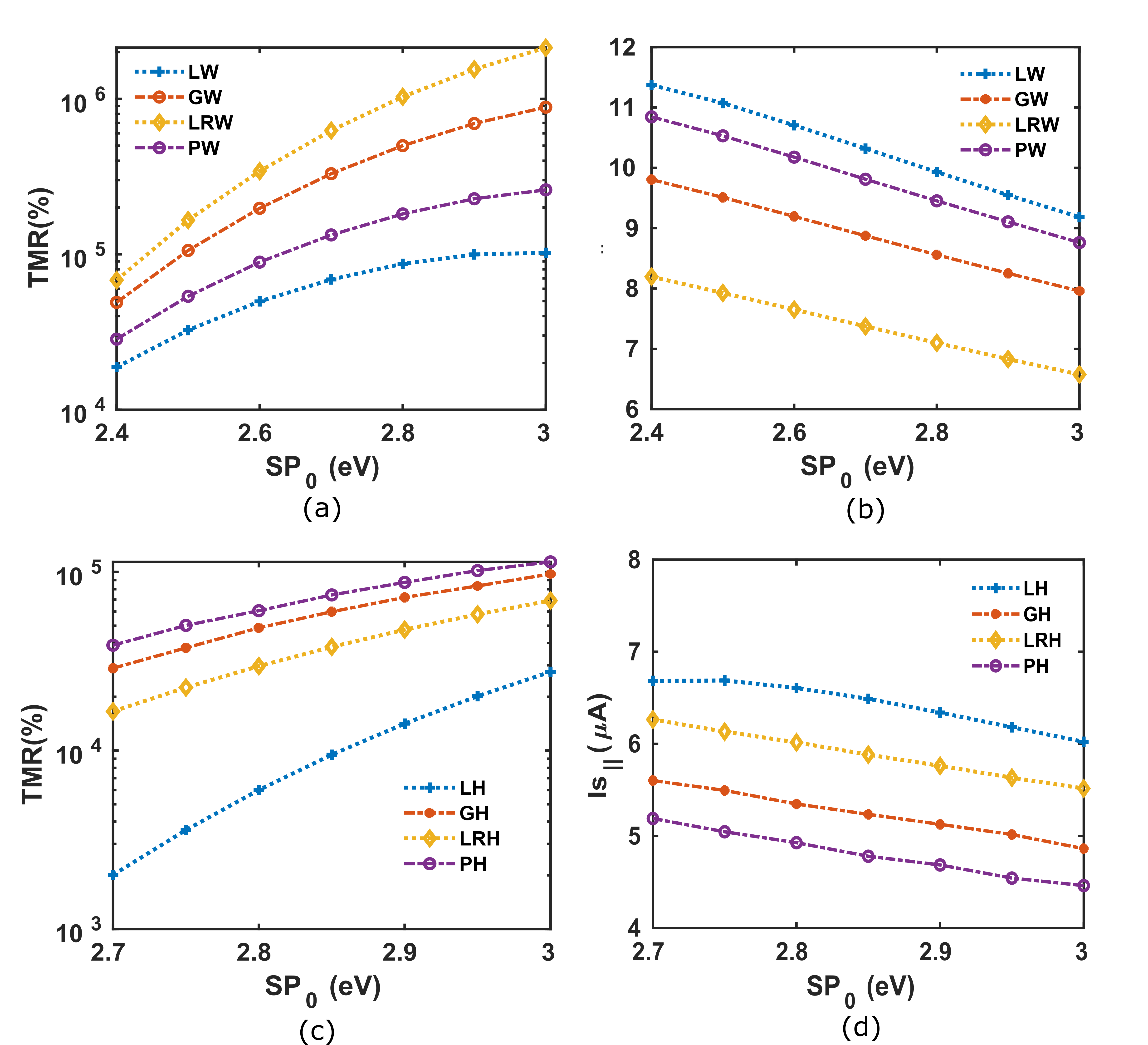}
	\caption{Device characteristics: The variation in the TMR(\%) of the (a)W-SLTJs and the (b)H-SLTJs, I$_{s||}$ of the (c)W-SLTJs and the (d)H-SLTJs with the $SP_0$ at an applied voltage of V=5 mV. }
	\label{fig:H_OX}
\end{figure}
 The results we obtain from Fig.\ref{fig:T_OX} demonstrate that an increase in the T$_{OX}^{min}$ eventuates a near monotonic upsurge of the TMR while manifesting an opposite impact on the I$_{s||}$. We may attribute such behavioral pattern to the fact that a wider T$_{OX}$ bestows an exacerbated A$_{PC}$ resulting a decimated I$_{s||}$, and at the same time boosts the ratio of the A$_{PC}$ and A$_{APC}$ leading to an increase in the TMR. Analogously, we also perform the analysis regarding the impact of varying scattering potentials($SP$) of the oxide barriers on the H-SLTJ and demonstrate the behavior of the I$_{s||}$ and the TMR to be entangled in a similar fashion. \\
\indent On the similar lines, we vary the $SP$s of the W-SLTJs from 2.4 eV to 3 eV and depict the device characteristics in Fig. \ref{fig:H_OX}(a)$\&$(b). Since the $SP$ of the terminal barriers($SP_{tb}$) depends on the central barrier($SP_0$) in the H-SLTJs, lowering the $SP_0$ significantly reduces the height of the $SP_{tb}$ far below the Fermi energy, thereby sabotaging the spin filtering near the FW.
To avoid this muddle, we keep the variation of the $SP_0$ within the limits of 2.7 eV to 3 eV for studying the characteristics of the I$_{s||}$ and the TMR as shown in Fig \ref{fig:H_OX}(c)$\&$(d). The results we obtain in this context conveys that an increase in the scattering potential reduces the I$_{s||}$ and improves the TMR which may again be attributed to the reduction in the A$_{PC}$ and improvement in the ratio of the A$_{PC}$ and A$_{APC}$, respectively.
\begin{figure}[h]
	\centering
	\includegraphics[scale=0.32]{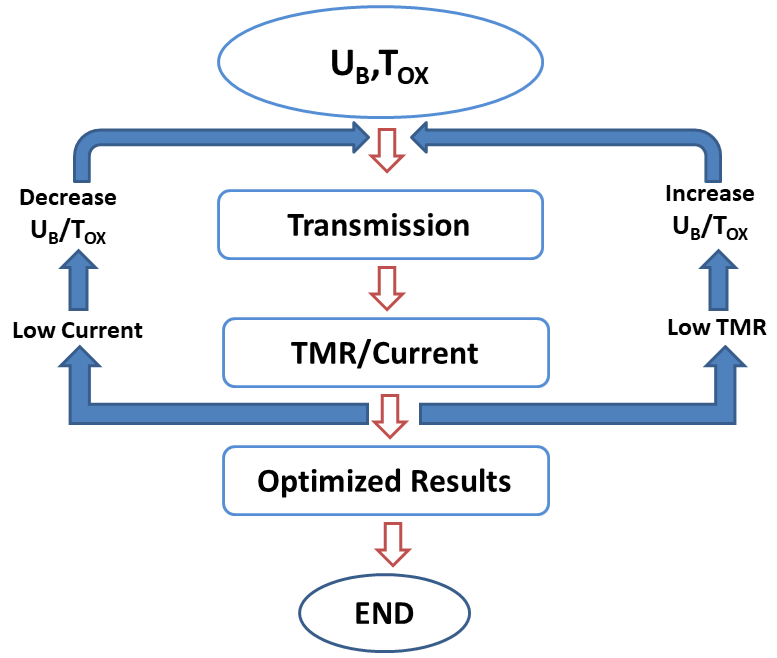}
	\caption{Optimization Algorithm }
	\label{fig:Optimization Algorithm}
\end{figure}
A qualitative analysis of Fig. \ref{fig:T_OX} and  Fig. \ref{fig:H_OX} reveal that the change in the width of the oxide barriers elicit more pronounced impact on the TMR and the I$_{s||}$ among the SLTJs compared to the variation in the height of the $SP$s. Therefore, we may use the variation of the T$_{OX}$ to design the framework of a Nu-SLTJ, while the alteration in the scattering potential might come in handy during the fine-tuning of the performance indices.\\
\indent It is worthwhile mentioning that the variation of TMR and I$_{s||}$ shown in Fig. \ref{fig:H_OX}(c)$\&$(d) establishes the fact that despite a significant variation due to the impact of error in the barrier heights of the H-SLTJs owing to the complexity involved in engineering the height variations for real samples, the TMR remains significantly above $10^3\%$ whereas the I$_{s||}$ doesn't suffer a significant decline. Consequently, the H-SLTJs promise a momentous improvement in the performance indices compared to a typical MTJ, irrespective of the errors in the barrier heights.\\
\indent The algorithm described in Fig. \ref{fig:Optimization Algorithm}, presents a generic guideline to modulate the performance indices of a typical  Nu-SLTJ in order to optimize it for various applications. While using the minimum number of oxide layers to moderate the fabrication complexity, a designer is left to alter only with the thickness of the oxide layers and the  scattering potential of the insulators as the U$_{BW}$ is fixed for a particular NM and the width of the quantum well($W$) is decided beforehand for the optimization purposes\cite{sharma2018band}. Therefore, by following the algorithm depicted in Fig. \ref{fig:Optimization Algorithm}, one may play around with both the $SP$ and T$_{ox}$ to meet the target specifications.
\nocite{*}
\bibliography{References}

\begin{thebibliography}{70}%
\makeatletter
\providecommand \@ifxundefined [1]{%
 \@ifx{#1\undefined}
}%
\providecommand \@ifnum [1]{%
 \ifnum #1\expandafter \@firstoftwo
 \else \expandafter \@secondoftwo
 \fi
}%
\providecommand \@ifx [1]{%
 \ifx #1\expandafter \@firstoftwo
 \else \expandafter \@secondoftwo
 \fi
}%
\providecommand \natexlab [1]{#1}%
\providecommand \enquote  [1]{``#1''}%
\providecommand \bibnamefont  [1]{#1}%
\providecommand \bibfnamefont [1]{#1}%
\providecommand \citenamefont [1]{#1}%
\providecommand \href@noop [0]{\@secondoftwo}%
\providecommand \href [0]{\begingroup \@sanitize@url \@href}%
\providecommand \@href[1]{\@@startlink{#1}\@@href}%
\providecommand \@@href[1]{\endgroup#1\@@endlink}%
\providecommand \@sanitize@url [0]{\catcode `\\12\catcode `\$12\catcode
  `\&12\catcode `\#12\catcode `\^12\catcode `\_12\catcode `\%12\relax}%
\providecommand \@@startlink[1]{}%
\providecommand \@@endlink[0]{}%
\providecommand \url  [0]{\begingroup\@sanitize@url \@url }%
\providecommand \@url [1]{\endgroup\@href {#1}{\urlprefix }}%
\providecommand \urlprefix  [0]{URL }%
\providecommand \Eprint [0]{\href }%
\providecommand \doibase [0]{http://dx.doi.org/}%
\providecommand \selectlanguage [0]{\@gobble}%
\providecommand \bibinfo  [0]{\@secondoftwo}%
\providecommand \bibfield  [0]{\@secondoftwo}%
\providecommand \translation [1]{[#1]}%
\providecommand \BibitemOpen [0]{}%
\providecommand \bibitemStop [0]{}%
\providecommand \bibitemNoStop [0]{.\EOS\space}%
\providecommand \EOS [0]{\spacefactor3000\relax}%
\providecommand \BibitemShut  [1]{\csname bibitem#1\endcsname}%
\let\auto@bib@innerbib\@empty
\bibitem [{\citenamefont {Parkin}\ \emph {et~al.}(2004)\citenamefont {Parkin},
  \citenamefont {Kaiser}, \citenamefont {Panchula}, \citenamefont {Rice},
  \citenamefont {Hughes}, \citenamefont {Samant},\ and\ \citenamefont
  {Yang}}]{parkin2004giant}%
  \BibitemOpen
  \bibfield  {author} {\bibinfo {author} {\bibfnamefont {S.~S.}\ \bibnamefont
  {Parkin}}, \bibinfo {author} {\bibfnamefont {C.}~\bibnamefont {Kaiser}},
  \bibinfo {author} {\bibfnamefont {A.}~\bibnamefont {Panchula}}, \bibinfo
  {author} {\bibfnamefont {P.~M.}\ \bibnamefont {Rice}}, \bibinfo {author}
  {\bibfnamefont {B.}~\bibnamefont {Hughes}}, \bibinfo {author} {\bibfnamefont
  {M.}~\bibnamefont {Samant}}, \ and\ \bibinfo {author} {\bibfnamefont {S.-H.}\
  \bibnamefont {Yang}},\ }\href@noop {} {\bibfield  {journal} {\bibinfo
  {journal} {Nature materials}\ }\textbf {\bibinfo {volume} {3}},\ \bibinfo
  {pages} {862} (\bibinfo {year} {2004})}\BibitemShut {NoStop}%
\bibitem [{\citenamefont {Berger}(1996)}]{berger1996emission}%
  \BibitemOpen
  \bibfield  {author} {\bibinfo {author} {\bibfnamefont {L.}~\bibnamefont
  {Berger}},\ }\href@noop {} {\bibfield  {journal} {\bibinfo  {journal}
  {Physical Review B}\ }\textbf {\bibinfo {volume} {54}},\ \bibinfo {pages}
  {9353} (\bibinfo {year} {1996})}\BibitemShut {NoStop}%
\bibitem [{\citenamefont {Slonczewski}(1996)}]{slonczewski1996current}%
  \BibitemOpen
  \bibfield  {author} {\bibinfo {author} {\bibfnamefont {J.~C.}\ \bibnamefont
  {Slonczewski}},\ }\href@noop {} {\bibfield  {journal} {\bibinfo  {journal}
  {Journal of Magnetism and Magnetic Materials}\ }\textbf {\bibinfo {volume}
  {159}},\ \bibinfo {pages} {L1} (\bibinfo {year} {1996})}\BibitemShut
  {NoStop}%
\bibitem [{\citenamefont {Le~Phan}\ \emph {et~al.}(2006)\citenamefont
  {Le~Phan}, \citenamefont {Boeve}, \citenamefont {Vanhelmont}, \citenamefont
  {Ikkink}, \citenamefont {De~Jong},\ and\ \citenamefont
  {De~Wilde}}]{le2006tunnel}%
  \BibitemOpen
  \bibfield  {author} {\bibinfo {author} {\bibfnamefont {K.}~\bibnamefont
  {Le~Phan}}, \bibinfo {author} {\bibfnamefont {H.}~\bibnamefont {Boeve}},
  \bibinfo {author} {\bibfnamefont {F.}~\bibnamefont {Vanhelmont}}, \bibinfo
  {author} {\bibfnamefont {T.}~\bibnamefont {Ikkink}}, \bibinfo {author}
  {\bibfnamefont {F.}~\bibnamefont {De~Jong}}, \ and\ \bibinfo {author}
  {\bibfnamefont {H.}~\bibnamefont {De~Wilde}},\ }\href@noop {} {\bibfield
  {journal} {\bibinfo  {journal} {Sensors and Actuators A: Physical}\ }\textbf
  {\bibinfo {volume} {129}},\ \bibinfo {pages} {69} (\bibinfo {year}
  {2006})}\BibitemShut {NoStop}%
\bibitem [{\citenamefont {Kanno}\ \emph {et~al.}(2022)\citenamefont {Kanno},
  \citenamefont {Nakasato}, \citenamefont {Oogane}, \citenamefont {Fujiwara},
  \citenamefont {Nakano}, \citenamefont {Arimoto}, \citenamefont {Matsuzaki},\
  and\ \citenamefont {Ando}}]{kanno2022scalp}%
  \BibitemOpen
  \bibfield  {author} {\bibinfo {author} {\bibfnamefont {A.}~\bibnamefont
  {Kanno}}, \bibinfo {author} {\bibfnamefont {N.}~\bibnamefont {Nakasato}},
  \bibinfo {author} {\bibfnamefont {M.}~\bibnamefont {Oogane}}, \bibinfo
  {author} {\bibfnamefont {K.}~\bibnamefont {Fujiwara}}, \bibinfo {author}
  {\bibfnamefont {T.}~\bibnamefont {Nakano}}, \bibinfo {author} {\bibfnamefont
  {T.}~\bibnamefont {Arimoto}}, \bibinfo {author} {\bibfnamefont
  {H.}~\bibnamefont {Matsuzaki}}, \ and\ \bibinfo {author} {\bibfnamefont
  {Y.}~\bibnamefont {Ando}},\ }\href@noop {} {\bibfield  {journal} {\bibinfo
  {journal} {Scientific reports}\ }\textbf {\bibinfo {volume} {12}},\ \bibinfo
  {pages} {1} (\bibinfo {year} {2022})}\BibitemShut {NoStop}%
\bibitem [{\citenamefont {Rudin}\ and\ \citenamefont
  {Bramble}(1997)}]{rudin1997investigative}%
  \BibitemOpen
  \bibfield  {author} {\bibinfo {author} {\bibfnamefont {L.~I.}\ \bibnamefont
  {Rudin}}\ and\ \bibinfo {author} {\bibfnamefont {S.~K.}\ \bibnamefont
  {Bramble}},\ }\href@noop {} {\bibfield  {journal} {\bibinfo  {journal}
  {Investigative Image Processing}\ }\textbf {\bibinfo {volume} {2942}}
  (\bibinfo {year} {1997})}\BibitemShut {NoStop}%
\bibitem [{\citenamefont {Chen}\ \emph {et~al.}(2016)\citenamefont {Chen},
  \citenamefont {Dumas}, \citenamefont {Eklund}, \citenamefont {Muduli},
  \citenamefont {Houshang}, \citenamefont {Awad}, \citenamefont
  {D{\"u}rrenfeld}, \citenamefont {Malm}, \citenamefont {Rusu},\ and\
  \citenamefont {{\AA}kerman}}]{chen2016spin}%
  \BibitemOpen
  \bibfield  {author} {\bibinfo {author} {\bibfnamefont {T.}~\bibnamefont
  {Chen}}, \bibinfo {author} {\bibfnamefont {R.~K.}\ \bibnamefont {Dumas}},
  \bibinfo {author} {\bibfnamefont {A.}~\bibnamefont {Eklund}}, \bibinfo
  {author} {\bibfnamefont {P.~K.}\ \bibnamefont {Muduli}}, \bibinfo {author}
  {\bibfnamefont {A.}~\bibnamefont {Houshang}}, \bibinfo {author}
  {\bibfnamefont {A.~A.}\ \bibnamefont {Awad}}, \bibinfo {author}
  {\bibfnamefont {P.}~\bibnamefont {D{\"u}rrenfeld}}, \bibinfo {author}
  {\bibfnamefont {B.~G.}\ \bibnamefont {Malm}}, \bibinfo {author}
  {\bibfnamefont {A.}~\bibnamefont {Rusu}}, \ and\ \bibinfo {author}
  {\bibfnamefont {J.}~\bibnamefont {{\AA}kerman}},\ }\href@noop {} {\bibfield
  {journal} {\bibinfo  {journal} {Proceedings of the IEEE}\ }\textbf {\bibinfo
  {volume} {104}},\ \bibinfo {pages} {1919} (\bibinfo {year}
  {2016})}\BibitemShut {NoStop}%
\bibitem [{\citenamefont {Braganca}\ \emph {et~al.}(2010)\citenamefont
  {Braganca}, \citenamefont {Gurney}, \citenamefont {Wilson}, \citenamefont
  {Katine}, \citenamefont {Maat},\ and\ \citenamefont
  {Childress}}]{braganca2010nanoscale}%
  \BibitemOpen
  \bibfield  {author} {\bibinfo {author} {\bibfnamefont {P.}~\bibnamefont
  {Braganca}}, \bibinfo {author} {\bibfnamefont {B.}~\bibnamefont {Gurney}},
  \bibinfo {author} {\bibfnamefont {B.}~\bibnamefont {Wilson}}, \bibinfo
  {author} {\bibfnamefont {J.}~\bibnamefont {Katine}}, \bibinfo {author}
  {\bibfnamefont {S.}~\bibnamefont {Maat}}, \ and\ \bibinfo {author}
  {\bibfnamefont {J.}~\bibnamefont {Childress}},\ }\href@noop {} {\bibfield
  {journal} {\bibinfo  {journal} {Nanotechnology}\ }\textbf {\bibinfo {volume}
  {21}},\ \bibinfo {pages} {235202} (\bibinfo {year} {2010})}\BibitemShut
  {NoStop}%
\bibitem [{\citenamefont {Robbes}(2006)}]{robbes2006highly}%
  \BibitemOpen
  \bibfield  {author} {\bibinfo {author} {\bibfnamefont {D.}~\bibnamefont
  {Robbes}},\ }\href@noop {} {\bibfield  {journal} {\bibinfo  {journal}
  {Sensors and Actuators A: Physical}\ }\textbf {\bibinfo {volume} {129}},\
  \bibinfo {pages} {86} (\bibinfo {year} {2006})}\BibitemShut {NoStop}%
\bibitem [{\citenamefont {Krivorotov}\ \emph {et~al.}(2005)\citenamefont
  {Krivorotov}, \citenamefont {Emley}, \citenamefont {Sankey}, \citenamefont
  {Kiselev}, \citenamefont {Ralph},\ and\ \citenamefont
  {Buhrman}}]{krivorotov2005time}%
  \BibitemOpen
  \bibfield  {author} {\bibinfo {author} {\bibfnamefont {I.}~\bibnamefont
  {Krivorotov}}, \bibinfo {author} {\bibfnamefont {N.}~\bibnamefont {Emley}},
  \bibinfo {author} {\bibfnamefont {J.}~\bibnamefont {Sankey}}, \bibinfo
  {author} {\bibfnamefont {S.}~\bibnamefont {Kiselev}}, \bibinfo {author}
  {\bibfnamefont {D.}~\bibnamefont {Ralph}}, \ and\ \bibinfo {author}
  {\bibfnamefont {R.}~\bibnamefont {Buhrman}},\ }\href@noop {} {\bibfield
  {journal} {\bibinfo  {journal} {Science}\ }\textbf {\bibinfo {volume}
  {307}},\ \bibinfo {pages} {228} (\bibinfo {year} {2005})}\BibitemShut
  {NoStop}%
\bibitem [{\citenamefont {Choi}\ \emph {et~al.}(2014)\citenamefont {Choi},
  \citenamefont {Kang}, \citenamefont {Cho}, \citenamefont {Oh}, \citenamefont
  {Shin}, \citenamefont {Park}, \citenamefont {Jang}, \citenamefont {Min},
  \citenamefont {Kim}, \citenamefont {Park} \emph {et~al.}}]{choi2014spin}%
  \BibitemOpen
  \bibfield  {author} {\bibinfo {author} {\bibfnamefont {H.~S.}\ \bibnamefont
  {Choi}}, \bibinfo {author} {\bibfnamefont {S.~Y.}\ \bibnamefont {Kang}},
  \bibinfo {author} {\bibfnamefont {S.~J.}\ \bibnamefont {Cho}}, \bibinfo
  {author} {\bibfnamefont {I.-Y.}\ \bibnamefont {Oh}}, \bibinfo {author}
  {\bibfnamefont {M.}~\bibnamefont {Shin}}, \bibinfo {author} {\bibfnamefont
  {H.}~\bibnamefont {Park}}, \bibinfo {author} {\bibfnamefont {C.}~\bibnamefont
  {Jang}}, \bibinfo {author} {\bibfnamefont {B.-C.}\ \bibnamefont {Min}},
  \bibinfo {author} {\bibfnamefont {S.-I.}\ \bibnamefont {Kim}}, \bibinfo
  {author} {\bibfnamefont {S.-Y.}\ \bibnamefont {Park}},  \emph {et~al.},\
  }\href@noop {} {\bibfield  {journal} {\bibinfo  {journal} {Scientific
  reports}\ }\textbf {\bibinfo {volume} {4}},\ \bibinfo {pages} {1} (\bibinfo
  {year} {2014})}\BibitemShut {NoStop}%
\bibitem [{\citenamefont {Cheng}\ \emph {et~al.}(2016)\citenamefont {Cheng},
  \citenamefont {Xiao},\ and\ \citenamefont {Brataas}}]{cheng2016terahertz}%
  \BibitemOpen
  \bibfield  {author} {\bibinfo {author} {\bibfnamefont {R.}~\bibnamefont
  {Cheng}}, \bibinfo {author} {\bibfnamefont {D.}~\bibnamefont {Xiao}}, \ and\
  \bibinfo {author} {\bibfnamefont {A.}~\bibnamefont {Brataas}},\ }\href@noop
  {} {\bibfield  {journal} {\bibinfo  {journal} {Physical review letters}\
  }\textbf {\bibinfo {volume} {116}},\ \bibinfo {pages} {207603} (\bibinfo
  {year} {2016})}\BibitemShut {NoStop}%
\bibitem [{\citenamefont {Bhattacharjee}\ \emph {et~al.}(2018)\citenamefont
  {Bhattacharjee}, \citenamefont {Sapozhnik}, \citenamefont {Bodnar},
  \citenamefont {Grigorev}, \citenamefont {Agustsson}, \citenamefont {Cao},
  \citenamefont {Dominko}, \citenamefont {Obergfell}, \citenamefont {Gomonay},
  \citenamefont {Sinova} \emph {et~al.}}]{bhattacharjee2018neel}%
  \BibitemOpen
  \bibfield  {author} {\bibinfo {author} {\bibfnamefont {N.}~\bibnamefont
  {Bhattacharjee}}, \bibinfo {author} {\bibfnamefont {A.}~\bibnamefont
  {Sapozhnik}}, \bibinfo {author} {\bibfnamefont {S.~Y.}\ \bibnamefont
  {Bodnar}}, \bibinfo {author} {\bibfnamefont {V.~Y.}\ \bibnamefont
  {Grigorev}}, \bibinfo {author} {\bibfnamefont {S.~Y.}\ \bibnamefont
  {Agustsson}}, \bibinfo {author} {\bibfnamefont {J.}~\bibnamefont {Cao}},
  \bibinfo {author} {\bibfnamefont {D.}~\bibnamefont {Dominko}}, \bibinfo
  {author} {\bibfnamefont {M.}~\bibnamefont {Obergfell}}, \bibinfo {author}
  {\bibfnamefont {O.}~\bibnamefont {Gomonay}}, \bibinfo {author} {\bibfnamefont
  {J.}~\bibnamefont {Sinova}},  \emph {et~al.},\ }\href@noop {} {\bibfield
  {journal} {\bibinfo  {journal} {Physical review letters}\ }\textbf {\bibinfo
  {volume} {120}},\ \bibinfo {pages} {237201} (\bibinfo {year}
  {2018})}\BibitemShut {NoStop}%
\bibitem [{\citenamefont {Jiang}\ \emph {et~al.}(2004)\citenamefont {Jiang},
  \citenamefont {Nozaki}, \citenamefont {Abe}, \citenamefont {Ochiai},
  \citenamefont {Hirohata}, \citenamefont {Tezuka},\ and\ \citenamefont
  {Inomata}}]{jiang2004substantial}%
  \BibitemOpen
  \bibfield  {author} {\bibinfo {author} {\bibfnamefont {Y.}~\bibnamefont
  {Jiang}}, \bibinfo {author} {\bibfnamefont {T.}~\bibnamefont {Nozaki}},
  \bibinfo {author} {\bibfnamefont {S.}~\bibnamefont {Abe}}, \bibinfo {author}
  {\bibfnamefont {T.}~\bibnamefont {Ochiai}}, \bibinfo {author} {\bibfnamefont
  {A.}~\bibnamefont {Hirohata}}, \bibinfo {author} {\bibfnamefont
  {N.}~\bibnamefont {Tezuka}}, \ and\ \bibinfo {author} {\bibfnamefont
  {K.}~\bibnamefont {Inomata}},\ }\href@noop {} {\bibfield  {journal} {\bibinfo
   {journal} {Nature materials}\ }\textbf {\bibinfo {volume} {3}},\ \bibinfo
  {pages} {361} (\bibinfo {year} {2004})}\BibitemShut {NoStop}%
\bibitem [{\citenamefont {Mangin}\ \emph {et~al.}(2006)\citenamefont {Mangin},
  \citenamefont {Ravelosona}, \citenamefont {Katine}, \citenamefont {Carey},
  \citenamefont {Terris},\ and\ \citenamefont {Fullerton}}]{mangin2006current}%
  \BibitemOpen
  \bibfield  {author} {\bibinfo {author} {\bibfnamefont {S.}~\bibnamefont
  {Mangin}}, \bibinfo {author} {\bibfnamefont {D.}~\bibnamefont {Ravelosona}},
  \bibinfo {author} {\bibfnamefont {J.}~\bibnamefont {Katine}}, \bibinfo
  {author} {\bibfnamefont {M.}~\bibnamefont {Carey}}, \bibinfo {author}
  {\bibfnamefont {B.}~\bibnamefont {Terris}}, \ and\ \bibinfo {author}
  {\bibfnamefont {E.~E.}\ \bibnamefont {Fullerton}},\ }\href@noop {} {\bibfield
   {journal} {\bibinfo  {journal} {Nature materials}\ }\textbf {\bibinfo
  {volume} {5}},\ \bibinfo {pages} {210} (\bibinfo {year} {2006})}\BibitemShut
  {NoStop}%
\bibitem [{\citenamefont {Diao}\ \emph {et~al.}(2007)\citenamefont {Diao},
  \citenamefont {Li}, \citenamefont {Wang}, \citenamefont {Ding}, \citenamefont
  {Panchula}, \citenamefont {Chen}, \citenamefont {Wang},\ and\ \citenamefont
  {Huai}}]{diao2007spin}%
  \BibitemOpen
  \bibfield  {author} {\bibinfo {author} {\bibfnamefont {Z.}~\bibnamefont
  {Diao}}, \bibinfo {author} {\bibfnamefont {Z.}~\bibnamefont {Li}}, \bibinfo
  {author} {\bibfnamefont {S.}~\bibnamefont {Wang}}, \bibinfo {author}
  {\bibfnamefont {Y.}~\bibnamefont {Ding}}, \bibinfo {author} {\bibfnamefont
  {A.}~\bibnamefont {Panchula}}, \bibinfo {author} {\bibfnamefont
  {E.}~\bibnamefont {Chen}}, \bibinfo {author} {\bibfnamefont {L.-C.}\
  \bibnamefont {Wang}}, \ and\ \bibinfo {author} {\bibfnamefont
  {Y.}~\bibnamefont {Huai}},\ }\href@noop {} {\bibfield  {journal} {\bibinfo
  {journal} {Journal of Physics: Condensed Matter}\ }\textbf {\bibinfo {volume}
  {19}},\ \bibinfo {pages} {165209} (\bibinfo {year} {2007})}\BibitemShut
  {NoStop}%
\bibitem [{\citenamefont {Santos}\ \emph {et~al.}(2020)\citenamefont {Santos},
  \citenamefont {Mihajlovi{\'c}}, \citenamefont {Smith}, \citenamefont {Li},
  \citenamefont {Carey}, \citenamefont {Katine},\ and\ \citenamefont
  {Terris}}]{santos2020ultrathin}%
  \BibitemOpen
  \bibfield  {author} {\bibinfo {author} {\bibfnamefont {T.~S.}\ \bibnamefont
  {Santos}}, \bibinfo {author} {\bibfnamefont {G.}~\bibnamefont
  {Mihajlovi{\'c}}}, \bibinfo {author} {\bibfnamefont {N.}~\bibnamefont
  {Smith}}, \bibinfo {author} {\bibfnamefont {J.-L.}\ \bibnamefont {Li}},
  \bibinfo {author} {\bibfnamefont {M.}~\bibnamefont {Carey}}, \bibinfo
  {author} {\bibfnamefont {J.~A.}\ \bibnamefont {Katine}}, \ and\ \bibinfo
  {author} {\bibfnamefont {B.~D.}\ \bibnamefont {Terris}},\ }\href@noop {}
  {\bibfield  {journal} {\bibinfo  {journal} {Journal of Applied Physics}\
  }\textbf {\bibinfo {volume} {128}},\ \bibinfo {pages} {113904} (\bibinfo
  {year} {2020})}\BibitemShut {NoStop}%
\bibitem [{\citenamefont {Ikeda}\ \emph {et~al.}(2007)\citenamefont {Ikeda},
  \citenamefont {Hayakawa}, \citenamefont {Lee}, \citenamefont {Matsukura},
  \citenamefont {Ohno}, \citenamefont {Hanyu},\ and\ \citenamefont
  {Ohno}}]{ikeda2007magnetic}%
  \BibitemOpen
  \bibfield  {author} {\bibinfo {author} {\bibfnamefont {S.}~\bibnamefont
  {Ikeda}}, \bibinfo {author} {\bibfnamefont {J.}~\bibnamefont {Hayakawa}},
  \bibinfo {author} {\bibfnamefont {Y.~M.}\ \bibnamefont {Lee}}, \bibinfo
  {author} {\bibfnamefont {F.}~\bibnamefont {Matsukura}}, \bibinfo {author}
  {\bibfnamefont {Y.}~\bibnamefont {Ohno}}, \bibinfo {author} {\bibfnamefont
  {T.}~\bibnamefont {Hanyu}}, \ and\ \bibinfo {author} {\bibfnamefont
  {H.}~\bibnamefont {Ohno}},\ }\href@noop {} {\bibfield  {journal} {\bibinfo
  {journal} {IEEE Transactions on Electron Devices}\ }\textbf {\bibinfo
  {volume} {54}},\ \bibinfo {pages} {991} (\bibinfo {year} {2007})}\BibitemShut
  {NoStop}%
\bibitem [{\citenamefont {Apalkov}\ \emph {et~al.}(2013)\citenamefont
  {Apalkov}, \citenamefont {Khvalkovskiy}, \citenamefont {Watts}, \citenamefont
  {Nikitin}, \citenamefont {Tang}, \citenamefont {Lottis}, \citenamefont
  {Moon}, \citenamefont {Luo}, \citenamefont {Chen}, \citenamefont {Ong} \emph
  {et~al.}}]{apalkov2013spin}%
  \BibitemOpen
  \bibfield  {author} {\bibinfo {author} {\bibfnamefont {D.}~\bibnamefont
  {Apalkov}}, \bibinfo {author} {\bibfnamefont {A.}~\bibnamefont
  {Khvalkovskiy}}, \bibinfo {author} {\bibfnamefont {S.}~\bibnamefont {Watts}},
  \bibinfo {author} {\bibfnamefont {V.}~\bibnamefont {Nikitin}}, \bibinfo
  {author} {\bibfnamefont {X.}~\bibnamefont {Tang}}, \bibinfo {author}
  {\bibfnamefont {D.}~\bibnamefont {Lottis}}, \bibinfo {author} {\bibfnamefont
  {K.}~\bibnamefont {Moon}}, \bibinfo {author} {\bibfnamefont {X.}~\bibnamefont
  {Luo}}, \bibinfo {author} {\bibfnamefont {E.}~\bibnamefont {Chen}}, \bibinfo
  {author} {\bibfnamefont {A.}~\bibnamefont {Ong}},  \emph {et~al.},\
  }\href@noop {} {\bibfield  {journal} {\bibinfo  {journal} {ACM Journal on
  Emerging Technologies in Computing Systems (JETC)}\ }\textbf {\bibinfo
  {volume} {9}},\ \bibinfo {pages} {1} (\bibinfo {year} {2013})}\BibitemShut
  {NoStop}%
\bibitem [{\citenamefont {Khvalkovskiy}\ \emph {et~al.}(2013)\citenamefont
  {Khvalkovskiy}, \citenamefont {Apalkov}, \citenamefont {Watts}, \citenamefont
  {Chepulskii}, \citenamefont {Beach}, \citenamefont {Ong}, \citenamefont
  {Tang}, \citenamefont {Driskill-Smith}, \citenamefont {Butler}, \citenamefont
  {Visscher} \emph {et~al.}}]{khvalkovskiy2013basic}%
  \BibitemOpen
  \bibfield  {author} {\bibinfo {author} {\bibfnamefont {A.}~\bibnamefont
  {Khvalkovskiy}}, \bibinfo {author} {\bibfnamefont {D.}~\bibnamefont
  {Apalkov}}, \bibinfo {author} {\bibfnamefont {S.}~\bibnamefont {Watts}},
  \bibinfo {author} {\bibfnamefont {R.}~\bibnamefont {Chepulskii}}, \bibinfo
  {author} {\bibfnamefont {R.}~\bibnamefont {Beach}}, \bibinfo {author}
  {\bibfnamefont {A.}~\bibnamefont {Ong}}, \bibinfo {author} {\bibfnamefont
  {X.}~\bibnamefont {Tang}}, \bibinfo {author} {\bibfnamefont {A.}~\bibnamefont
  {Driskill-Smith}}, \bibinfo {author} {\bibfnamefont {W.}~\bibnamefont
  {Butler}}, \bibinfo {author} {\bibfnamefont {P.}~\bibnamefont {Visscher}},
  \emph {et~al.},\ }\href@noop {} {\bibfield  {journal} {\bibinfo  {journal}
  {Journal of Physics D: Applied Physics}\ }\textbf {\bibinfo {volume} {46}},\
  \bibinfo {pages} {074001} (\bibinfo {year} {2013})}\BibitemShut {NoStop}%
\bibitem [{\citenamefont {Bhatti}\ \emph {et~al.}(2017)\citenamefont {Bhatti},
  \citenamefont {Sbiaa}, \citenamefont {Hirohata}, \citenamefont {Ohno},
  \citenamefont {Fukami},\ and\ \citenamefont
  {Piramanayagam}}]{bhatti2017spintronics}%
  \BibitemOpen
  \bibfield  {author} {\bibinfo {author} {\bibfnamefont {S.}~\bibnamefont
  {Bhatti}}, \bibinfo {author} {\bibfnamefont {R.}~\bibnamefont {Sbiaa}},
  \bibinfo {author} {\bibfnamefont {A.}~\bibnamefont {Hirohata}}, \bibinfo
  {author} {\bibfnamefont {H.}~\bibnamefont {Ohno}}, \bibinfo {author}
  {\bibfnamefont {S.}~\bibnamefont {Fukami}}, \ and\ \bibinfo {author}
  {\bibfnamefont {S.}~\bibnamefont {Piramanayagam}},\ }\href@noop {} {\bibfield
   {journal} {\bibinfo  {journal} {Materials Today}\ }\textbf {\bibinfo
  {volume} {20}},\ \bibinfo {pages} {530} (\bibinfo {year} {2017})}\BibitemShut
  {NoStop}%
\bibitem [{\citenamefont {Prenat}\ \emph {et~al.}(2007)\citenamefont {Prenat},
  \citenamefont {El~Baraji}, \citenamefont {Guo}, \citenamefont {Sousa},
  \citenamefont {Buda-Prejbeanu}, \citenamefont {Dieny}, \citenamefont
  {Javerliac}, \citenamefont {Nozieres}, \citenamefont {Zhao},\ and\
  \citenamefont {Belhaire}}]{prenat2007cmos}%
  \BibitemOpen
  \bibfield  {author} {\bibinfo {author} {\bibfnamefont {G.}~\bibnamefont
  {Prenat}}, \bibinfo {author} {\bibfnamefont {M.}~\bibnamefont {El~Baraji}},
  \bibinfo {author} {\bibfnamefont {W.}~\bibnamefont {Guo}}, \bibinfo {author}
  {\bibfnamefont {R.}~\bibnamefont {Sousa}}, \bibinfo {author} {\bibfnamefont
  {L.}~\bibnamefont {Buda-Prejbeanu}}, \bibinfo {author} {\bibfnamefont
  {B.}~\bibnamefont {Dieny}}, \bibinfo {author} {\bibfnamefont
  {V.}~\bibnamefont {Javerliac}}, \bibinfo {author} {\bibfnamefont {J.-P.}\
  \bibnamefont {Nozieres}}, \bibinfo {author} {\bibfnamefont {W.}~\bibnamefont
  {Zhao}}, \ and\ \bibinfo {author} {\bibfnamefont {E.}~\bibnamefont
  {Belhaire}},\ }in\ \href@noop {} {\emph {\bibinfo {booktitle} {2007 14th IEEE
  international conference on electronics, circuits and systems}}}\ (\bibinfo
  {organization} {IEEE},\ \bibinfo {year} {2007})\ pp.\ \bibinfo {pages}
  {190--193}\BibitemShut {NoStop}%
\bibitem [{\citenamefont {Bonetti}\ \emph {et~al.}(2010)\citenamefont
  {Bonetti}, \citenamefont {Tiberkevich}, \citenamefont {Consolo},
  \citenamefont {Finocchio}, \citenamefont {Muduli}, \citenamefont {Mancoff},
  \citenamefont {Slavin},\ and\ \citenamefont
  {{\AA}kerman}}]{bonetti2010experimental}%
  \BibitemOpen
  \bibfield  {author} {\bibinfo {author} {\bibfnamefont {S.}~\bibnamefont
  {Bonetti}}, \bibinfo {author} {\bibfnamefont {V.}~\bibnamefont
  {Tiberkevich}}, \bibinfo {author} {\bibfnamefont {G.}~\bibnamefont
  {Consolo}}, \bibinfo {author} {\bibfnamefont {G.}~\bibnamefont {Finocchio}},
  \bibinfo {author} {\bibfnamefont {P.}~\bibnamefont {Muduli}}, \bibinfo
  {author} {\bibfnamefont {F.}~\bibnamefont {Mancoff}}, \bibinfo {author}
  {\bibfnamefont {A.}~\bibnamefont {Slavin}}, \ and\ \bibinfo {author}
  {\bibfnamefont {J.}~\bibnamefont {{\AA}kerman}},\ }\href@noop {} {\bibfield
  {journal} {\bibinfo  {journal} {Physical review letters}\ }\textbf {\bibinfo
  {volume} {105}},\ \bibinfo {pages} {217204} (\bibinfo {year}
  {2010})}\BibitemShut {NoStop}%
\bibitem [{\citenamefont {Villard}\ \emph {et~al.}(2009)\citenamefont
  {Villard}, \citenamefont {Ebels}, \citenamefont {Houssameddine},
  \citenamefont {Katine}, \citenamefont {Mauri}, \citenamefont {Delaet},
  \citenamefont {Vincent}, \citenamefont {Cyrille}, \citenamefont {Viala},
  \citenamefont {Michel} \emph {et~al.}}]{villard2009ghz}%
  \BibitemOpen
  \bibfield  {author} {\bibinfo {author} {\bibfnamefont {P.}~\bibnamefont
  {Villard}}, \bibinfo {author} {\bibfnamefont {U.}~\bibnamefont {Ebels}},
  \bibinfo {author} {\bibfnamefont {D.}~\bibnamefont {Houssameddine}}, \bibinfo
  {author} {\bibfnamefont {J.}~\bibnamefont {Katine}}, \bibinfo {author}
  {\bibfnamefont {D.}~\bibnamefont {Mauri}}, \bibinfo {author} {\bibfnamefont
  {B.}~\bibnamefont {Delaet}}, \bibinfo {author} {\bibfnamefont
  {P.}~\bibnamefont {Vincent}}, \bibinfo {author} {\bibfnamefont {M.-C.}\
  \bibnamefont {Cyrille}}, \bibinfo {author} {\bibfnamefont {B.}~\bibnamefont
  {Viala}}, \bibinfo {author} {\bibfnamefont {J.-P.}\ \bibnamefont {Michel}},
  \emph {et~al.},\ }\href@noop {} {\bibfield  {journal} {\bibinfo  {journal}
  {IEEE Journal of solid-state circuits}\ }\textbf {\bibinfo {volume} {45}},\
  \bibinfo {pages} {214} (\bibinfo {year} {2009})}\BibitemShut {NoStop}%
\bibitem [{\citenamefont {Zeng}\ \emph {et~al.}(2013)\citenamefont {Zeng},
  \citenamefont {Finocchio},\ and\ \citenamefont {Jiang}}]{zeng2013spin}%
  \BibitemOpen
  \bibfield  {author} {\bibinfo {author} {\bibfnamefont {Z.}~\bibnamefont
  {Zeng}}, \bibinfo {author} {\bibfnamefont {G.}~\bibnamefont {Finocchio}}, \
  and\ \bibinfo {author} {\bibfnamefont {H.}~\bibnamefont {Jiang}},\
  }\href@noop {} {\bibfield  {journal} {\bibinfo  {journal} {Nanoscale}\
  }\textbf {\bibinfo {volume} {5}},\ \bibinfo {pages} {2219} (\bibinfo {year}
  {2013})}\BibitemShut {NoStop}%
\bibitem [{\citenamefont {Mizushima}\ \emph {et~al.}(2010)\citenamefont
  {Mizushima}, \citenamefont {Kudo}, \citenamefont {Nagasawa},\ and\
  \citenamefont {Sato}}]{mizushima2010signal}%
  \BibitemOpen
  \bibfield  {author} {\bibinfo {author} {\bibfnamefont {K.}~\bibnamefont
  {Mizushima}}, \bibinfo {author} {\bibfnamefont {K.}~\bibnamefont {Kudo}},
  \bibinfo {author} {\bibfnamefont {T.}~\bibnamefont {Nagasawa}}, \ and\
  \bibinfo {author} {\bibfnamefont {R.}~\bibnamefont {Sato}},\ }\href@noop {}
  {\bibfield  {journal} {\bibinfo  {journal} {Journal of Applied Physics}\
  }\textbf {\bibinfo {volume} {107}},\ \bibinfo {pages} {063904} (\bibinfo
  {year} {2010})}\BibitemShut {NoStop}%
\bibitem [{\citenamefont {Butler}\ \emph {et~al.}(2001)\citenamefont {Butler},
  \citenamefont {Zhang}, \citenamefont {Schulthess},\ and\ \citenamefont
  {MacLaren}}]{butler2001spin}%
  \BibitemOpen
  \bibfield  {author} {\bibinfo {author} {\bibfnamefont {W.}~\bibnamefont
  {Butler}}, \bibinfo {author} {\bibfnamefont {X.-G.}\ \bibnamefont {Zhang}},
  \bibinfo {author} {\bibfnamefont {T.}~\bibnamefont {Schulthess}}, \ and\
  \bibinfo {author} {\bibfnamefont {J.}~\bibnamefont {MacLaren}},\ }\href@noop
  {} {\bibfield  {journal} {\bibinfo  {journal} {Physical Review B}\ }\textbf
  {\bibinfo {volume} {63}},\ \bibinfo {pages} {054416} (\bibinfo {year}
  {2001})}\BibitemShut {NoStop}%
\bibitem [{\citenamefont {Meo}\ \emph {et~al.}(2022)\citenamefont {Meo},
  \citenamefont {Chureemart}, \citenamefont {Chantrell},\ and\ \citenamefont
  {Chureemart}}]{meo2022magnetisation}%
  \BibitemOpen
  \bibfield  {author} {\bibinfo {author} {\bibfnamefont {A.}~\bibnamefont
  {Meo}}, \bibinfo {author} {\bibfnamefont {J.}~\bibnamefont {Chureemart}},
  \bibinfo {author} {\bibfnamefont {R.~W.}\ \bibnamefont {Chantrell}}, \ and\
  \bibinfo {author} {\bibfnamefont {P.}~\bibnamefont {Chureemart}},\
  }\href@noop {} {\bibfield  {journal} {\bibinfo  {journal} {Scientific
  reports}\ }\textbf {\bibinfo {volume} {12}},\ \bibinfo {pages} {1} (\bibinfo
  {year} {2022})}\BibitemShut {NoStop}%
\bibitem [{\citenamefont {Slaughter}\ \emph {et~al.}(2016)\citenamefont
  {Slaughter}, \citenamefont {Nagel}, \citenamefont {Whig}, \citenamefont
  {Deshpande}, \citenamefont {Aggarwal}, \citenamefont {DeHerrera},
  \citenamefont {Janesky}, \citenamefont {Lin}, \citenamefont {Chia},
  \citenamefont {Hossain} \emph {et~al.}}]{slaughter2016technology}%
  \BibitemOpen
  \bibfield  {author} {\bibinfo {author} {\bibfnamefont {J.}~\bibnamefont
  {Slaughter}}, \bibinfo {author} {\bibfnamefont {K.}~\bibnamefont {Nagel}},
  \bibinfo {author} {\bibfnamefont {R.}~\bibnamefont {Whig}}, \bibinfo {author}
  {\bibfnamefont {S.}~\bibnamefont {Deshpande}}, \bibinfo {author}
  {\bibfnamefont {S.}~\bibnamefont {Aggarwal}}, \bibinfo {author}
  {\bibfnamefont {M.}~\bibnamefont {DeHerrera}}, \bibinfo {author}
  {\bibfnamefont {J.}~\bibnamefont {Janesky}}, \bibinfo {author} {\bibfnamefont
  {M.}~\bibnamefont {Lin}}, \bibinfo {author} {\bibfnamefont {H.-J.}\
  \bibnamefont {Chia}}, \bibinfo {author} {\bibfnamefont {M.}~\bibnamefont
  {Hossain}},  \emph {et~al.},\ }in\ \href@noop {} {\emph {\bibinfo {booktitle}
  {2016 IEEE International Electron Devices Meeting (IEDM)}}}\ (\bibinfo
  {organization} {IEEE},\ \bibinfo {year} {2016})\ pp.\ \bibinfo {pages}
  {21--5}\BibitemShut {NoStop}%
\bibitem [{\citenamefont {Endoh}\ \emph {et~al.}(2020)\citenamefont {Endoh},
  \citenamefont {Honjo}, \citenamefont {Nishioka},\ and\ \citenamefont
  {Ikeda}}]{endoh2020recent}%
  \BibitemOpen
  \bibfield  {author} {\bibinfo {author} {\bibfnamefont {T.}~\bibnamefont
  {Endoh}}, \bibinfo {author} {\bibfnamefont {H.}~\bibnamefont {Honjo}},
  \bibinfo {author} {\bibfnamefont {K.}~\bibnamefont {Nishioka}}, \ and\
  \bibinfo {author} {\bibfnamefont {S.}~\bibnamefont {Ikeda}},\ }in\ \href@noop
  {} {\emph {\bibinfo {booktitle} {2020 IEEE Symposium on VLSI Technology}}}\
  (\bibinfo {organization} {IEEE},\ \bibinfo {year} {2020})\ pp.\ \bibinfo
  {pages} {1--2}\BibitemShut {NoStop}%
\bibitem [{\citenamefont {Byun}\ \emph {et~al.}(2021)\citenamefont {Byun},
  \citenamefont {Kang},\ and\ \citenamefont {Shin}}]{byun2021switching}%
  \BibitemOpen
  \bibfield  {author} {\bibinfo {author} {\bibfnamefont {J.}~\bibnamefont
  {Byun}}, \bibinfo {author} {\bibfnamefont {D.-H.}\ \bibnamefont {Kang}}, \
  and\ \bibinfo {author} {\bibfnamefont {M.}~\bibnamefont {Shin}},\ }\href@noop
  {} {\bibfield  {journal} {\bibinfo  {journal} {AIP Advances}\ }\textbf
  {\bibinfo {volume} {11}},\ \bibinfo {pages} {015035} (\bibinfo {year}
  {2021})}\BibitemShut {NoStop}%
\bibitem [{\citenamefont {Sharma}\ \emph {et~al.}(2017)\citenamefont {Sharma},
  \citenamefont {Tulapurkar},\ and\ \citenamefont
  {Muralidharan}}]{PhysRevApplied.8.064014}%
  \BibitemOpen
  \bibfield  {author} {\bibinfo {author} {\bibfnamefont {A.}~\bibnamefont
  {Sharma}}, \bibinfo {author} {\bibfnamefont {A.~A.}\ \bibnamefont
  {Tulapurkar}}, \ and\ \bibinfo {author} {\bibfnamefont {B.}~\bibnamefont
  {Muralidharan}},\ }\href {\doibase 10.1103/PhysRevApplied.8.064014}
  {\bibfield  {journal} {\bibinfo  {journal} {Phys. Rev. Applied}\ }\textbf
  {\bibinfo {volume} {8}},\ \bibinfo {pages} {064014} (\bibinfo {year}
  {2017})}\BibitemShut {NoStop}%
\bibitem [{\citenamefont {Sharma}\ \emph {et~al.}(2016)\citenamefont {Sharma},
  \citenamefont {Tulapurkar},\ and\ \citenamefont {Muralidharan}}]{7571106}%
  \BibitemOpen
  \bibfield  {author} {\bibinfo {author} {\bibfnamefont {A.}~\bibnamefont
  {Sharma}}, \bibinfo {author} {\bibfnamefont {A.}~\bibnamefont {Tulapurkar}},
  \ and\ \bibinfo {author} {\bibfnamefont {B.}~\bibnamefont {Muralidharan}},\
  }\href {\doibase 10.1109/TED.2016.2606354} {\bibfield  {journal} {\bibinfo
  {journal} {IEEE Transactions on Electron Devices}\ }\textbf {\bibinfo
  {volume} {63}},\ \bibinfo {pages} {4527} (\bibinfo {year}
  {2016})}\BibitemShut {NoStop}%
\bibitem [{\citenamefont {Sharma}\ \emph
  {et~al.}(2018{\natexlab{a}})\citenamefont {Sharma}, \citenamefont
  {Tulapurkar},\ and\ \citenamefont {Muralidharan}}]{sharma2018role}%
  \BibitemOpen
  \bibfield  {author} {\bibinfo {author} {\bibfnamefont {A.}~\bibnamefont
  {Sharma}}, \bibinfo {author} {\bibfnamefont {A.~A.}\ \bibnamefont
  {Tulapurkar}}, \ and\ \bibinfo {author} {\bibfnamefont {B.}~\bibnamefont
  {Muralidharan}},\ }\href@noop {} {\bibfield  {journal} {\bibinfo  {journal}
  {AIP Advances}\ }\textbf {\bibinfo {volume} {8}},\ \bibinfo {pages} {055913}
  (\bibinfo {year} {2018}{\natexlab{a}})}\BibitemShut {NoStop}%
\bibitem [{\citenamefont {Chen}\ and\ \citenamefont
  {Hsueh}(2014)}]{chen2014enhancement}%
  \BibitemOpen
  \bibfield  {author} {\bibinfo {author} {\bibfnamefont {C.}~\bibnamefont
  {Chen}}\ and\ \bibinfo {author} {\bibfnamefont {W.}~\bibnamefont {Hsueh}},\
  }\href@noop {} {\bibfield  {journal} {\bibinfo  {journal} {Applied Physics
  Letters}\ }\textbf {\bibinfo {volume} {104}},\ \bibinfo {pages} {042405}
  (\bibinfo {year} {2014})}\BibitemShut {NoStop}%
\bibitem [{\citenamefont {Chen}\ \emph {et~al.}(2015)\citenamefont {Chen},
  \citenamefont {Cheng},\ and\ \citenamefont {Hsueh}}]{chen2015ultrahigh}%
  \BibitemOpen
  \bibfield  {author} {\bibinfo {author} {\bibfnamefont {C.~H.}\ \bibnamefont
  {Chen}}, \bibinfo {author} {\bibfnamefont {Y.~H.}\ \bibnamefont {Cheng}}, \
  and\ \bibinfo {author} {\bibfnamefont {W.~J.}\ \bibnamefont {Hsueh}},\
  }\href@noop {} {\bibfield  {journal} {\bibinfo  {journal} {EPL (Europhysics
  Letters)}\ }\textbf {\bibinfo {volume} {111}},\ \bibinfo {pages} {47005}
  (\bibinfo {year} {2015})}\BibitemShut {NoStop}%
\bibitem [{\citenamefont {Devaraj}\ and\ \citenamefont
  {Tarafder}(2021)}]{devaraj2021large}%
  \BibitemOpen
  \bibfield  {author} {\bibinfo {author} {\bibfnamefont {N.}~\bibnamefont
  {Devaraj}}\ and\ \bibinfo {author} {\bibfnamefont {K.}~\bibnamefont
  {Tarafder}},\ }\href@noop {} {\bibfield  {journal} {\bibinfo  {journal}
  {Physical Review B}\ }\textbf {\bibinfo {volume} {103}},\ \bibinfo {pages}
  {165407} (\bibinfo {year} {2021})}\BibitemShut {NoStop}%
\bibitem [{\citenamefont {Zhang}\ \emph {et~al.}(2022)\citenamefont {Zhang},
  \citenamefont {Liu}, \citenamefont {Deng}, \citenamefont {Shi}, \citenamefont
  {Tang}, \citenamefont {Chen},\ and\ \citenamefont
  {Yuan}}]{zhang2022electronic}%
  \BibitemOpen
  \bibfield  {author} {\bibinfo {author} {\bibfnamefont {Y.}~\bibnamefont
  {Zhang}}, \bibinfo {author} {\bibfnamefont {J.}~\bibnamefont {Liu}}, \bibinfo
  {author} {\bibfnamefont {R.}~\bibnamefont {Deng}}, \bibinfo {author}
  {\bibfnamefont {X.}~\bibnamefont {Shi}}, \bibinfo {author} {\bibfnamefont
  {H.}~\bibnamefont {Tang}}, \bibinfo {author} {\bibfnamefont {H.}~\bibnamefont
  {Chen}}, \ and\ \bibinfo {author} {\bibfnamefont {H.}~\bibnamefont {Yuan}},\
  }\href@noop {} {\bibfield  {journal} {\bibinfo  {journal} {RSC advances}\
  }\textbf {\bibinfo {volume} {12}},\ \bibinfo {pages} {28533} (\bibinfo {year}
  {2022})}\BibitemShut {NoStop}%
\bibitem [{\citenamefont {Zhao}\ \emph {et~al.}(2018)\citenamefont {Zhao},
  \citenamefont {Li}, \citenamefont {Jin}, \citenamefont {Yu}, \citenamefont
  {Huang},\ and\ \citenamefont {Ying}}]{zhao2018designing}%
  \BibitemOpen
  \bibfield  {author} {\bibinfo {author} {\bibfnamefont {P.}~\bibnamefont
  {Zhao}}, \bibinfo {author} {\bibfnamefont {J.}~\bibnamefont {Li}}, \bibinfo
  {author} {\bibfnamefont {H.}~\bibnamefont {Jin}}, \bibinfo {author}
  {\bibfnamefont {L.}~\bibnamefont {Yu}}, \bibinfo {author} {\bibfnamefont
  {B.}~\bibnamefont {Huang}}, \ and\ \bibinfo {author} {\bibfnamefont
  {D.}~\bibnamefont {Ying}},\ }\href@noop {} {\bibfield  {journal} {\bibinfo
  {journal} {Physical Chemistry Chemical Physics}\ }\textbf {\bibinfo {volume}
  {20}},\ \bibinfo {pages} {10286} (\bibinfo {year} {2018})}\BibitemShut
  {NoStop}%
\bibitem [{\citenamefont {Tao}\ \emph {et~al.}(2019)\citenamefont {Tao},
  \citenamefont {Wan}, \citenamefont {Tang}, \citenamefont {Feng},
  \citenamefont {Wei}, \citenamefont {Wang}, \citenamefont {Andrieu},
  \citenamefont {Yang}, \citenamefont {Chshiev}, \citenamefont {Devaux} \emph
  {et~al.}}]{tao2019coherent}%
  \BibitemOpen
  \bibfield  {author} {\bibinfo {author} {\bibfnamefont {B.}~\bibnamefont
  {Tao}}, \bibinfo {author} {\bibfnamefont {C.}~\bibnamefont {Wan}}, \bibinfo
  {author} {\bibfnamefont {P.}~\bibnamefont {Tang}}, \bibinfo {author}
  {\bibfnamefont {J.}~\bibnamefont {Feng}}, \bibinfo {author} {\bibfnamefont
  {H.}~\bibnamefont {Wei}}, \bibinfo {author} {\bibfnamefont {X.}~\bibnamefont
  {Wang}}, \bibinfo {author} {\bibfnamefont {S.}~\bibnamefont {Andrieu}},
  \bibinfo {author} {\bibfnamefont {H.}~\bibnamefont {Yang}}, \bibinfo {author}
  {\bibfnamefont {M.}~\bibnamefont {Chshiev}}, \bibinfo {author} {\bibfnamefont
  {X.}~\bibnamefont {Devaux}},  \emph {et~al.},\ }\href@noop {} {\bibfield
  {journal} {\bibinfo  {journal} {Nano Letters}\ }\textbf {\bibinfo {volume}
  {19}},\ \bibinfo {pages} {3019} (\bibinfo {year} {2019})}\BibitemShut
  {NoStop}%
\bibitem [{\citenamefont {Bhattacharjee}\ \emph {et~al.}(2016)\citenamefont
  {Bhattacharjee}, \citenamefont {Nemade},\ and\ \citenamefont
  {Bandyopadhyay}}]{bhattacharjee2016effect}%
  \BibitemOpen
  \bibfield  {author} {\bibinfo {author} {\bibfnamefont {M.}~\bibnamefont
  {Bhattacharjee}}, \bibinfo {author} {\bibfnamefont {H.}~\bibnamefont
  {Nemade}}, \ and\ \bibinfo {author} {\bibfnamefont {D.}~\bibnamefont
  {Bandyopadhyay}},\ }in\ \href@noop {} {\emph {\bibinfo {booktitle} {Journal
  of Physics: Conference Series}}},\ Vol.\ \bibinfo {volume} {759}\ (\bibinfo
  {organization} {IOP Publishing},\ \bibinfo {year} {2016})\ p.\ \bibinfo
  {pages} {012051}\BibitemShut {NoStop}%
\bibitem [{\citenamefont {Tseng}\ \emph {et~al.}(2020)\citenamefont {Tseng},
  \citenamefont {Chen},\ and\ \citenamefont {Hsueh}}]{tseng2020superlattice}%
  \BibitemOpen
  \bibfield  {author} {\bibinfo {author} {\bibfnamefont {P.}~\bibnamefont
  {Tseng}}, \bibinfo {author} {\bibfnamefont {Z.-Y.}\ \bibnamefont {Chen}}, \
  and\ \bibinfo {author} {\bibfnamefont {W.-J.}\ \bibnamefont {Hsueh}},\
  }\href@noop {} {\bibfield  {journal} {\bibinfo  {journal} {New Journal of
  Physics}\ }\textbf {\bibinfo {volume} {22}},\ \bibinfo {pages} {093005}
  (\bibinfo {year} {2020})}\BibitemShut {NoStop}%
\bibitem [{\citenamefont {Diez}\ \emph {et~al.}(2000)\citenamefont {Diez},
  \citenamefont {G{\'o}mez}, \citenamefont {Dom{\i}nguez-Adame}, \citenamefont
  {Hey}, \citenamefont {Bellani},\ and\ \citenamefont
  {Parravicini}}]{diez2000gaussian}%
  \BibitemOpen
  \bibfield  {author} {\bibinfo {author} {\bibfnamefont {E.}~\bibnamefont
  {Diez}}, \bibinfo {author} {\bibfnamefont {I.}~\bibnamefont {G{\'o}mez}},
  \bibinfo {author} {\bibfnamefont {F.}~\bibnamefont {Dom{\i}nguez-Adame}},
  \bibinfo {author} {\bibfnamefont {R.}~\bibnamefont {Hey}}, \bibinfo {author}
  {\bibfnamefont {V.}~\bibnamefont {Bellani}}, \ and\ \bibinfo {author}
  {\bibfnamefont {G.}~\bibnamefont {Parravicini}},\ }\href@noop {} {\bibfield
  {journal} {\bibinfo  {journal} {Physica E: Low-dimensional Systems and
  Nanostructures}\ }\textbf {\bibinfo {volume} {7}},\ \bibinfo {pages} {832}
  (\bibinfo {year} {2000})}\BibitemShut {NoStop}%
\bibitem [{\citenamefont {G{\'o}mez}\ \emph {et~al.}(1999)\citenamefont
  {G{\'o}mez}, \citenamefont {Dom{\i}nguez-Adame}, \citenamefont {Diez},\ and\
  \citenamefont {Bellani}}]{gomez1999electron}%
  \BibitemOpen
  \bibfield  {author} {\bibinfo {author} {\bibfnamefont {I.}~\bibnamefont
  {G{\'o}mez}}, \bibinfo {author} {\bibfnamefont {F.}~\bibnamefont
  {Dom{\i}nguez-Adame}}, \bibinfo {author} {\bibfnamefont {E.}~\bibnamefont
  {Diez}}, \ and\ \bibinfo {author} {\bibfnamefont {V.}~\bibnamefont
  {Bellani}},\ }\href@noop {} {\bibfield  {journal} {\bibinfo  {journal}
  {Journal of Applied Physics}\ }\textbf {\bibinfo {volume} {85}},\ \bibinfo
  {pages} {3916} (\bibinfo {year} {1999})}\BibitemShut {NoStop}%
\bibitem [{\citenamefont {S{\'a}nchez-Arellano}\ \emph
  {et~al.}(2019)\citenamefont {S{\'a}nchez-Arellano}, \citenamefont
  {Madrigal-Melchor},\ and\ \citenamefont
  {Rodr{\'\i}guez-Vargas}}]{sanchez2019non}%
  \BibitemOpen
  \bibfield  {author} {\bibinfo {author} {\bibfnamefont {A.}~\bibnamefont
  {S{\'a}nchez-Arellano}}, \bibinfo {author} {\bibfnamefont {J.}~\bibnamefont
  {Madrigal-Melchor}}, \ and\ \bibinfo {author} {\bibfnamefont
  {I.}~\bibnamefont {Rodr{\'\i}guez-Vargas}},\ }\href@noop {} {\bibfield
  {journal} {\bibinfo  {journal} {Scientific reports}\ }\textbf {\bibinfo
  {volume} {9}},\ \bibinfo {pages} {1} (\bibinfo {year} {2019})}\BibitemShut
  {NoStop}%
\bibitem [{\citenamefont {Tian}\ \emph {et~al.}(2014)\citenamefont {Tian},
  \citenamefont {Duan}, \citenamefont {Li}, \citenamefont {Chen}, \citenamefont
  {Sha}, \citenamefont {Zhao}, \citenamefont {Liu},\ and\ \citenamefont
  {Cui}}]{tian2014miscibility}%
  \BibitemOpen
  \bibfield  {author} {\bibinfo {author} {\bibfnamefont {F.}~\bibnamefont
  {Tian}}, \bibinfo {author} {\bibfnamefont {D.}~\bibnamefont {Duan}}, \bibinfo
  {author} {\bibfnamefont {D.}~\bibnamefont {Li}}, \bibinfo {author}
  {\bibfnamefont {C.}~\bibnamefont {Chen}}, \bibinfo {author} {\bibfnamefont
  {X.}~\bibnamefont {Sha}}, \bibinfo {author} {\bibfnamefont {Z.}~\bibnamefont
  {Zhao}}, \bibinfo {author} {\bibfnamefont {B.}~\bibnamefont {Liu}}, \ and\
  \bibinfo {author} {\bibfnamefont {T.}~\bibnamefont {Cui}},\ }\href@noop {}
  {\bibfield  {journal} {\bibinfo  {journal} {Scientific reports}\ }\textbf
  {\bibinfo {volume} {4}},\ \bibinfo {pages} {1} (\bibinfo {year}
  {2014})}\BibitemShut {NoStop}%
\bibitem [{\citenamefont {Li}\ \emph {et~al.}(2014)\citenamefont {Li},
  \citenamefont {Ma}, \citenamefont {Wang}, \citenamefont {Ward}, \citenamefont
  {Hesjedal}, \citenamefont {Zhang}, \citenamefont {Kohn}, \citenamefont
  {Amsellem}, \citenamefont {Yang}, \citenamefont {Liu} \emph
  {et~al.}}]{li2014controlling}%
  \BibitemOpen
  \bibfield  {author} {\bibinfo {author} {\bibfnamefont {D.}~\bibnamefont
  {Li}}, \bibinfo {author} {\bibfnamefont {Q.}~\bibnamefont {Ma}}, \bibinfo
  {author} {\bibfnamefont {S.}~\bibnamefont {Wang}}, \bibinfo {author}
  {\bibfnamefont {R.}~\bibnamefont {Ward}}, \bibinfo {author} {\bibfnamefont
  {T.}~\bibnamefont {Hesjedal}}, \bibinfo {author} {\bibfnamefont {X.-G.}\
  \bibnamefont {Zhang}}, \bibinfo {author} {\bibfnamefont {A.}~\bibnamefont
  {Kohn}}, \bibinfo {author} {\bibfnamefont {E.}~\bibnamefont {Amsellem}},
  \bibinfo {author} {\bibfnamefont {G.}~\bibnamefont {Yang}}, \bibinfo {author}
  {\bibfnamefont {J.}~\bibnamefont {Liu}},  \emph {et~al.},\ }\href@noop {}
  {\bibfield  {journal} {\bibinfo  {journal} {Scientific Reports}\ }\textbf
  {\bibinfo {volume} {4}},\ \bibinfo {pages} {1} (\bibinfo {year}
  {2014})}\BibitemShut {NoStop}%
\bibitem [{\citenamefont {Sharma}\ \emph {et~al.}(2021)\citenamefont {Sharma},
  \citenamefont {Tulapurkar},\ and\ \citenamefont
  {Muralidharan}}]{sharma2021proposal}%
  \BibitemOpen
  \bibfield  {author} {\bibinfo {author} {\bibfnamefont {A.}~\bibnamefont
  {Sharma}}, \bibinfo {author} {\bibfnamefont {A.~A.}\ \bibnamefont
  {Tulapurkar}}, \ and\ \bibinfo {author} {\bibfnamefont {B.}~\bibnamefont
  {Muralidharan}},\ }\href@noop {} {\bibfield  {journal} {\bibinfo  {journal}
  {Journal of Applied Physics}\ }\textbf {\bibinfo {volume} {129}},\ \bibinfo
  {pages} {233901} (\bibinfo {year} {2021})}\BibitemShut {NoStop}%
\bibitem [{\citenamefont {Sharma}\ \emph
  {et~al.}(2018{\natexlab{b}})\citenamefont {Sharma}, \citenamefont
  {Tulapurkar},\ and\ \citenamefont {Muralidharan}}]{sharma2018band}%
  \BibitemOpen
  \bibfield  {author} {\bibinfo {author} {\bibfnamefont {A.}~\bibnamefont
  {Sharma}}, \bibinfo {author} {\bibfnamefont {A.~A.}\ \bibnamefont
  {Tulapurkar}}, \ and\ \bibinfo {author} {\bibfnamefont {B.}~\bibnamefont
  {Muralidharan}},\ }\href@noop {} {\bibfield  {journal} {\bibinfo  {journal}
  {Applied Physics Letters}\ }\textbf {\bibinfo {volume} {112}},\ \bibinfo
  {pages} {192404} (\bibinfo {year} {2018}{\natexlab{b}})}\BibitemShut
  {NoStop}%
\bibitem [{\citenamefont {Datta}(2018)}]{datta2018lessons}%
  \BibitemOpen
  \bibfield  {author} {\bibinfo {author} {\bibfnamefont {S.}~\bibnamefont
  {Datta}},\ }\href@noop {} {\emph {\bibinfo {title} {Lessons from
  Nanoelectronics: A New Perspective on Transport—Part B: Quantum
  Transport}}}\ (\bibinfo  {publisher} {World Scientific},\ \bibinfo {year}
  {2018})\BibitemShut {NoStop}%
\bibitem [{\citenamefont {Ralph}\ and\ \citenamefont
  {Stiles}(2008)}]{ralph2008spin}%
  \BibitemOpen
  \bibfield  {author} {\bibinfo {author} {\bibfnamefont {D.~C.}\ \bibnamefont
  {Ralph}}\ and\ \bibinfo {author} {\bibfnamefont {M.~D.}\ \bibnamefont
  {Stiles}},\ }\href@noop {} {\bibfield  {journal} {\bibinfo  {journal}
  {Journal of Magnetism and Magnetic Materials}\ }\textbf {\bibinfo {volume}
  {320}},\ \bibinfo {pages} {1190} (\bibinfo {year} {2008})}\BibitemShut
  {NoStop}%
\bibitem [{\citenamefont {Datta}(2005)}]{datta2005quantum}%
  \BibitemOpen
  \bibfield  {author} {\bibinfo {author} {\bibfnamefont {S.}~\bibnamefont
  {Datta}},\ }\href@noop {} {\emph {\bibinfo {title} {Quantum transport: atom
  to transistor}}}\ (\bibinfo  {publisher} {Cambridge university press},\
  \bibinfo {year} {2005})\BibitemShut {NoStop}%
\bibitem [{\citenamefont {Datta}(1997)}]{datta1997electronic}%
  \BibitemOpen
  \bibfield  {author} {\bibinfo {author} {\bibfnamefont {S.}~\bibnamefont
  {Datta}},\ }\href@noop {} {\emph {\bibinfo {title} {Electronic transport in
  mesoscopic systems}}}\ (\bibinfo  {publisher} {Cambridge university press},\
  \bibinfo {year} {1997})\BibitemShut {NoStop}%
\bibitem [{\citenamefont {Ralph}\ \emph {et~al.}(2011)\citenamefont {Ralph},
  \citenamefont {Cui}, \citenamefont {Liu}, \citenamefont {Moriyama},
  \citenamefont {Wang},\ and\ \citenamefont {Buhrman}}]{ralph2011spin}%
  \BibitemOpen
  \bibfield  {author} {\bibinfo {author} {\bibfnamefont {D.}~\bibnamefont
  {Ralph}}, \bibinfo {author} {\bibfnamefont {Y.-T.}\ \bibnamefont {Cui}},
  \bibinfo {author} {\bibfnamefont {L.}~\bibnamefont {Liu}}, \bibinfo {author}
  {\bibfnamefont {T.}~\bibnamefont {Moriyama}}, \bibinfo {author}
  {\bibfnamefont {C.}~\bibnamefont {Wang}}, \ and\ \bibinfo {author}
  {\bibfnamefont {R.}~\bibnamefont {Buhrman}},\ }\href@noop {} {\bibfield
  {journal} {\bibinfo  {journal} {Philosophical Transactions of the Royal
  Society A: Mathematical, Physical and Engineering Sciences}\ }\textbf
  {\bibinfo {volume} {369}},\ \bibinfo {pages} {3617} (\bibinfo {year}
  {2011})}\BibitemShut {NoStop}%
\bibitem [{\citenamefont {Tudu}\ and\ \citenamefont
  {Tiwari}(2017)}]{tudu2017recent}%
  \BibitemOpen
  \bibfield  {author} {\bibinfo {author} {\bibfnamefont {B.}~\bibnamefont
  {Tudu}}\ and\ \bibinfo {author} {\bibfnamefont {A.}~\bibnamefont {Tiwari}},\
  }\href@noop {} {\bibfield  {journal} {\bibinfo  {journal} {Vacuum}\ }\textbf
  {\bibinfo {volume} {146}},\ \bibinfo {pages} {329} (\bibinfo {year}
  {2017})}\BibitemShut {NoStop}%
\bibitem [{\citenamefont {Prakash}\ \emph {et~al.}(2017)\citenamefont
  {Prakash}, \citenamefont {Rathore},\ and\ \citenamefont
  {Kaur}}]{prakash2017effect}%
  \BibitemOpen
  \bibfield  {author} {\bibinfo {author} {\bibfnamefont {R.}~\bibnamefont
  {Prakash}}, \bibinfo {author} {\bibfnamefont {B.~P.~S.}\ \bibnamefont
  {Rathore}}, \ and\ \bibinfo {author} {\bibfnamefont {D.}~\bibnamefont
  {Kaur}},\ }\href@noop {} {\bibfield  {journal} {\bibinfo  {journal} {Journal
  of Alloys and Compounds}\ }\textbf {\bibinfo {volume} {726}},\ \bibinfo
  {pages} {693} (\bibinfo {year} {2017})}\BibitemShut {NoStop}%
\bibitem [{\citenamefont {Ikeda}\ \emph {et~al.}(2010)\citenamefont {Ikeda},
  \citenamefont {Miura}, \citenamefont {Yamamoto}, \citenamefont {Mizunuma},
  \citenamefont {Gan}, \citenamefont {Endo}, \citenamefont {Kanai},
  \citenamefont {Hayakawa}, \citenamefont {Matsukura},\ and\ \citenamefont
  {Ohno}}]{ikeda2010perpendicular}%
  \BibitemOpen
  \bibfield  {author} {\bibinfo {author} {\bibfnamefont {S.}~\bibnamefont
  {Ikeda}}, \bibinfo {author} {\bibfnamefont {K.}~\bibnamefont {Miura}},
  \bibinfo {author} {\bibfnamefont {H.}~\bibnamefont {Yamamoto}}, \bibinfo
  {author} {\bibfnamefont {K.}~\bibnamefont {Mizunuma}}, \bibinfo {author}
  {\bibfnamefont {H.}~\bibnamefont {Gan}}, \bibinfo {author} {\bibfnamefont
  {M.}~\bibnamefont {Endo}}, \bibinfo {author} {\bibfnamefont {S.}~\bibnamefont
  {Kanai}}, \bibinfo {author} {\bibfnamefont {J.}~\bibnamefont {Hayakawa}},
  \bibinfo {author} {\bibfnamefont {F.}~\bibnamefont {Matsukura}}, \ and\
  \bibinfo {author} {\bibfnamefont {H.}~\bibnamefont {Ohno}},\ }\href@noop {}
  {\bibfield  {journal} {\bibinfo  {journal} {Nature materials}\ }\textbf
  {\bibinfo {volume} {9}},\ \bibinfo {pages} {721} (\bibinfo {year}
  {2010})}\BibitemShut {NoStop}%
\bibitem [{\citenamefont {Miao}\ \emph {et~al.}(2006)\citenamefont {Miao},
  \citenamefont {Chetry}, \citenamefont {Gupta}, \citenamefont {Butler},
  \citenamefont {Tsunekawa}, \citenamefont {Djayaprawira},\ and\ \citenamefont
  {Xiao}}]{miao2006inelastic}%
  \BibitemOpen
  \bibfield  {author} {\bibinfo {author} {\bibfnamefont {G.-X.}\ \bibnamefont
  {Miao}}, \bibinfo {author} {\bibfnamefont {K.~B.}\ \bibnamefont {Chetry}},
  \bibinfo {author} {\bibfnamefont {A.}~\bibnamefont {Gupta}}, \bibinfo
  {author} {\bibfnamefont {W.~H.}\ \bibnamefont {Butler}}, \bibinfo {author}
  {\bibfnamefont {K.}~\bibnamefont {Tsunekawa}}, \bibinfo {author}
  {\bibfnamefont {D.}~\bibnamefont {Djayaprawira}}, \ and\ \bibinfo {author}
  {\bibfnamefont {G.}~\bibnamefont {Xiao}},\ }\href@noop {} {\bibfield
  {journal} {\bibinfo  {journal} {Journal of applied physics}\ }\textbf
  {\bibinfo {volume} {99}},\ \bibinfo {pages} {08T305} (\bibinfo {year}
  {2006})}\BibitemShut {NoStop}%
\bibitem [{\citenamefont {Datta}\ \emph {et~al.}(2011)\citenamefont {Datta},
  \citenamefont {Behin-Aein}, \citenamefont {Datta},\ and\ \citenamefont
  {Salahuddin}}]{datta2011voltage}%
  \BibitemOpen
  \bibfield  {author} {\bibinfo {author} {\bibfnamefont {D.}~\bibnamefont
  {Datta}}, \bibinfo {author} {\bibfnamefont {B.}~\bibnamefont {Behin-Aein}},
  \bibinfo {author} {\bibfnamefont {S.}~\bibnamefont {Datta}}, \ and\ \bibinfo
  {author} {\bibfnamefont {S.}~\bibnamefont {Salahuddin}},\ }\href@noop {}
  {\bibfield  {journal} {\bibinfo  {journal} {IEEE Transactions on
  Nanotechnology}\ }\textbf {\bibinfo {volume} {11}},\ \bibinfo {pages} {261}
  (\bibinfo {year} {2011})}\BibitemShut {NoStop}%
\bibitem [{\citenamefont {Datta}(2012)}]{datta2012behin}%
  \BibitemOpen
  \bibfield  {author} {\bibinfo {author} {\bibfnamefont {B.}~\bibnamefont
  {Datta}},\ }\href@noop {} {\bibfield  {journal} {\bibinfo  {journal} {IEEE
  Transactions on Nanotechnology}\ }\textbf {\bibinfo {volume} {11}},\ \bibinfo
  {pages} {261} (\bibinfo {year} {2012})}\BibitemShut {NoStop}%
\bibitem [{\citenamefont {Yang}\ \emph {et~al.}(2015)\citenamefont {Yang},
  \citenamefont {Ryu},\ and\ \citenamefont {Parkin}}]{yang2015domain}%
  \BibitemOpen
  \bibfield  {author} {\bibinfo {author} {\bibfnamefont {S.-H.}\ \bibnamefont
  {Yang}}, \bibinfo {author} {\bibfnamefont {K.-S.}\ \bibnamefont {Ryu}}, \
  and\ \bibinfo {author} {\bibfnamefont {S.}~\bibnamefont {Parkin}},\
  }\href@noop {} {\bibfield  {journal} {\bibinfo  {journal} {Nature
  nanotechnology}\ }\textbf {\bibinfo {volume} {10}},\ \bibinfo {pages} {221}
  (\bibinfo {year} {2015})}\BibitemShut {NoStop}%
\bibitem [{\citenamefont {Deac}\ \emph {et~al.}(2008)\citenamefont {Deac},
  \citenamefont {Fukushima}, \citenamefont {Kubota}, \citenamefont {Maehara},
  \citenamefont {Suzuki}, \citenamefont {Yuasa}, \citenamefont {Nagamine},
  \citenamefont {Tsunekawa}, \citenamefont {Djayaprawira},\ and\ \citenamefont
  {Watanabe}}]{deac2008bias}%
  \BibitemOpen
  \bibfield  {author} {\bibinfo {author} {\bibfnamefont {A.~M.}\ \bibnamefont
  {Deac}}, \bibinfo {author} {\bibfnamefont {A.}~\bibnamefont {Fukushima}},
  \bibinfo {author} {\bibfnamefont {H.}~\bibnamefont {Kubota}}, \bibinfo
  {author} {\bibfnamefont {H.}~\bibnamefont {Maehara}}, \bibinfo {author}
  {\bibfnamefont {Y.}~\bibnamefont {Suzuki}}, \bibinfo {author} {\bibfnamefont
  {S.}~\bibnamefont {Yuasa}}, \bibinfo {author} {\bibfnamefont
  {Y.}~\bibnamefont {Nagamine}}, \bibinfo {author} {\bibfnamefont
  {K.}~\bibnamefont {Tsunekawa}}, \bibinfo {author} {\bibfnamefont {D.~D.}\
  \bibnamefont {Djayaprawira}}, \ and\ \bibinfo {author} {\bibfnamefont
  {N.}~\bibnamefont {Watanabe}},\ }\href@noop {} {\bibfield  {journal}
  {\bibinfo  {journal} {Nature Physics}\ }\textbf {\bibinfo {volume} {4}},\
  \bibinfo {pages} {803} (\bibinfo {year} {2008})}\BibitemShut {NoStop}%
\bibitem [{\citenamefont {Sato}\ \emph {et~al.}(2014)\citenamefont {Sato},
  \citenamefont {Enobio}, \citenamefont {Yamanouchi}, \citenamefont {Ikeda},
  \citenamefont {Fukami}, \citenamefont {Kanai}, \citenamefont {Matsukura},\
  and\ \citenamefont {Ohno}}]{sato2014properties}%
  \BibitemOpen
  \bibfield  {author} {\bibinfo {author} {\bibfnamefont {H.}~\bibnamefont
  {Sato}}, \bibinfo {author} {\bibfnamefont {E.}~\bibnamefont {Enobio}},
  \bibinfo {author} {\bibfnamefont {M.}~\bibnamefont {Yamanouchi}}, \bibinfo
  {author} {\bibfnamefont {S.}~\bibnamefont {Ikeda}}, \bibinfo {author}
  {\bibfnamefont {S.}~\bibnamefont {Fukami}}, \bibinfo {author} {\bibfnamefont
  {S.}~\bibnamefont {Kanai}}, \bibinfo {author} {\bibfnamefont
  {F.}~\bibnamefont {Matsukura}}, \ and\ \bibinfo {author} {\bibfnamefont
  {H.}~\bibnamefont {Ohno}},\ }\href@noop {} {\bibfield  {journal} {\bibinfo
  {journal} {Applied Physics Letters}\ }\textbf {\bibinfo {volume} {105}},\
  \bibinfo {pages} {062403} (\bibinfo {year} {2014})}\BibitemShut {NoStop}%
\bibitem [{\citenamefont {Gajek}\ \emph {et~al.}(2012)\citenamefont {Gajek},
  \citenamefont {Nowak}, \citenamefont {Sun}, \citenamefont {Trouilloud},
  \citenamefont {O’sullivan}, \citenamefont {Abraham}, \citenamefont
  {Gaidis}, \citenamefont {Hu}, \citenamefont {Brown}, \citenamefont {Zhu}
  \emph {et~al.}}]{gajek2012spin}%
  \BibitemOpen
  \bibfield  {author} {\bibinfo {author} {\bibfnamefont {M.}~\bibnamefont
  {Gajek}}, \bibinfo {author} {\bibfnamefont {J.}~\bibnamefont {Nowak}},
  \bibinfo {author} {\bibfnamefont {J.}~\bibnamefont {Sun}}, \bibinfo {author}
  {\bibfnamefont {P.}~\bibnamefont {Trouilloud}}, \bibinfo {author}
  {\bibfnamefont {E.}~\bibnamefont {O’sullivan}}, \bibinfo {author}
  {\bibfnamefont {D.}~\bibnamefont {Abraham}}, \bibinfo {author} {\bibfnamefont
  {M.}~\bibnamefont {Gaidis}}, \bibinfo {author} {\bibfnamefont
  {G.}~\bibnamefont {Hu}}, \bibinfo {author} {\bibfnamefont {S.}~\bibnamefont
  {Brown}}, \bibinfo {author} {\bibfnamefont {Y.}~\bibnamefont {Zhu}},  \emph
  {et~al.},\ }\href@noop {} {\bibfield  {journal} {\bibinfo  {journal} {Applied
  Physics Letters}\ }\textbf {\bibinfo {volume} {100}},\ \bibinfo {pages}
  {132408} (\bibinfo {year} {2012})}\BibitemShut {NoStop}%
\bibitem [{\citenamefont {Zhao}\ \emph {et~al.}(2022)\citenamefont {Zhao},
  \citenamefont {Wang}, \citenamefont {Shao}, \citenamefont {Chen},
  \citenamefont {Fu}, \citenamefont {Xia}, \citenamefont {Wang}, \citenamefont
  {Li}, \citenamefont {Dong}, \citenamefont {Zhou} \emph
  {et~al.}}]{zhao2022temperature}%
  \BibitemOpen
  \bibfield  {author} {\bibinfo {author} {\bibfnamefont {D.}~\bibnamefont
  {Zhao}}, \bibinfo {author} {\bibfnamefont {Y.}~\bibnamefont {Wang}}, \bibinfo
  {author} {\bibfnamefont {J.}~\bibnamefont {Shao}}, \bibinfo {author}
  {\bibfnamefont {Y.}~\bibnamefont {Chen}}, \bibinfo {author} {\bibfnamefont
  {Z.}~\bibnamefont {Fu}}, \bibinfo {author} {\bibfnamefont {Q.}~\bibnamefont
  {Xia}}, \bibinfo {author} {\bibfnamefont {S.}~\bibnamefont {Wang}}, \bibinfo
  {author} {\bibfnamefont {X.}~\bibnamefont {Li}}, \bibinfo {author}
  {\bibfnamefont {G.}~\bibnamefont {Dong}}, \bibinfo {author} {\bibfnamefont
  {M.}~\bibnamefont {Zhou}},  \emph {et~al.},\ }\href@noop {} {\bibfield
  {journal} {\bibinfo  {journal} {AIP Advances}\ }\textbf {\bibinfo {volume}
  {12}},\ \bibinfo {pages} {055114} (\bibinfo {year} {2022})}\BibitemShut
  {NoStop}%
\bibitem [{\citenamefont {Kou}\ \emph {et~al.}(2006)\citenamefont {Kou},
  \citenamefont {Schmalhorst}, \citenamefont {Thomas},\ and\ \citenamefont
  {Reiss}}]{kou2006temperature}%
  \BibitemOpen
  \bibfield  {author} {\bibinfo {author} {\bibfnamefont {X.}~\bibnamefont
  {Kou}}, \bibinfo {author} {\bibfnamefont {J.}~\bibnamefont {Schmalhorst}},
  \bibinfo {author} {\bibfnamefont {A.}~\bibnamefont {Thomas}}, \ and\ \bibinfo
  {author} {\bibfnamefont {G.}~\bibnamefont {Reiss}},\ }\href@noop {}
  {\bibfield  {journal} {\bibinfo  {journal} {Applied physics letters}\
  }\textbf {\bibinfo {volume} {88}},\ \bibinfo {pages} {212115} (\bibinfo
  {year} {2006})}\BibitemShut {NoStop}%
\bibitem [{\citenamefont {Drewello}\ \emph {et~al.}(2008)\citenamefont
  {Drewello}, \citenamefont {Schmalhorst}, \citenamefont {Thomas},\ and\
  \citenamefont {Reiss}}]{PhysRevB.77.014440}%
  \BibitemOpen
  \bibfield  {author} {\bibinfo {author} {\bibfnamefont {V.}~\bibnamefont
  {Drewello}}, \bibinfo {author} {\bibfnamefont {J.}~\bibnamefont
  {Schmalhorst}}, \bibinfo {author} {\bibfnamefont {A.}~\bibnamefont {Thomas}},
  \ and\ \bibinfo {author} {\bibfnamefont {G.}~\bibnamefont {Reiss}},\ }\href
  {\doibase 10.1103/PhysRevB.77.014440} {\bibfield  {journal} {\bibinfo
  {journal} {Phys. Rev. B}\ }\textbf {\bibinfo {volume} {77}},\ \bibinfo
  {pages} {014440} (\bibinfo {year} {2008})}\BibitemShut {NoStop}%
\bibitem [{\citenamefont {Wang}\ \emph {et~al.}(2015)\citenamefont {Wang},
  \citenamefont {Cai}, \citenamefont {Naviner}, \citenamefont {Zhang},
  \citenamefont {Klein},\ and\ \citenamefont {Zhao}}]{WANG20151649}%
  \BibitemOpen
  \bibfield  {author} {\bibinfo {author} {\bibfnamefont {Y.}~\bibnamefont
  {Wang}}, \bibinfo {author} {\bibfnamefont {H.}~\bibnamefont {Cai}}, \bibinfo
  {author} {\bibfnamefont {L.}~\bibnamefont {Naviner}}, \bibinfo {author}
  {\bibfnamefont {Y.}~\bibnamefont {Zhang}}, \bibinfo {author} {\bibfnamefont
  {J.}~\bibnamefont {Klein}}, \ and\ \bibinfo {author} {\bibfnamefont
  {W.}~\bibnamefont {Zhao}},\ }\href {\doibase
  https://doi.org/10.1016/j.microrel.2015.06.029} {\bibfield  {journal}
  {\bibinfo  {journal} {Microelectronics Reliability}\ }\textbf {\bibinfo
  {volume} {55}},\ \bibinfo {pages} {1649} (\bibinfo {year} {2015})},\ \bibinfo
  {note} {proceedings of the 26th European Symposium on Reliability of Electron
  Devices, Failure Physics and Analysis}\BibitemShut {NoStop}%
\bibitem [{\citenamefont {Cao}\ \emph {et~al.}(2019)\citenamefont {Cao},
  \citenamefont {Li}, \citenamefont {Cai}, \citenamefont {Wei}, \citenamefont
  {Wang}, \citenamefont {Hu}, \citenamefont {Jiang}, \citenamefont {Cui},
  \citenamefont {Zhao},\ and\ \citenamefont {Zhao}}]{8532138}%
  \BibitemOpen
  \bibfield  {author} {\bibinfo {author} {\bibfnamefont {K.}~\bibnamefont
  {Cao}}, \bibinfo {author} {\bibfnamefont {H.}~\bibnamefont {Li}}, \bibinfo
  {author} {\bibfnamefont {W.}~\bibnamefont {Cai}}, \bibinfo {author}
  {\bibfnamefont {J.}~\bibnamefont {Wei}}, \bibinfo {author} {\bibfnamefont
  {L.}~\bibnamefont {Wang}}, \bibinfo {author} {\bibfnamefont {Y.}~\bibnamefont
  {Hu}}, \bibinfo {author} {\bibfnamefont {Q.}~\bibnamefont {Jiang}}, \bibinfo
  {author} {\bibfnamefont {H.}~\bibnamefont {Cui}}, \bibinfo {author}
  {\bibfnamefont {C.}~\bibnamefont {Zhao}}, \ and\ \bibinfo {author}
  {\bibfnamefont {W.}~\bibnamefont {Zhao}},\ }\href {\doibase
  10.1109/TMAG.2018.2877446} {\bibfield  {journal} {\bibinfo  {journal} {IEEE
  Transactions on Magnetics}\ }\textbf {\bibinfo {volume} {55}},\ \bibinfo
  {pages} {1} (\bibinfo {year} {2019})}\BibitemShut {NoStop}%
\bibitem [{\citenamefont {Prenat}\ \emph {et~al.}(2015)\citenamefont {Prenat},
  \citenamefont {Jabeur}, \citenamefont {Vanhauwaert}, \citenamefont
  {Di~Pendina}, \citenamefont {Oboril}, \citenamefont {Bishnoi}, \citenamefont
  {Ebrahimi}, \citenamefont {Lamard}, \citenamefont {Boulle}, \citenamefont
  {Garello} \emph {et~al.}}]{prenat2015ultra}%
  \BibitemOpen
  \bibfield  {author} {\bibinfo {author} {\bibfnamefont {G.}~\bibnamefont
  {Prenat}}, \bibinfo {author} {\bibfnamefont {K.}~\bibnamefont {Jabeur}},
  \bibinfo {author} {\bibfnamefont {P.}~\bibnamefont {Vanhauwaert}}, \bibinfo
  {author} {\bibfnamefont {G.}~\bibnamefont {Di~Pendina}}, \bibinfo {author}
  {\bibfnamefont {F.}~\bibnamefont {Oboril}}, \bibinfo {author} {\bibfnamefont
  {R.}~\bibnamefont {Bishnoi}}, \bibinfo {author} {\bibfnamefont
  {M.}~\bibnamefont {Ebrahimi}}, \bibinfo {author} {\bibfnamefont
  {N.}~\bibnamefont {Lamard}}, \bibinfo {author} {\bibfnamefont
  {O.}~\bibnamefont {Boulle}}, \bibinfo {author} {\bibfnamefont
  {K.}~\bibnamefont {Garello}},  \emph {et~al.},\ }\href@noop {} {\bibfield
  {journal} {\bibinfo  {journal} {IEEE Transactions on Multi-Scale Computing
  Systems}\ }\textbf {\bibinfo {volume} {2}},\ \bibinfo {pages} {49} (\bibinfo
  {year} {2015})}\BibitemShut {NoStop}%
\end{thebibliography}%

\end{document}